\documentclass[12pt]{article}
 \usepackage{ifpdf}
 \usepackage{epsfig}
  \usepackage[latin1]{inputenc}
 \usepackage{amsfonts}
 \usepackage{amsthm}
 \usepackage{amssymb, latexsym, amsmath}
 \usepackage{amsbsy}
\usepackage{amstext}
\usepackage{amsopn}
\usepackage{amsgen}
\usepackage[dvips]{color}
 \usepackage{graphicx}
  \usepackage{color}
   \usepackage{wasysym}
   \usepackage{deluxetable}
  \usepackage{verbatim}
  \usepackage[T1]{fontenc}

 \newcommand{\inc}{{\it i}}

 \newcommand{\be}{\begin{equation}}
 \newcommand{\ee}{\end{equation}}
 \newcommand{\ba}{\begin{eqnarray}}
 \newcommand{\ea}{\end{eqnarray}}
 \newcommand{\bs}{\begin{subequations}}
 \newcommand{\es}{\end{subequations}}

 \newcommand{\erbold}{\mbox{{\boldmath $
 r$}}}
  \newcommand{\smallerbold}{\mbox{{\boldmath $
 _r$}}}
 \newcommand{\rbold}{\mbox{{\boldmath $
 r$}}}

 \newcommand{\omegabold}{\mbox{{\boldmath $
 {\omega}$}}}
 \newcommand{\dotomegabold}{\mbox{{\boldmath $\dot{
 {\omega}}$}}}
  \newcommand{\smallfirst}{\mbox{\small{\boldmath $_{{\small{{2}\omega\times}\dot{\small{u}}}}$}}}
   \newcommand{\smallsecond}{\mbox{\small{\boldmath $_{{\small{\omega\times(\omega\times r)}}}$}}}
  \newcommand{\smalldotomegabold}{\mbox{\small{\boldmath $_{\dot{\small{\omega}}}$}}}
  \newcommand{\smalllambdabold}{\mbox{\small{\boldmath $_{{\small{\lambda}}}$}}}
  
       \newcommand{\er}{\hat{{\bf{e}}}_{\mbox{\small{\boldmath $_{{\small{r}}}$}}}}
       
       \newcommand{\elambda}{\hat{{\bf{e}}}_{\mbox{\small{\boldmath $_{{\small{\lambda}}}$}}}}
       \newcommand{\evarphi}{\hat{{\bf{e}}}_{\mbox{\small{\boldmath $_{{\small{\varphi}}}$}}}}

 \newcommand{\xbold}{\mbox{{\boldmath $
 x$}}}

 \newcommand{\ubold}{\mbox{{\boldmath $
 u$}}}

 \newcommand{\ddotubold}{\stackrel{\bf{\centerdot\,\centerdot}}{\textbf {\mbox{\boldmath $
 {\boldmath u}$}}}}
 \newcommand{\dotubold}{\stackrel{\bf{\centerdot}}{\textbf {\mbox{\boldmath $
 {\boldmath u}$}}}}

 \newcommand{\Fbold}{\mbox{\boldmath $
 {\boldmath{F}}$}}

\begin{document}
  \title{
         ${{~~~~~~~~~~~~~~~~~~~~}^{^{^{
     %   To~be~
       {\rm   Published~in\,}
     %            Submitted~to
     %              ~~Celestial~Mechanics~and~Dynamical~Astronomy
     %              ~the~Astronomical~Journal~
                  ~Icarus
      \rm  \,,~Vol.~306\,,~pp.~328~-~354\,~(2018)
                  }}}}$\\
     %             ~
 {\Large{\textbf{{{
 Dissipation in a tidally perturbed body\\
 librating in longitude
 % I. Theory
 }
 ~\\
 \vspace{2mm}}
            }}}}
 \author{
            %  {\Large{Author 1~,}}\\
                                     {\Large{Michael Efroimsky}}\\
                                     {\small{US Naval Observatory, Washington DC 20392}}\\
                                     {\small{e-mail: ~michael.efroimsky$\,$@$\,$navy.mil~$\,$}}\\
            %                         ~\\
            %                         {\Large{and}}\\
            %                        ~\\
            %  {\Large{Author 2}}\\
            %                        {\Large{   }}\\
            %                        {\small{US Naval Observatory, Washington DC 20392}}\\
            %                        {\small{e-mail: ~  $\,$@$\,$  ~$\,$}}
 }
     \date{}

 \maketitle

 \begin{abstract}
 Internal dissipation in a tidally perturbed librating body differs in several respects from the tidal dissipation in a steadily spinning rotator.
 First, libration changes the spectral distribution of tidal damping across the tidal modes, as compared to the case of steady spin.
 This changes both the tidal heating rate and the tidal torque.
 Second, while a non-librating rotator experiences alternating deformation only due to the potential force exerted on it by the perturber, a librating body is also subject to
 a toroidal force proportional to the angular acceleration.
 Third, while the centrifugal force in a steadily spinning body renders only a permanent deformation (which defines the oblateness when the body cools down), in a librating body this force contains two alternating components~---~one purely radial, another a degree-2 potential force. Both contribute to heating, as well as to the tidal torque and potential (and, thereby, to the orbital evolution).

 We develop a formalism needed to describe dissipation in a homogeneous terrestrial body performing small-amplitude libration in longitude. This formalism incorporates as its part a linear rheological law defining the response of the rotator's material to forcing. While the developed formalism can work with an arbitrary linear rheology, we consider a simple example of a Maxwell material.

 We demonstrate that, independent of the rheology, forced libration in longitude can provide a considerable and even leading~---~and sometimes overwhelming~---~input in the tidal heating. Based on the observed parameters, this input amounts to 52\% in Phobos, 33\% in Mimas, 23\% in Enceladus, and 96\% in Epimetheus. This supports the hypothesis by Makarov \& Efroimsky (2014) that the additional tidal damping due to  forced libration may have participated in the early heating up of some of the large moons.
 As one possibility, such a moon could have been chipped by collisions~---~whereby it acquired a higher permanent triaxiality and, therefore, a higher forced-libration magnitude and, consequently, a higher heating rate. After the moon warms up, its permanent triaxiality decreases, and so does the tidal heating rate.
 \end{abstract}

 \section{Preliminaries\label{section1}}

 This paper continues the line of research started in our preceding works (Efroimsky \& Makarov 2014) and (Makarov \& Efroimsky 2014). In the former publication, we developed in detail the formalism to calculate tidal damping in a homogeneous near-spherical body. The main burden of that work was to generalise the classical theory of Peale \& Cassen (1978) to the case of a body spinning outside synchronism. In the paper by Makarov \& Efroimsky (2014), we provided several practical examples and pointed out a reason why libration may sometimes cause a noticeable increase in tidal heating. That conclusion, however, was qualitative and lacked rigour. In the current paper, we provide such an example based on an accurate calculation.

 We develop a theory of tidal dissipation in a homogeneous near-spherical object librating in longitude about an arbitrary (not necessarily 1:1) spin-orbit resonance. Since many celestial bodies are differentiated, the possibility of modeling such bodies with a homogeneous sphere depends on a goal set. For example, the subtleties of librational dynamics (the magnitudes and phases of the librational harmonics) are largely defined by the presence or absence of a global ocean (Thomas et al. 2016). So analysis of these magnitudes and phases requires that the ocean be taken into account. On the other hand, in calculations of dissipation the presence of such an ocean may be neglected. This was established by Chen, Nimmo and Glatzmaier (2014, Table 3) who demonstrated that in Enceladus the dissipation associated with the ocean is $\,23.2$ kW , a negligible fraction in the overall heat budget.

 The overall dissipation rate comprises four inputs. One is caused by toroidal displacements. The second is due to the alternating part of the purely radial deformation. The third is produced by the alternating part of the degree-two (quadrupole) component of the centrifugal force. The fourth is due to the ordinary tides caused directly by the gravity of the external perturber (the host star, if we are considering tides in a planet; or the host planet, if we are studying tides in a satellite).

 The power exerted by the quadrupole part of the centrifugal force can be calculated in the same manner as the power exerted by the quadrupole part of the external perturbing potential~---~which somewhat simplifies our work. On the other hand, the calculation of the input from the gravitational tides becomes more involved because in a librating body the spectral distribution of tidal damping changes across the tidal modes, as compared to a steadily spinning rotator. This alters (usually increases) the tidal heating rate and provides an addition to the tidal torque. As was demonstrated by Frouard \& Efroimsky (2017\,a), under libration in longitude the usual Kaula spectrum $\,\omega_{lmpq}\,$ must be substituted with a spectrum $\,\beta_{lmpqs}\,$ where the extra ``quantum number'' $\,s\,$ runs from $\;-\,\infty\,$ through $\,\infty\,$. This new spectrum incorporates the Kaula spectrum as its part: $\,\omega_{lmpq}\,=\,\beta_{lmpq0}\,$. For weak libration, it is sufficient to take into account just several terms with $\,s\,$ not very distant from zero.\,\footnote{~The paper by Frouard \& Efroimsky (2017\,a) dealt with the gravitational tides solely.
 The other types of deformation in a librating body were not considered there.}

 Our calculation is valid libration of a small magnitude (less than $\,\sim\,0.2$ rad).\,\footnote{~Forced libration can be large in amplitude for small satellites with the permanent triaxiality $\,(B-A)/C\,$ approaching or exceeding $\,0.3\,$. This situation lies outside the scope of our treatment.}

 Aside from the ramifications for the heat budget, the knowledge of the libration-caused dissipation allows for the calculation of the ensuing corrections to the orbit evolution.
 %  , a topic to be addressed in the next paper (Efroimsky 2018).

 \section{Setting}

 We consider a near-spherical extended body of a mean radius $\,R\,$, mass $\,M\,$, and the principal moments of inertia $\,A\leq B<C\,$. The body is spinning about its major-inertia axis (the one related to the maximal moment of inertia $\,C\,$), and is captured in one of the spin-orbit resonances with a perturber of mass $\,M^*\,$. The body is always trying to align its long (minimal-inertia) axis with the instantaneous direction to the perturber~---~which gives birth to a restoring torque proportional to the permanent triaxiality $\,(B-A)/C\,$. This causes forced libration imposed upon the regular spin. Our study addresses libration in longitude. Libration in latitude may be neglected at low obliquities. We also neglect the episodes of wobble.\,\footnote{~Although such episodes may have happened to satellites due to intense collisions during bombardment, estimates by Frouard \& Efroimsky (2017\,b, Section 5.2) show that wobble relaxation of a near-oblate Enceladus-size icy body requires a relatively short time span (thousands to hundreds of thousands years, dependent on the ice viscosity). Validity of this conclusion for silicate planets and moons should, however, be checked on a case-to-case basis, because their mean viscosity may exceed that of ice by orders of magnitude.}

 Here we present some key formulae to be needed in the subsequent sections. More detail on this material can be found in Frouard \& Efroimsky (2017\,a).

 \subsection{The equation of motion}

\begin{figure}[h]
 \begin{minipage}{135mm}
 \includegraphics[width=135mm]{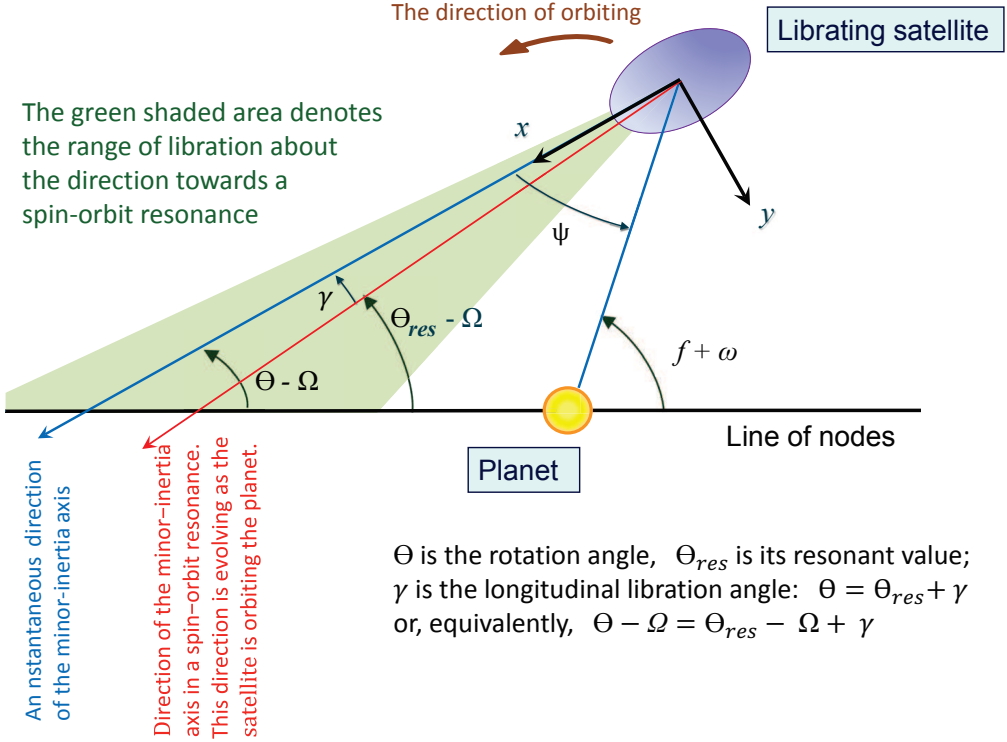}
  \end{minipage}
 \caption{\small{The principal axes $\,x\,$ and $\,y\,$ of the extended body relate to the minimal and middle moments of inertia, correspondingly. The horizontal line is that of nodes. The true anomaly is denoted with $\,f\,$, the argument of the pericentre with $\,\omega\,$; so their sum $\,f+\omega\,$ is the angle between the line connecting the bodies and the line of nodes. The rotation angle of the body, $\,\theta\,$, is reckoned from the same fiducial direction as the node $\,\Omega\,$; so the difference $\,\theta-\Omega\,$ is the angle between the minimal-inertia axis $\,x\,$ and the line of nodes. For a finite inclination (obliquity) $\,i\,$, the angles $\,f+\omega\,$ and $\,\theta-\Omega\,$ belong to different planes. However, in the limit of vanishing $\,i\,$, they are in the same plane, and their difference
 %  In neglect of the apsidal precession, $\,\theta\,$ is the sidereal angle of the planet.
 $\,\psi=(f+\omega)-(\theta-\Omega)\,$ gives the separation between the minimal-inertia axis $\,x\,$ and the direction towards the perturber. (For details, see Frouard \& Efroimsky 2017\,a.)
 The red line denotes the minor-inertia axis' direction in a spin-orbit resonance. The blue line denotes this axis' instantaneous direction under libration. The angular separation between these directions is the libration angle $\,\gamma\,$ which can assume values within the green shaded area.\vspace{4mm}}
 \label{Figure}}
 \end{figure}

 In Figure \ref{Figure}, the rotation angle $\,\theta\,$ of an extended body is the separation between some fiducial direction and the largest-elongation axis $\,x\,$, the one corresponding to the minimal moment of inertia $\,A\,$.
 The external perturber exerts on the body a torque $\,\vec{\mathcal{T}}^{^{(TRI)}}$  due to the body's permanent triaxiality,
 and a torque $\,\vec{\mathcal{T}}^{^{(TIDE)}}\,$ due to tides in the body. The longitudinal motion is then defined by these torques' polar components, so the rotation angle $\,\theta\,$ obeys the equation
  \ba
 \stackrel{\bf\centerdot\,\centerdot}{\theta~}\,=~\frac{\,{\cal{T}}^{\rm{^{\,(TRI)}}}_{polar}\,+~{\cal{T}}^{\rm{^{\,(TIDE)}}}_{polar}}{C~~}~\,.
 %=~\frac{{\cal{T}}^{\rm{^{\,(TRI)}}}_z\,+~
 %{\cal{T}}^{\rm{^{\,(TIDE)}}}_z}{\xi ~M_{2}~R^{\,2}}\,~,
 \label{eq.eq}
 \label{201}
 \label{1}
 \label{eq:despinning}
 \ea
 An approximate description of the longitudinal libration in a spin-orbit resonance is derived by neglecting the tidal torque and reducing the ensuing equation $\,C \ddot{\theta} = {\cal{T}}^{\rm{^{\,(TRI)}}}_{polar}\,$ to a forced oscillator (which becomes harmonic at not too large magnitudes, see Frouard \& Efroimsky 2017\,a). The tidal torque is then introduced as a small correction:
 \ba
 |\,\vec{\cal{T}}^{\rm{^{\,(TIDE)}}}\,|\,\ll\;|\,\vec{\cal{T}}^{\rm{^{\,(TRI)}}}\,|\;\,.
 \label{inequality}
 \ea
 Obeyed in realistic situations (see Appendix \ref{comparison}), this inequality justifies our setting the moments of inertia to be constant in time.$\,$\footnote{~Tides induce periodic variations of the components of the matrix of inertia (and, therefore, periodic variations of the direction of the principal axes). We, however, assume these effects to be of a higher order of smallness, owing to the \,{\it{a priori}}\, assumption of the tides being much weaker than the permanent triaxiality.}

 Although the tidal torque is usually small, its importance lies in its participation in production of tidal heat. We say ``participation'' because in a librating body energy is dissipated also by the alternating parts of the toroidal and centrifugal deformation.

 \subsection{Definition of a spin-orbit resonance\label{definition}}

 Usually in physics a resonance is a situation where two frequencies of a system are commensurable. In the case of spin-orbit coupling, however, a resonance imposes a condition not only on the angles' rates but also on their actual $\,${\it{values}}. For example, when we say that the Moon is presently in a 1:1 spin-orbit resonance, not only do we imply that it spins in agreement with orbiting, but also we mean that its longer axis points (approximately) towards the Earth centre. So we write the resonance condition through the angles, not their rates.

 In a Kaula-type expansion of the ellipsoidal part of the permanent-triaxiality-caused torque (Frouard \& Efroimsky 2017\,a), an $\,lmpq\,$ term vanishes when the rotation angle assumes the value of
  \ba
 {\theta}_{\rm{res}}\,=~\frac{l\,-\,2\,p\,+\,q}{m}\;{\cal{M}}\;+\;\frac{l\,-\,2\,p}{m}\;{\omega}\,+\,\Omega\,+\,\frac{N}{m}\,\pi\,~,
 \label{}
 \ea
  where $\,{\cal{M}}\,$ is the mean anomaly, $\,\omega\,$ is the argument of the pericentre, $\,\Omega\,$ is the longitude of the node. The rotation angle $\,\theta\,$ and the node $\,\Omega\,$ must be reckoned from the same fiducial direction, usually the vernal equinox. The number $\,N\,$ is integer.

 When a triaxial rotator gets captured in this resonance and is librating in it, as shown in Figure \ref{Figure}, the rotation angle $\,\theta\,$ becomes a sum of the resonant part $\,\theta_{\rm{res}}\,$ and a longitudinal libration angle $\,\gamma\;$:
 \ba
 {\theta}\,=\,\theta_{\rm{res}}\,+\,\gamma\,=~z\,{\cal{M}}\;+\;z\,'\,{\omega}\,+\,\Omega\,+\,\frac{N}{m}\,\pi\,+\,{{\gamma}}\,~,
 \label{condition}
 \label{2}
 \ea
 where the rational numbers $\,z\,$ and $\,z\,'\,$ are related to the integers $\,lmpq\,$ from Kaula (1964) via
 \ba
 z\;=\;\frac{l\,-\,2\,p\,+\,q}{m}\quad,\quad z\,'\,=\;\frac{l\,-\,2\,p}{m}\;\;,
 \label{3}
 \ea
 while $\,\gamma\,$  includes a small constant bias due to the $\,l>2\,$ terms, see Williams et al. (2014).

 Also, the vanishing of the time-derivative of the condition (\ref{condition}) warrants the vanishing of the $\,lmpq\,$ term in the Kaula series for the tidal torque. So the condition (\ref{condition}) can serve as the definition of a resonance for both torques.

 When the inclination is small, only the terms with $\,p=0\,$ are important.
 If, along with that, only the quadrupole parts of the torques are needed ($\,l=m=2\,$), the definition of a resonance becomes
 \ba
 {\theta}~=~z\,{\cal{M}}\,+\,{\omega}\,+\,\Omega\,+\,N\,\frac{\pi}{2}\,+\,{{\gamma}}\,~,
 \label{gora}
 \label{4}
 \ea
 with $\,z\,$ now being integer or semiinteger: $\,z=1\,+\,\frac{\textstyle q}{\textstyle 2}\,$.

 % We prefer to write the resonance condition in the form of equalities (\ref{condition}) or (\ref{gora}), though it is more common in literature to write this condition as a time derivative
 % of these equalities.

 The writing of the resonance condition not through the angles' rates but through the angles' values, as in equations (\ref{condition}) and (\ref{gora}), enables us to keep the term $\,N\pi/m\,$. This term serves to remind that, dependent upon the eccentricity, libration may be taking place about different values of the angle $\,\theta\;$. \footnote{~For example, at high eccentricities, a rotation regime is possible wherein the body traverses the pericentre being sidewards to the perturber, see Section 2.2.2 in Frouard \& Efroimsky (2017\,a) for details. Mentioned for the first time by Goldreich \& Peale (1968), this possibility was also addressed by Makarov (2012) who explored the case of an initially retrograde Moon getting into the \,1:2\, and (seldom) the \,1:1\, resonances, with its long axis librating about $\,\pi/2\,$ (sidewise) with respect to the direction towards the Earth.

 Also, it would be important to emphasise that of a special importance are the angles' values at the moment of passing the pericentre. As was pointed out by Makarov (2012), it is the value of $\,\theta\,$ in the pericentre that determines whether the rotator transcends a spin-orbit resonance or gets stuck in it.}

 \subsection{Examples of forced libration in longitude\label{examples}}

 In what follows, we always assume that the body has already been caught in a spin-orbit resonance $\,z\,$. In this resonance, the forced libration in longitude can be expressed as a sum over positive integers:
 \ba
 \gamma(t)\;=\;\sum_{j=1}^{\infty}\;^{(z)}{\cal{A}}_j\;\sin(j n t)\,\;,
 \label{follows}
 \label{5}
 \ea
 where $\,n\,\equiv\,\stackrel{\bf\centerdot}{\cal{M}\,}$ is the anomalistic mean motion, while the superscript $\;^{(z)}\;$ serves to emphasise that the magnitude $\,^{(z)}{\cal{A}}_j\,$ of the $\,j$-th mode will be different in a different spin-orbit resonance.

 In a spin-orbit resonance $\,z\,$, the magnitudes of the libration modes are given by
 \ba
 ^{(z)}{\cal{A}}_j & = & \;\omega_0^2\;\frac{G_{20(j+2z-2)}(e) - G_{20(-j+2z-2)}(e)}{^{(z)}\chi^2\,-\,j^2\,n^2}
 %  \label{} ~\\  \nonumber\\  & \approx & -\;\omega_0^2\;\frac{G_{20(j+2z-2)}(e) - G_{20(-j+2z-2)}(e)}{j^2\,n^2}
 \,\;.
 \label{eq2}
 \ea
 The standard notation $\,\omega_0^2\,$ stands for the quantity
 \bs
 \ba
 \omega_0^2 & \equiv & \frac{3}{2}~\frac{B-A}{C}~\frac{G\;M^*}{a^3}
 \label{7a}\\
 \nonumber\\
 &= &\frac{3}{2}~\frac{B-A}{C}~~\frac{G\,(M^*\,+\,M)}{a^3}\;\frac{M^*}{M^*\,+\,M}\;\approx\;\frac{3}{2}~\frac{B-A}{C}~n^2
 \;\,,
 \label{7b}
 \ea
 \label{omega0}
 \label{7}
 \es
 where $\,C>B\geq A\,$ are the moments of inertia, $\,M^{\,*}\,$ is the perturber's mass; and the osculating mean motion $\,\sqrt{G(M^*+M)/a^3\,}
 % \approx\sqrt{GM^*/a^3\,}
 \,$ is approximated with its anomalistic counterpart $\,n\equiv\,\stackrel{\bf{\centerdot}}{\cal{M}\,}$.

 The notation $\,^{(z)}\chi\,$ stands for the quantity
 \ba
 ^{(z)}\chi\;=\;\omega_0\;\sqrt{2\,G_{20(2z-2)}(e)\,}
 %  &=&\sqrt{\,3~\frac{B-A}{C}~\frac{G\;M^*}{a^3}\;G_{20(2z-2)}(e)\,}
 %  \label{}\\ \nonumber\\
 \;\approx \; n\,\sqrt{\,3~\frac{B-A}{C}\;G_{20(2z-2)}(e)\,}
 \,\;,\;\;\;
 \label{8}
 \label{publish}
 \ea
 which happens to be the frequency of weak sinusoidal free libration in a spin-orbit resonance $\,z\,$.

 Epimetheus being a rare exception (Tiscareno et al. 2009), in most practical situations $\,^{(z)}\chi\,$ obeys the strong inequality $\,^{(z)}\chi\ll n\,$. Under this condition, in the 1:1 spin-orbit resonance ($z=1$), the forced libration angle evolves as
 \ba
 \begin{split}
^{(1:1)}\gamma({\cal{M}}) \simeq  \frac{\omega_0^2}{n^2} \bigg[
\bigg( -4 e  + \frac{31}{4} e^3 + \mathcal{O}(e^5) \bigg) \sin {\cal{M}}
+ \bigg( - \frac{17}{8} e^2  + \mathcal{O}(e^4) \bigg) \sin 2{\cal{M}} \\
+ \bigg( - \frac{211}{108} e^3 + \mathcal{O}(e^5) \bigg) \sin 3{\cal{M}}
+ \mathcal{O}(e^4) \bigg]\,\;,\;\;\;\;
\end{split}
\label{1:1}
\ea
 while in the 3:2 situation ($z=3/2$) it is given by
\ba
\begin{split}
^{(3:2)}\gamma({\cal{M}}) \simeq  \frac{\omega_0^2}{n^2} \bigg[
\bigg( 1 - 11 e^2  + \mathcal{O}(e^4) \bigg) \sin {\cal{M}}
+ \bigg( - \frac{1}{8} e  - \frac{421}{96} e^3 + \mathcal{O}(e^5) \bigg) \sin 2{\cal{M}} \\
+ \bigg( \mathcal{O}(e^4) \bigg) \sin 3{\cal{M}}
+ \bigg(  \frac{1}{768} e^3   + \mathcal{O}(e^5) \bigg) \sin 4{\cal{M}}
+ \mathcal{O}(e^4) \bigg]\,\;.\;\;
\end{split}
\label{3:2}
\ea

 We see that in the 1:1 spin-orbit state the principal mode of forced libration is
 \ba
 ^{(1:1)}{\cal{A}}_1\;=\;-\;6\;e\;\frac{B-A}{C}\;+\;O(e^3)\,\;,
 \label{10}
 \ea
 while in the 3:2 resonance it is given by
 \ba
 ^{(3:2)}{\cal{A}}_1\;=\;\frac{3}{2}\;\frac{B-A}{C}\;+\;O(e^2)\,\;,
 \label{11}
 \ea
  both these expressions being valid under the approximation of $\,^{(z)}\chi\ll n\,$.

  The ``minus'' sign in the expression for $\;^{(1:1)}{\cal{A}}_1\;$ indicates that under synchronism the principal (first) harmonic of forced libration is in a counterphase with the corresponding harmonic of the permanent-triaxiality-caused torque. For details, see, e.g., Frouard \& Efroimsky (2017\,a) and references therein.

  \begin{deluxetable}{lr}
 \tablecaption{Symbol key \label{Table}}
 \tablewidth{0pt}
 \tablehead{
 \multicolumn{1}{c}{Notation}  &   \multicolumn{1}{c}{Description}
 }
 \startdata
 $\rho$ & \dotfill the mean density of the extended body \\
 $M$ & \dotfill the mass of the extended body  \\
 $M^*$ & \dotfill the mass of the perturber \\
 $C>B\geq A$ & \dotfill the moments of inertia of the extended body \\
    %    $\xi$ & \dotfill the moment of inertia coefficient of the extended body\\
 $R$ & \dotfill the mean radius of the extended body \\
 ${\cal{T}}^{\rm{^{\,(TRI)}}}_{polar}$ & \dotfill
  $\,.\,$
 the polar component of the permanent-triaxiality-caused torque on the extended body\\
  ${\cal{T}}^{\rm{^{\,(TIDE)}}}_{polar}$ & \dotfill the polar component of the tidal torque acting on the extended body\\
 $a$ & \dotfill the semimajor axis \\
   %   $f$ & \dotfill the true anomaly \\
 $e$ & \dotfill the orbital eccentricity \\
 ${\cal{M}}$ & \dotfill the mean anomaly \\
 $n\equiv\,\stackrel{\bf\centerdot}{\cal{M}\,}$ & \dotfill the anomalistic mean motion ~\\
 $\theta$ & \dotfill the rotation angle of the extended body\\
 $\stackrel{\bf{\centerdot}}{\theta\,}$ & \dotfill the rotation rate of the extended body\\
 $\omegabold$ & \dotfill the angular velocity (so $\,|\,\omegabold\,|\,=\,\stackrel{\bf{\centerdot}}{\theta\,}$) \\
 $z$ & \dotfill the number of a spin-orbit resonance ~\\
 $G$ & \dotfill Newton's gravitational constant \\
 $\gamma$ & \dotfill the libration angle of the extended body\\
 ${\cal{A}}$ & \dotfill the magnitude of libration of the extended body\\
 $\ubold_{\textstyle{_{\rm{tide}}}}$ & \dotfill the alternating component of the tidal part of the displacement field\\
 $\ubold_{\textstyle{_{\rm{rad}}}}$ & \dotfill the alternating component of the radial part of the displacement field\\
 $\ubold_{\textstyle{_{\rm{tor}}}}$ & \dotfill the toroidal part of the displacement field\\
 $\ubold$ & \dotfill the alternating component of the total displacement
 % in the extended body
 % $~$      & \dotfill
 (so $\;\ubold=\ubold_{\textstyle{_{\rm{tide}}}}+\ubold_{\textstyle{_{\rm{rad}}}}+\ubold_{\textstyle{_{\rm{tor}}}}\,$)\\
 $\rbold$ & \dotfill position in the centre-of-mass frame of the extended body\\
 $W_l$ & \dotfill the degree-$l$ Legendre harmonic of the tide-raising potential\\
 $U_l$ & \dotfill the degree-$l$ Legendre harmonic of the additional potential due to tidal deformation\\
 $W_2^{(cent)}$ & \dotfill the oscillating quadrupole part of the centrifugal potential\\
 $U_2^{(cent)}$ & \dotfill the oscillating quadrupole part of the additional potential \\
 $\,$ &   due to the tidal deformation caused by the centrifugal force\\
 $W_{\rm{(rad)}}$ & \dotfill the oscillating radial part of the potential due to the centrifugal force\\
 $U_{\rm{(rad)}}$ & \dotfill the oscillating radial part of the additional potential due\\
 $\,$ &  to the tidal deformation caused by the centrifugal force\\
 $\,U_{\smallsecond}\,$ & \dotfill the oscillating additional potential due to the deformation caused\\
 $\,$ & by the centrifugal force (so $\;U_{\smallsecond}\,=\,U_2^{(cent)}\,+\,U_{\rm{(rad)}}\;$)\\
 $\,U_{\smallfirst}\,$ & \dotfill the oscillating component of the additional potential due to\\
 $\,$ & the deformation caused by the Coriolis force\\
 $\mu,\;K$ & \dotfill the shear and bulk moduli of the extended body\\
 $\eta,\;\zeta$ & \dotfill the shear and bulk viscosities of the extended body\\
 \enddata
 \end{deluxetable}

 \section{Forces acting in a rotating body\label{section3}}

 \subsection{Generalities\label{section3.1}}

 As demonstrated in Appendix \ref{A}, in the reference frame of the centre of mass the equation of motion for a small parcel of material of a rotating body is
 %  Denoting the libration frequency with $\,\chi\,$, we see that the terms on the left-hand side are of the order of $\,\rho\,\chi^2\,u~$,
 %  $~\rho\,\omega\,\chi\,u~$, $~\rho\,\omega^2\,r\,$, ~and $~\rho\,\dot{\chi}\,\omega\,r~$, respectively. From this, we observe that in realistic
 %  settings the first two terms are negligible. Omitting them, we arrive at
 \ba
 \nonumber
 \rho\,\ddotubold&=&\nabla{\mathbb{S}}
 %  \,-\,\nabla p
 \,-\,\rho\,\sum_{l=2}^\infty\nabla(U_l\,+\;W_l)~-~\rho\,\dotomegabold\times\rbold\\
 \label{eq}
 &-&2\,\rho~\omegabold\times\dotubold\;-\;\rho\,\nabla U_{\smallfirst}
 ~-~\rho~\omegabold\times\left(\omegabold\times\rbold\right)^{(oscill)}\;-\;\rho\,\nabla U_{\smallsecond}~~,\quad~
 \ea
 where $\,\rho\,$ is the density, $\,\rbold\,$ is the position in the centre-of-mass frame, $\,\omegabold\,$ is the angular velocity, $\,\ubold\,$ is the oscillating part of the displacement, $\,{\mathbb{S}}\,$ is the oscillating part of the stress,
 %  $\,p\,$ is the oscillating part of the pressure,
 and $\;-\,\rho\,\omegabold\times\left(\omegabold\times\rbold\right)^{(oscill)}\,$ is the oscillating part of the centrifugal force (see equation \ref{centrip} below). For purely longitudinal libration, the angular velocity is aimed along the polar axis of the body and is related to the rotation angle though $\,|\omegabold|=\,\stackrel{\bf\centerdot}{\theta\,}$.

 The notations $\,W_l\,$ and $\,U_l\,$ stand for the degree-$l\,$ Legendre harmonics of the external potential and the additional tidal potential due to the tidal deformation. They are related by
 \ba
 U_l\,=\,\hat{k}_l\,W_l\,\;,
 \label{oper}
 \ea
 where $\,\hat{k}_l\,$ is an integral operator called the $\,${\it{Love operator}} (Peltier 1974, Wu \& Peltier 1982, Sabadini \& Vermeersen 2016, Efroimsky 2012$\,$a,$\,$b).

 The toroidal force $\,~-\,\rho\,\dotomegabold\times\rbold\,$ produces no change of shape and no incremental potential either, provided the body's shape has rotational symmetry (see Footnote \ref{Thomson} below).
 In our case, the body has a small permanent triaxiality, so the toroidal-force-generated changes of the shape and potential may be neglected.
 The Coriolis force $\;-\,2\,\rho\,\omegabold\times\dotubold\,$ generates a shape variation resulting in an additional potential $\,U_{\smallfirst}\,$.

 The centrifugal force $\;-\,\rho\,\omegabold\times\left(\omegabold\times\rbold\right)\,$, too, gives birth to a shape distortion rendering an additional potential. This force (and, likewise, the ensuing distortion and potential) comprises a constant part and an oscillating part. The constant parts
 of both the centrifugal force and the force due to the extra potential are compensated by the constant part of the stress; so all these constant inputs may be omitted in our analysis. We shall be interested only in the oscillating part of the centrifugal force, $\;-\,\rho\,\omegabold\times\left(\omegabold\times\rbold\right)^{(oscill)}\;$, as well as in the corresponding oscillating change of shape, and in the resulting oscillating extra potential $\,U_{\smallsecond}\,$.

 \subsection{Deviatoric and volumetric stresses}

 The stress comprises a deviatoric and a volumetric component (e.g., Landau \& Lifshitz 1986):
 \ba
 {\mathbb{S}}\;=\;{\mathbb{S}}_{dev}\,+\;{\mathbb{S}}_{vol}\,\;.
 \label{}
 \ea

 The deviatoric stress $\,{\mathbb{S}}_{dev}\,$ is related to the deviatoric strain $\,\varepsilon_{dev}\,$ via $\;{\mathbb{S}}_{dev}\;=\;2\;\mu\;\varepsilon_{dev}\;$. While generally the rigidity modulus $\,\mu\,$ is a twice covariant and twice contravariant tensor, in isotropic media it gets reduced to a scalar. Under an additional assumption of homogeneity of the medium, $\,\mu\,$ can be moved outside the gradient operator:
 \ba
 \nabla{\mathbb{S}}_{dev}\;=\;2\;\mu\;\nabla\varepsilon_{dev}\,\;.
 \label{sec}
 \ea
 The volumetric stress (related to the pressure through $\,p\,{\mathbb{I}}\,=\,-\,{\mathbb{S}}_{vol}\,$) is linked to the volumetric strain by $\;{\mathbb{S}}_{vol}\;=\;3\;K\;\varepsilon_{vol}\;$, $\,$where $\,K\,$ is the bulk rigidity modulus. Accordingly,
 \ba
 \nabla{\mathbb{S}}_{vol}\,=\;3\;K\;\nabla\varepsilon_{vol}\,\;.
 \label{fir}
 \ea
 Being numbers in static problems, both $\mu$ and $K$ become operators in settings evolving in time.

 %  In a static problem, these potentials can be expressed as
 %  \ba
 %  W(\rbold,\,\rbold^{\,*})\;=\;\sum_{l=2}^{\infty}W_l(\rbold,\,\rbold^{\,*})\quad,\qquad
 %  W_l(\erbold,\,\erbold^*)\;=\;-\;\frac{G\,M_{sec}}{r^*}\;\left(\frac{r}{r^*}\right)^l\,P_l(\cos\gamma)~~~,
 %  \label{w}
 %  \ea
 %  \ba
 %  U(\rbold,\,\rbold^{\,*})\;=\;\sum_{l=2}^{\infty}U_l(\rbold,\,\rbold^{\,*})\quad,\qquad
 %  U_l(\erbold,\,\erbold^*)\;=\;k_l\;\left(\frac{\,r}{R}\right)^{l+1}\;W_l(\Rbold,\,\erbold^*)~~~,
 %  \label{u}
 %  \ea
 %  $M^{\,*}\,$ being the mass of the perturber, $\,\erbold^{\,*}$ being the vector pointing from the centre of mass of the body towards the perturber;
 %  $\,\rbold\,$ is the point where a potential is measured; and $\,\gamma\,$ being the angle between $\,\erbold\,$ and $\,\erbold^{\,*}\,$, subtended at
 %  the centre of mass of the body. The notation $\,k_l\,$ stands for the static Love numbers. Be mindful that equation (\ref{u}) contains the value
 %  of the perturbing potential at the surface point $\,\Rbold\,$ located right beneath the exterior point $\,\rbold\,$.

 \subsection{The oscillating part of the centrifugal force\label{centripetal}}

 It is explained in Appendix \ref{cen} that the oscillating part of the centrifugal force is
 \ba
 -\;\rho~\omegabold\times\left(\omegabold\times\rbold\right)^{(oscill)}=\;-\;\rho~\omegabold\times\left(\omegabold\times\rbold\right)\;-\;
 \langle\;-\;\rho~\omegabold\times\left(\omegabold\times\rbold\right)\;\rangle
 % \qquad\qquad\qquad\qquad\qquad\qquad\qquad\qquad\qquad\qquad
 \qquad\qquad\qquad\qquad
 \label{pell}\\
 \nonumber\\
 \nonumber
 =\;-\;\nabla\left[~\frac{\rho}{3}~\left(\omegabold^{\,2}\,-\;\langle\,\omegabold^{\,2}\,\rangle
 \right)
 \,\erbold^{\,2}
 \,P_2\left(\cos\varphi\right)~\right]
 ~+~\nabla\;\left[~\frac{\rho}{3}~\left(\omegabold^{\,2}\,-\;\langle\,\omegabold^{\,2}\,\rangle
 \right)\,\erbold^{\,2}~\right]~~~,
  \ea
 where angular brackets denote time averaging.

 For purely longitudinal libration, both the angular velocity $\,\omegabold\,$ and its rate $\,\dotomegabold\,$ are aimed along the polar axis.
 Consider a particular Fourier mode $\,\chi\,$ of the libration angle $\,\gamma\,$:
 \ba
 \gamma\;=\;{\cal{A}}\;\sin \chi t\,\;,\qquad
 \omegabold\;\equiv\;\hat{\bf{e}}_z\stackrel{\bf\centerdot}{\theta\,}\,=\,
 \hat{\bf{e}}_z\,(\,\stackrel{\bf\centerdot}{\theta\,}_{res}\,+\,\stackrel{\bf\centerdot}{\gamma\,})\,
 =\;\hat{\bf{e}}_z\,[\,\stackrel{\bf\centerdot}{\theta\,}_{res}+\,\chi\;{\cal{A}}\;\cos \chi t\,]\,\;.
 \label{main}
 \ea
 As shown in Appendix \ref{cen}, libration at the frequency $\,\chi\,$ makes expression (\ref{pell}) look as
 \ba
 -\;\rho~\omegabold\times\left(\omegabold\times\rbold\right)^{(oscill)}=\,-\,\rho\,\nabla W_2^{(cent)}
 \,-\,\rho\,\nabla W_{\textstyle{_{\rm{rad}}}}\;~~,\;\;\;
 \label{centrip}
 \ea
 where the time-variable part of the centrifugal degree-2 perturbation is
 \ba
 W_2^{(cent)}=\;\frac{1}{6}~{\cal{A}}^2\;\chi^2\,\erbold^{\,2}~P_2\left(\cos\varphi\right)\;\cos 2 \chi t
 \;+\;\frac{2}{3}~{\cal{A}}\;\chi\;\dot{\theta}_{res}\,\erbold^{\,2}~P_2\left(\cos\varphi\right)~\cos \chi t
 \,\;,
  \label{}
 \ea
 while the time-variable part of the purely radial perturbation is given by
 \ba
 W_{\textstyle{_{\rm{rad}}}}=\;-\;\frac{1}{6}\;{\cal{A}}^2\;\chi^2\;\erbold^{\,2}\;\cos 2 \chi t
 \;-\;\frac{2}{3}~{\cal{A}}\;\chi\;\dot{\theta}_{res}\,\erbold^{\,2}~\cos \chi t
 \,\;.
  \label{radrad}
 \ea
 (We call $\,W_{\textstyle{_{\rm{rad}}}}\,$ purely radial, for it gives birth to a force that has a radial component only.)
 %  This way, libration at the frequency $\,\chi\,$ creates in expression (\ref{pell}) two harmonics, one at the frequency $\,\chi\,$ and one at $\,2\chi\,$.

 Both perturbations distort the shape and create incremental potentials. The degree-2 incremental potential $\,U_2^{(cent)}\,$ is related to $\,W^{(cent)}_2\,$ through the degree-2 Love operator
 $\;\hat{k}_2\;$:
 \ba
 U_2^{(cent)}\,=\,\hat{k}_2\,W_2^{(cent)}\,~.
 \label{}
 \ea
 The radial incremental potential is expressed via $\,W_{\textstyle{_{\rm{rad}}}}\,$ by the radial Love operator $\;\hat{k}_{\textstyle{_{\rm{rad}}}}\;$:
  \ba
 U_{\textstyle{_{\rm{rad}}}}\,=\,\hat{k}_{\textstyle{_{\rm{rad}}}}\,W_{\textstyle{_{\rm{rad}}}}\,\;.
 \label{}
 \ea
 In the time domain, both $\,\hat{k}_2\,$ and $\,\hat{k}_{\textstyle{_{\rm{rad}}}}\,$ are convolution operators, while in the frequency domain the above two expressions are simply products of Fourier components  (Efroimsky 2012$\,$a,b).

 \subsection{Splitting of the deformation into components}

 With the above developments taken into account, the equation of motion becomes:
 \ba
 \nonumber
 \rho\,\ddotubold&=&2\,\mu\,\nabla\varepsilon_{dev}\,+\,3\,K\,\nabla\varepsilon_{vol}\,-\,\rho\,\nabla\,\left[\,\left(\,U_2^{(cent)}\,+\,W_2^{(cent)}\,\right) \;+\;\sum_{l=2}^\infty(U_l\,+\;W_l)\,\right] ~\\
 \label{eq}
 &-&\rho~\nabla\left(U_{\textstyle{_{\rm{rad}}}}\,+\,W_{\textstyle{_{\rm{rad}}}}\right)
 ~-~\rho\,\dotomegabold\times\rbold\,-\,2\,\rho~\omegabold\times\dotubold\;-\;\rho\,\nabla U_{\smallfirst}
 ~~.\quad~
 \ea
 In the linear approximation, the displacement field expands into
 \ba
 \ubold\;=\;\ubold_{\textstyle{_{\rm{tide}}}}\,+\;\ubold_{\textstyle{_{\rm{rad}}}}\,+\;\ubold_{\textstyle{_{\rm{tor}}}}\,\;,
 \label{}
 \ea
 $\ubold_{\textstyle{_{\rm{tide}}}}\,$, $\,\ubold_{\textstyle{_{\rm{rad}}}}\,$ $\,\ubold_{\textstyle{_{\rm{tor}}}}\,$ being the tidal, radial, and toroidal components.
 With $\,\varepsilon_{\textstyle{_{\rm{tide}}}}\,$, $\,\varepsilon_{\textstyle{_{\rm{rad}}}}\,$ $\,\varepsilon_{\textstyle{_{\rm{tor}}}}\,$ being the corresponding components of the strain,
 we then obtain:
 \ba
 \label{}
 \varepsilon\;=\;
 % \overbrace
 {{\varepsilon}_{\textstyle{_{\rm{tide}}}}\,+\,{\varepsilon}_{\textstyle{_{\rm{rad}}}}}
 % ^{\stackrel{\textstyle\varepsilon_{dev}}{}}
 \,+\,
 % \overbrace
 {{\varepsilon}_{\textstyle{_{\rm{tor}}}}}
 %  ^{\stackrel{\textstyle\varepsilon_{vol}}{}}
 \,\;.
 \ea
 The tidal and toroidal deformations are deviatoric, because they do not change the volume.\,\footnote{~We shall use the standard solution to the tidal problem. Traditionally misattributed to Love, though actually pioneered by Thomson, that solution was written for a homogeneous incompressible sphere~---~for which reason the resulting expressions for the Love numbers contain only the shear modulus $\,\mu\,$, but not the bulk modulus $\,K\,$.

 The toroidal part of the deformation is, by definition, a part expandable over the functions $\,T^m_n\,\equiv\,{\bf{r}}\times\nabla Y^m_n\,$. For such functions, $\,\nabla\cdot T^m_n\,=\,0\,$, so the volume stays constant under such deformation.
 \label{Thomson}}
 The radial deformation changes the volume and, therefore, includes a bulk part:\,\footnote{~Along with a bulk part, the radial deformation incorporates also a deviatoric part. Its presence is indicated by the emergence of the shear modulus $\,\mu\,$ in the solution (\ref{urad}) for $\,\ubold_{\textstyle{_{\rm{rad}}}}\,$.}
 \ba
 \varepsilon_{\textstyle{_{\rm{rad}}}}\,=\,\varepsilon_{\textstyle{_{\rm{rad}}}_{\;({dev})}}\,+\,\varepsilon_{\textstyle{_{\rm{rad}}}_{\;({vol})}}\,\;.
 \label{}
 \ea
 The law of motion will now split into three:
 \ba
 \nonumber
 \rho\,\ddotubold_{\textstyle{_{\rm{tide}}}}\;=\;2\;\mu\;\nabla\varepsilon_{\textstyle{_{\rm{tide}}}}&-&\rho\,\nabla\,\left[\,\left(\,U_2^{(cent)}\,+\,W_2^{(cent)}\,\right) \;+\;\sum_{l=2}^\infty(U_l\,+\;W_l)\,\right] ~\\
 \label{raz}
 &-&\,2\,\rho~\omegabold\times\dotubold\;-\;\rho\,\nabla U_{\smallfirst}
 ~~,\quad~
 \ea
 \ba
 \rho\,\ddotubold_{\textstyle{_{\rm{rad}}}}\,=\,2\,\mu\;\nabla\varepsilon_{\textstyle{_{\rm{rad}}}_{\;({dev})}}\,+\;3\;K\;\nabla\varepsilon_{\textstyle{_{\rm{rad}}}_{\;({vol})}}\;-\;\rho~\nabla\left(U_{\textstyle{_{\rm{rad}}}}\,+\,W_{\textstyle{_{\rm{rad}}}}\right)
 ~~,\,\quad
 \label{dva}
 \ea
 \ba
 \rho\,\ddotubold_{\textstyle{_{\rm{tor}}}}\;=\,2\;\mu\;\nabla\varepsilon_{\textstyle{_{\rm{tor}}}}~-~\rho\,\dotomegabold\times\rbold\,\;.\qquad\qquad\qquad\qquad\qquad\;\qquad\qquad\qquad
 \label{tri}
 \ea
 Be mindful that our $\,\ubold_{\textstyle{_{\rm{tide}}}}\,$ and $\,\varepsilon_{\textstyle{_{\rm{tide}}}}\,$ comprise not only the deformation due to the gravity of the external perturber, but also that due to the quadrupole part of the centrifugal force.

 Rigorously speaking, equations (\ref{raz}) and (\ref{dva}) are not fully independent, because the Coriolis force in the first equation includes inputs with both the tidal and radial $\,\dotubold\,$. However, we shall show now that this coupling is unimportant and the Coriolis force may be completely neglected, along with the potential $\,U_{\smallfirst}\,$ and all the acceleration terms.

 \subsection{Further simplification}

 The harmonics of forced libration are equal to the mean motion $\,n\,$ multiplied by integers.$\,$\footnote{~Except in a high-triaxiality case (which we shall not address here), the frequencies of free libration are lower than $\,n\,$. So whatever we now prove for forced libration will surely work for free libration also.} So the acceleration term $\,\rho\,\ddotubold\,$ in each of the three equations (\ref{raz} - \ref{tri}), as well as the Coriolis force $\,-\,\rho\,\omegabold\times\dotubold\,$ in equation (\ref{raz}) are of the order of $\,\rho\,n^2\,u\,$. At the same time, in both equations (\ref{raz}) and (\ref{tri}), the term  $\,2\mu\nabla{\mathbb{\varepsilon}}\,$ is of the order of $\,|\bar{\mu}(n)|\,u/R^{\,2}\,$. Here $\,R\,$ is the radius of the body, that may vary from dozens of kilometers (for Phobos) to thousands of kilometers (for Mercury or the Moon). The notation $\,|\bar{\mu}(n)|\,$ stands for the absolute value of the complex shear rigidity at the frequency $\,n\,$. Although this value may be much lower than the unrelaxed or relaxed shear rigidity $\,\mu\,$, a very conservative estimate of $\,|\bar{\mu}(n)|\propto 10^{\,5}$ Pa$\,$ is sufficient to observe that $\;\rho\,n^2\ll|\bar{\mu}|\,R^{\,-\,2}\;$ for a broad range of realistic values of $\,n\,$ and $\,R\,$ and a nominal value of $\,\rho=2\times 10^3\,$ kg m$^{-3}\,$.
    This justifies our neglect of the acceleration and Coriolis terms (and of the Coriolis-caused additional potential $\,U_{\smallfirst}\,$) in equations
 (\ref{raz}) and (\ref{tri}).
     Usually, $\,|\bar{K}(n)|\,>\,|\bar{\mu}(n)|\,$. So the afore-justified neglect, being valid for equations (\ref{raz}) and (\ref{tri}), is surely correct also for the  equation (\ref{dva}).  Thus we arrive at
 \ba
 0\;=\;2\;\mu\;\nabla\varepsilon_{\textstyle{_{\rm{tide}}}}\;-\;\rho\,\nabla\,\left[\,\left(\,U_2^{(cent)}\,+\,W_2^{(cent)}\,\right) \;+\;\sum_{l=2}^\infty(U_l\,+\;W_l)\,\right]~~,\qquad\;
 \label{pe}
 \ea
 \ba
 0\;=\;2\,\mu\;\nabla\varepsilon_{\textstyle{_{\rm{rad}}}_{\;({dev})}}\,+\;3\;K\;\nabla\varepsilon_{\textstyle{_{\rm{rad}}}_{\;({vol})}}\;-\;\rho~\nabla\left(U_{\textstyle{_{\rm{rad}}}}\,+\,W_{\textstyle{_{\rm{rad}}}}\right)
 ~~,\qquad\qquad\,\quad
 \label{vt}
 \ea
 \ba
 0\;=\;2\;\mu\;\nabla\varepsilon_{\textstyle{_{\rm{tor}}}}~-~\rho\,\dotomegabold\times\rbold\,\;.\qquad\qquad\qquad\qquad\qquad\,\qquad\qquad\qquad\qquad\qquad
 \label{tretij}
 \label{tr}
 \ea

 \section{Dissipation due to the radial deformation $\;\ubold_{\textstyle{_{\rm{rad}}}}\,$}

 The radial part of the centrifugal potential generates a change in the potential and, thereby, a radial deformation that evolves due to variations of spin. While this deformation contributes to heating, the chances of direct measurement of the radial centrifugal elevation look slim.$\,$\footnote{~If we take for the Earth a $2$ ms variation of the length of the day over a ten year span, the resulting radial displacement will be of the order of $\,10^{-4}$ m only. For Phobos, the radial centrifugal elevation for its average spin is also of the order of $\,10^{-4}$ m, provided its rigidity is like that of the Earth's crust. Hence the variations due to the libration should be even smaller. (Tim Van Hoolst, private communication.)}
 %  The emergence of the purely radial deformation gives birth to the radial Love number (Dahlen 1976, Matsuyama \& Bills 2010).
 %  Using Dahlen's results, Yoder (1982, eqns 21 - 22) demonstrated that the contribution of the radial part of the centrifugal potential to the change in mean motion of Phobos
 %  is about 3\%, which is smaller than the uncertainty in our knowledge of Phobos' $\,k_2/Q\,$. It should be mentioned, however, that the calculations by Dahlen (1976) and
 %  Matsuyama \& Bills (2010) were performed for steady (or slowly changing) rotation, and not for libration. This means that Yoder's application of Dahlen's result to Phobos
 %  requires extra justification.

 \subsection{A harmonic of the radial force}

 It ensues from expression (\ref{radrad}) that a harmonic $\,\chi\,$ of the longitudinal libration gives birth to both the harmonic $\,\chi\,$
 and the double harmonic $\,2\chi\,$ in the radial force:
 \ba
 \Fbold_{\textstyle{_{\rm{rad}}}}\,=\;-\;\rho\;\nabla W_{\textstyle{_{\rm{rad}}}}\;=\;
 \frac{4}{3}\;{\cal{A}}\;\rho\;\chi\,\stackrel{\bf\centerdot}{\theta}_{res}\,\erbold\;\cos \chi t
 \;+\;
 \frac{1}{3}\;{\cal{A}}^2\,\rho\;\chi^2\;\erbold\;\cos 2 \chi t
 \,\;
 \label{29}
 \ea
 or, in complex notation:
 \ba
 {F}_{\textstyle{_{\rm{rad}}}}\;=\;^{(1)}{F}_{\textstyle{_{\rm{rad}}}}(\chi)\,+\,^{(2)}{F}_{\textstyle{_{\rm{rad}}}}(2\chi)\,=\;
 \frac{4}{3}\;{\cal{A}}\,\rho\;\chi\,\stackrel{\bf\centerdot}{\theta}_{res}\,r\;e^{i \chi t}
 \;+\;
 \frac{1}{3}\;{\cal{A}}^2\,\rho\;\chi^2\;r\;e^{2 i \chi t}
 \,\;,
 \label{fcomp}
 \ea
 where the left superscripts $\,^{(1)}\,$ and $\,^{(2)}\,$ accompanying $\,^{(1)}{F}_{\textstyle{_{\rm{rad}}}}(\chi)\,$ and $\,^{(2)}{F}_{\textstyle{_{\rm{rad}}}}(2\chi)\,$
 serve to denote the components oscillating with the frequencies $\,\chi\,$ and $\,2\chi\,$, correspondingly.

 Recall that an actual physical quantity is the real part of the appropriate complex quantity.

 \subsection{A harmonic of the radial velocity\label{radial}}

 A radial displacement generates a change in the density. Therefore, in distinction from equations (\ref{pe}) and (\ref{tretij}), equation (\ref{vt}) cannot be solved under the incompressibility assumption. Instead, it must be solved together with the continuity equation $\;\partial \rho(\rbold,\,t)/\partial t\,+\,\nabla\cdot\left[\rho(\rbold)\,\dot{\ubold}_{\textstyle{_{\rm{rad}}}}\right]\,=\,0\;$, where $\,\rho(\rbold)\,$ is now the $\,${\it{mean}}$\,$ density.

 Just like in the case of the radial force, a harmonic $\,\chi\,$ of the libration spectrum produces a double harmonic $\,2\chi\,$ in the velocity. A somewhat cumbersome expression for that velocity harmonic is given by equation (\ref{velocity}) in Appendix \ref{AppendixRad}. It is also explained in that Appendix that, dependent on the libration frequency (as compared to the inverse Maxwell times) and on the hypothesised porosity of the body, that expression may be simplified to a shorter form of
 \ba
 \nonumber
 \stackrel{\bf\centerdot}{{u}}_{\textstyle{_{\rm{rad}}}}&=&^{(1)}\stackrel{\bf\centerdot}{{u}}_{\textstyle{_{\rm{rad}}}}\;+\;^{(2)}\stackrel{\bf\centerdot}{{u}}_{\textstyle{_{\rm{rad}}}}\\
 \nonumber\\
 \nonumber
 &=&i\;e^{i\chi t}\;\chi^2\;{\cal{A}}(\chi)\;\stackrel{\,\bf\centerdot}{\theta}_{res}\;\frac{\,2\,\rho\,R^{\,3}\,}{15\;\bar{K}(\chi)}\;
 \left[\,\frac{5}{3}\;\frac{r}{R}\;-\;\frac{r^3}{R^{\,3}}\,\right]\;+\;O\left(\;{\,\bar{\mu}(\chi)}/{\bar{K}(\chi)\,}\;\right)\\
 \label{approximation}\\
 &+&i\;e^{2i\chi t}\;\chi^2\;{\cal{A}}\; \frac{\chi\;{\cal{A}}}{2} \;\frac{\,2\,\rho\,R^{\,3}\,}{15\;\bar{K}(2\chi)}\;
 \left[\,\frac{5}{3}\;\frac{r}{R}\;-\;\frac{r^3}{R^{\,3}}\,\right]\;+\;O\left(\;{\,\bar{\mu}(2\chi)}/{\bar{K}(2\chi)\,}\;\right)\,\;,
 \nonumber
 \ea
 $\bar{\mu}\,$ and $\,\bar{K}\,$ being the Fourier components of the shear and bulk complex elasticity moduli. Similarly to equation (\ref{fcomp}), here the left superscripts $\,^{(1)}\,$ and $\,^{(2)}\,$ denote the velocity components oscillating with the frequencies $\,\chi\,$ and $\,2\chi\,$, correspondingly.

 \subsection{Power}

 The rate of working of the radial force $\,\Fbold_{\textstyle{_{\rm{rad}}}}\,$ is given by the standard formula
 \ba
 \langle\,P\,\rangle_{\textstyle{_{\rm{rad}}}}\;=\;
 \left\langle\; \int\,\dot{\ubold}_{\textstyle{_{\rm{rad}}}}\,\cdot\,\Fbold_{\textstyle{_{\rm{rad}}}}\;d^3\rbold\;\right\rangle
  \;\,,
 \label{power}
 \ea
 % where the Eulerian position $\,\rbold\,$, the velocity $\,\dot{\ubold}_{\textstyle{_{\rm{tor}}}}\,$, and the force $\,\Fbold_{\textstyle{_{\rm{tor}}}}\,$ are defined in the centre-of-mass frame. The
 angular brackets standing for time average.
 In the complex notation, this becomes:
 \bs
 \ba
 \langle\,P\,\rangle_{\textstyle{_{\rm{rad}}}}\;=\;\frac{1}{2}\;\Re
 \int\,
 \left(\,^{(1)}\dot{\ubold}_{\textstyle{_{\rm{rad}}}}\,\cdot\,^{(1)}\Fbold^*_{\textstyle{_{\rm{rad}}}}\,+\,
 ^{(2)}\dot{\ubold}_{\textstyle{_{\rm{rad}}}}\,\cdot\,^{(2)}\Fbold^*_{\textstyle{_{\rm{rad}}}}\,\right)\;d^3r
  \;\,,
 \label{begom}
 \ea
 where asterisk denotes complex conjugation, while $\,\Re\,$ signifies the real part.

  % Recall that the left superscripts $\,^{(1)}\,$ and $\,^{(2)}\,$ serve to denote the components oscillating with the frequencies $\,\chi\,$ and $\,2\chi\,$, correspondingly.
  As follows from equation (\ref{approximation}), the magnitude of the first harmonic of the velocity relates to the magnitude of the second harmonic as $\,\stackrel{\,\bf\centerdot}{\theta}_{res}\,$ to $\,\chi\,{\cal{A}}/2\,$. By equation (\ref{fcomp}), the first harmonic of the force relates to the second harmonic as $\,\stackrel{\,\bf\centerdot}{\theta}_{res}\,$ to $\,\chi\,{\cal{A}}/4\,$. Hence the first term in the expression (\ref{begom}) for the power relates to the second term as $\,\dot{\theta}_{res}^{\,2}\,$ to $\,\chi^2\,{\cal{A}}^2/8\,$. So for not too strong libration the power input from the second harmonic may be safely omitted:
  \ba
  \langle\,P\,\rangle_{\textstyle{_{\rm{rad}}}}\;\approx\;\frac{1}{2}\;\Re\int\,
  ^{(1)}\dot{\ubold}_{\textstyle{_{\rm{rad}}}}\,\cdot\,^{(1)}\Fbold^*_{\textstyle{_{\rm{rad}}}}\;d^3r\;\,.
  \label{powercomp}
  \ea
  \es
  Inserting therein expressions (\ref{fcomp}) and (\ref{approximation}), we arrive at
 \bs
 \ba
 \langle\,P\,\rangle_{\textstyle{_{\rm{rad}}}}&=&-\;\frac{4}{45}\;\int d^3\rbold\;\rho^2\,\chi^3\,
 \dot{\theta}_{\rm{res}}^{\,2}\;{\cal{A}}^2 \,R^{\,4}\,\left[\,\frac{5}{3}\;\frac{r^2}{R^{\,2}}\;-\;\frac{r^4}{R^{\,4}}\,\right]
 \,\;\Im\left[\,\bar{J}^{(bulk)}(\chi)\,\right]
 ~\\
 \nonumber\\
 &=&\frac{4}{45}\;\int d^3\rbold\;\rho^2\,\chi^3\,\dot{\theta}_{\rm{res}}^{\,2}\;{\cal{A}}^2 \,R^{\,4}\,\left[\,\frac{5}{3}\;\frac{r^2}{R^{\,2}}\;-\;\frac{r^4}{R^{\,4}}\,\right]\;\,|\,\bar{J}^{(bulk)}(\chi)\,|\;\sin\delta^{(bulk)}(\chi)\,\;,\;\;\qquad
 \label{work}
 \ea
 \es
 where $\,\Im\,$ signifies the imaginary part. The notation $\,\bar{J}^{(bulk)}(\chi)\,=\,1/\bar{K}(\chi)\,$ is for the complex compliance operator. We equip it with the superscript $\,${\it{`$\,(bulk)\,$'}}$\,$ to emphasise that this is a bulk, not shear compliance. Also recall that we are using the convention
 \ba
 \bar{J}^{(bulk)}\;=\;|\,\bar{J}^{(bulk)}\,|\;e^{\,-\;i\delta^{(bulk)}}\,\;,
 \label{}
 \ea
 where a ``minus'' sign endows the angle with the meaning of phase lag.

 For a Maxwell material with a volumetric (bulk) viscosity $\,\zeta\,$,  we have:
 \ba
 -\;\Im\left[\,\bar{J}^{(bulk)}(\chi)\,\right]\;\equiv\;|\,\bar{J}^{(bulk)}(\chi)\,|\;\sin\delta^{(bulk)}(\chi)\;=\;\frac{1}{\zeta\;\chi}\,\;,
 \label{}
 \ea
 wherefrom, under the homogeneity assumption,
  \bs
 \ba
 \langle\,P\,\rangle_{\textstyle{_{\rm{rad}}}}&=&\frac{4}{45}\;\frac{\rho^2\;\chi^2\;\dot{\theta}_{\rm{res}}^{\,2}\;{\cal{A}}(\chi)^2 }{\zeta}\;R^{\,4}\;
 \int d^3\rbold
 \left[\,\frac{5}{3}\;\frac{r^2}{R^{\,2}}\;-\;\frac{r^4}{R^{\,4}}\,\right]
 ~\\
 \nonumber\\
 &=&\frac{64\,\pi}{945}\;\frac{\rho^2\;\chi^2\;\dot{\theta}_{\rm{res}}^{\,2}\;{\cal{A}}^2 }{\zeta}\;\pi\;R^{\,7}\;\approx\;\frac{\rho^2\;\chi^2\;\dot{\theta}_{\rm{res}}^{\,2}\;{\cal{A}}^2 }{4.7\;\zeta}\;R^{\,7}\;
 \,\;.\;\;\qquad
 \label{}
 \ea
 \es
 This formula furnishes the input in the damped power, provided by a librational harmonic $\,\chi\,$ with a magnitude $\,{\cal{A}}\,$. It does not matter whether this librational harmonic is free or forced.

 In the presence of several libration modes $\,\chi_j\,$ with magnitudes $\,{\cal{A}}_j\,$, we obtain:
 \ba
 \langle\,P\,\rangle_{\textstyle{_{\rm{rad}}}}\;=\;\frac{64\,\pi}{945}\;R^{\,7}\;\rho^2
 \;\dot{\theta}_{\rm{res}}^{\,2}\;\sum_{j=1}^{\infty} \chi_j^3\;{\cal{A}}_j^2\;|\,\bar{J}^{(bulk)}(\chi_j)\,|\;\sin\delta^{(bulk)}(\chi_j)\,\;.
 \label{gert}
 \ea
 For a Maxwell material, this series takes the form of
 \ba
 \langle\,P\,\rangle_{\textstyle{_{\rm{rad}}}}\;=
 \;\frac{64\,\pi}{945\;\zeta}\;R^{\,7}\;\rho^2\;\dot{\theta}_{\rm{res}}^{\,2}\;\sum_{j=1}^{\infty} \chi_j^2\;{\cal{A}}_j^2\;\approx\;
 \;\frac{R^{\,7}\;\rho^2}{4.7\;\zeta}\;\dot{\theta}_{\rm{res}}^{\,2}\;\sum_{j=1}^{\infty} \chi_j^2\;{\cal{A}}_j^2
 \,\;.
 \label{power_radial}
 \ea
 The number of terms to be kept in series (\ref{power_radial}) depends on how quickly the magnitude $\,{\cal{A}}_j\,$ decreases with the increase of $\,j\,$.
 In Appendix \ref{check}, we demonstrate that in the case of forced libration it is sufficient to leave the first term, provided the eccentricity is not very high:
 \ba
 \nonumber
 \mbox{Forced libration, ~~$\;e\,\lesssim\,0.5\;\,:$} \qquad \qquad \qquad \qquad \qquad \qquad \qquad \qquad  \qquad \qquad \qquad \qquad \qquad \qquad  \;\;\;  \\
 \langle\,P\,\rangle_{\textstyle{_{\rm{rad}}}}
 \approx
  \frac{\textstyle R^{\,7}\;\rho^2}{\textstyle 4.7\;\zeta}\;\dot{\theta}_{\rm{res}}^{\,2}\;n^2\;{\cal{A}}_1^2
 =\left\{
 \begin{array}{ll}
  ^{(1:1)}{\cal{A}}_1^2\;\frac{\textstyle R^{\,7}\;\rho^2}{\textstyle 4.7\;\zeta}\;n^4
  \quad,\quad & \mbox{in~the~1:1~resonance,}\\
 ~\\
 ~\\
 ^{(3:2)}{\cal{A}}_1^2\;\frac{\textstyle R^{\,7}\;\rho^2}{\textstyle 2.1\;\zeta}\;n^4      \,\;,
 & \mbox{in~the~3:2~resonance.} \qquad\qquad
 \end{array}
 \right.
 \label{appa}
 \ea
 On the right-hand side of this equation, we endowed the the forced-libration magnitude $\,^{(z)}{\cal{A}}_j\,$ with a superscript $\,{(z)}\,$, to specify in which spin-orbit resonance the rotator is. The subscript $\,1\,$ says that we are considering the principal mode, the one with the frequency $\,n\,$. The expressions for the magnitudes $\,^{(1:1)}{\cal{A}}_1\,$ and $\,^{(3:2)}{\cal{A}}_1\,$ are given by equations (\ref{10}) and (\ref{11}), correspondingly.

 \section{Dissipation due to the toroidal deformation $\;\ubold_{\textstyle{_{\rm{tor}}}}\,$}

   The inertial force $\;-\,\rho\,\dotomegabold\times\rbold\,=\,\rho\,\erbold\times\nabla(\dotomegabold\cdot\erbold)\;$ is toroidal of degree 1. Negligible for a steadily despinning rotator, it becomes important under libration.

 In spherically-symmetric solid bodies, the toroidal force produces toroidal deformation only. Being divergence-free (see Footnote \ref{Thomson}), this deformation causes no contraction or expansion, i.e., is purely shear.

 Toroidal deformation causes neither radial uplifts nor changes of the gravitational potential. Therefore, its presence does not influence the expressions for the Love numbers.
 Somewhat naively, we can also say that, since toroidal deformation renders no change in the gravitational potential of the tidally-perturbed body, it makes no contribution into the tidal torque. In reality, the body does have a small permanent triaxiality (lest there would be no libration). Hence a very small tidal torque and, consequently, a weak contribution in the orbital evolution.

 Most importantly, toroidal deformation contributes to internal friction and therefore to energy damping.

 %  To estimate the dissipation caused by the toroidal rotational force, Yoder (1982) introduced an equivalent effective torque.
 %  He pointed out that the force gets important when the magnitude of the physical libration is comparable to that of the optical
 %  libration. According to Yoder (1982), the toroidal force contributes to the change of Phobos' mean motion about 1.6\%, which
 %  is less than the input from the purely radial part.

 \subsection{The toroidal displacement field}

 As demonstrated in Appendix \ref{tor}, the toroidal force per unit volume,
 $\;
 %  \ba
 \Fbold_{\textstyle{_{\rm{tor}}}}\,=\;-\,\rho\,\dotomegabold\times\rbold\;,
 %  \label{for} \ea
 $
 and the resulting toroidal displacement, $\,\ubold_{\textstyle{_{\rm{tor}}}}\,$, can be cast into the form of
 \ba
 \Fbold_{\textstyle{_{\rm{tor}}}}\,=\;\rho\;r\,\stackrel{\bf\centerdot}{\omega\,}\,\sin\varphi\;\,
 \hat{\bf{e}}_{\smalllambdabold}\,\;,
 \label{F}
 \ea
 and
 \ba
 \ubold_{\textstyle{_{\rm{tor}}}}\;=\;D(r,\,t)\;\sin\varphi\;\,
 \hat{\bf{e}}_{\smalllambdabold}\,\;,
 \label{u}
 \ea
 $\varphi\,$ being the colatitude of the point $\rbold$, and $\hat{\bf{e}}_{\smalllambdabold}$ being the unit vector pointing in the direction of increasing longitude $\,\lambda\,$ in that point. Naturally, the toroidal deformation vanishes at the pole.

 From expressions (\ref{condition}) and (\ref{main}) we see that in the complex notation a harmonic $\,\chi\,$ of forced libration and the corresponding angular velocity and acceleration are
 %  $\,\stackrel{\bf\centerdot}{\omega\,}\,$ is
 given by
 \ba
 \gamma\;=\;{\cal{A}}\;e^{i \chi t}\;\;,\qquad\omega\;\equiv\;\stackrel{\bf\centerdot}{\theta\,}\;=\;\stackrel{\bf\centerdot}{\theta\,}_{res}\,+\;i\;\chi\;{\cal{A}}\;e^{i\chi t}\;\;,\qquad
 \stackrel{\bf\centerdot}{\omega\,}\;=\;-\;\chi^2\;{\cal{A}}\;e^{i\chi t}\,\;.
 \label{osa}
 \ea
 Consequently, formula (\ref{F}) in the complex notation is
 \ba
 {\Fbold}_{\textstyle{_{\rm{tor}}}}\,=\;-\;\rho\;r\;\chi^2\;{\cal{A}}\;\sin\varphi\;e^{i\chi t}\;\,\hat{\bf{e}}_{\smalllambdabold}\,\;.
 \label{complexforce}
 \ea
 A Fourier component of the displacement looks as
 \ba
 {\ubold}_{\textstyle{_{\rm{tor}}}}\;=\;\bar{D}(r,\,\chi)\;\sin\varphi\;e^{i\chi t}\;\,
 \hat{\bf{e}}_{\smalllambdabold}\,\;,
 \label{complexu}
 \ea
 where the complex magnitude $\,\bar{D}(r,\,\chi)\,$ contains a phase factor describing the lagging of deformation relative to forcing. In the presence of several frequencies, each of the three above formulae would have a Fourier series on its right-hand side.

 \subsection{Strain and displacement}

 To solve equation (\ref{tretij}), we need: (a) to insert expression (\ref{u}) for the displacement into the expression for the strain; (b) to calculate the divergence of the strain; (c) to insert the divergence into equation (\ref{tretij}). It turns out (see equation (\ref{divv}) in Appendix \ref{tor}) that in our setting the only nonzero component of the divergence of the strain is the longitudinal one:
 \ba
 (\nabla{\epsilon}_{\textstyle{_{\rm{tor}}}}\,)_{\smalllambdabold}\,=\;\frac{\sin\varphi}{2}\;\frac{1}{r^3}\frac{\partial }{\partial r}\;\left[
 r^4\;\frac{\partial }{\partial r}\left(\frac{D}{r}\right)\right]
 \;\;.
 \label{strain1}
 \ea
 In the complex notation, both $\,\epsilon_{\,tor}\,$ and $\,D\,$ acquire overbar.
 The simplified equation of motion (\ref{tretij}) for toroidal deformations can be written as
 \ba
 -\;2\;{\mu}\;\nabla\varepsilon_{\textstyle{_{\rm{tor}}}}\;=\;~-~\rho\,\dotomegabold\times\rbold\,\;,
 \label{}
 \ea
 where the left-hand side is found from expression (\ref{strain1}), while the right-hand side is the toroidal force
 (\ref{F}). Insertion of those formulae yields:
  \ba
 -\;2\;\mu\;\frac{\sin\varphi}{2}\;\frac{1}{r^3}\frac{\partial }{\partial r}\;\left[
 r^4\;\frac{\partial }{\partial r}\left(\frac{D}{r}\right)\right]\;=\;\rho\;r\,\stackrel{\bf\centerdot}{\omega\,}\,\sin\varphi\,\;,
 \label{}
 \ea
 whence
 \ba
 D(r,\,t)\;=\;-\;\frac{\,\rho\;\stackrel{\bf\centerdot}{\omega\,}\,r^3}{10\;\mu}\;-\;C\;r\;+\;\frac{B}{r^2}\,\;.
 \label{W}
 \ea
 This coincides with equation (A16) from Yoder (1982), except for an unimportant misprint made by Yoder in the last term. That term has to be dropped anyway by setting the integration constant $\,B\,$ equal to zero, so that we avoid an infinity at $\,r=0\,$. Yoder (1982) also suggested to set $\,C=0\,$, in order to nullify the displacement in the absence of libration.

 At this point, we deviate from Yoder's treatment. While in his development $\,\mu\,$ was the unrelaxed rigidity modulus, we now switch to complex notation and make the rigidity complex and frequency-dependent. This is needed in order to take into account the phase lag and to calculate the damped power. Combining expressions (\ref{osa}) and (\ref{W}), we get:
  \ba
 \bar{D}(r,\,\chi)\;=\;\frac{\rho\;\chi^2\;{\cal{A}}\;r^3}{10\;\bar{\mu}(\chi)}
 \label{}
 \ea
 or, equivalently:
 \ba
 \bar{D}(r,\,\chi)\;=\;\frac{\rho\;\chi^2\;{\cal{A}}\;r^3}{10}\;\bar{J}(\chi)\,\;,
 \label{}
 \ea
 where the complex compliance is the inverse of the complex rigidity:
 \ba
 \bar{J}(\chi)\;=\;1/\bar{\mu}(\chi)\,\;.
 \label{}
 \ea
 Owing to expression (\ref{complexu}), the complex displacement and velocity then become:
 \ba
 {\ubold}_{\textstyle{_{\rm{tor}}}}\;=\;\frac{\,\rho\;\chi^2\;{\cal{A}}\;r^3\,}{10}\;\bar{J}(\chi)\;\sin\varphi\;\,e^{i\chi t}\;\,
 \hat{\bf{e}}_{\smalllambdabold}\,\;,
 \label{}
 \ea
 \ba
 \stackrel{\bf\centerdot}{\ubold}_{\textstyle{_{\rm{tor}}}}\;=\;\frac{\,\rho\;\chi^3\;i\;{\cal{A}}\;r^3\,}{10}\;\bar{J}(\chi)\;\sin\varphi\;\,e^{i\chi t}\;\,
 \hat{\bf{e}}_{\smalllambdabold}\,\;.
 \label{complexuu}
 \ea

 \subsection{Power}

 %  With the Eulerian position $\,\rbold\,$, the deformation $\,\ubold_{\textstyle{_{\rm{tor}}}}\,$, and the force $\,\Fbold_{\textstyle{_{\rm{tor}}}}\,$
 %  introduced in the centre-of-mass frame,
 The power exerted by the toroidal inertial force $\,\Fbold_{\textstyle{_{\rm{tor}}}}\,$ is calculated through
 \ba
 \langle\,P\,\rangle_{\textstyle{_{\rm{tor}}}}\;=\;
 \left\langle\; \int\,\dot{\ubold}_{\textstyle{_{\rm{tor}}}}\,\cdot\,\Fbold_{\textstyle{_{\rm{tor}}}}\;d^3\rbold\;\right\rangle
  \;\,
 \label{}
 \ea
 or, in the complex notation:
 \ba
 \langle\,P\,\rangle_{\textstyle{_{\rm{tor}}}}\;=\;\frac{1}{2}\;\Re
 % \left\langle\;
 \int\,\dot{\ubold}_{\textstyle{_{\rm{tor}}}}\,\cdot\,\Fbold^*_{\textstyle{_{\rm{tor}}}}\;d^3\rbold
 % \;\right\rangle
  \;\,.
 \label{P}
 \ea
 % where asterisk denotes complex conjugation, while $\,\Re\,$ signifies the real part.
 Plugging formulae (\ref{complexforce}) and (\ref{complexuu}) in equation (\ref{P}), we arrive at:
 \ba
 \langle\,P\,\rangle_{\textstyle{_{\rm{tor}}}}&=&-\;\int d^3\rbold\;\frac{\rho^2\;\chi^5\;{\cal{A}}\,{\cal{A}}^*\;r^4\;\sin^2\varphi}{20}\,\;\Im\left[\,\bar{J}(\chi)\,\right]
 \ea
 with $\,\Im\,$ denoting the imaginary part. For an homogeneous near-spherical body, we get:
 \footnote{~The integral over a homogeneous sphere is: $\;\int d^3\rbold\;r^4\;\sin^2\varphi\;=\;2\,\pi\,\int_0^R r^6 dr\;\int_{0}^{\pi} \sin^3\varphi\,d\varphi\;=\;2\,\pi\;\frac{\textstyle R^{\,7}}{\textstyle 7\;}\;\frac{\textstyle 4}{\textstyle 3}\;$.}
 \bs
 \ba
 \langle\,P\,\rangle_{\textstyle{_{\rm{tor}}}}&=&-\;\frac{2\,\pi}{105}\;R^{\,7}\;\rho^2\;\chi^5\;{\cal{A}}\,{\cal{A}}^*\,\;\Im\left[\,\bar{J}(\chi)\,\right]
 ~\\
 \nonumber\\
 &=&\frac{2\,\pi}{105}\;R^{\,7}\;\rho^2\;\chi^5\;{\cal{A}}(\chi)^2\;|\,\bar{J}(\chi)\,|\;\sin\delta(\chi)\,\;,
 \label{work}
 \ea
 \es
 where we used the convention
 \ba
 \bar{J}(\chi)\;=\;|\,\bar{J}(\chi)\,|\;e^{\,-\;i\delta(\chi)}\,\;.
 \label{}
 \ea
 Recall that within this convention the angle $\,\delta(\chi)\,$ is introduced with a ``minus'' sign in order to impart it with the meaning of phase lag
 (lag of deformation relative to forcing). Consequently, the change of energy (\ref{work}) comes out negative for a positive phase lag $\,\delta(\chi)\,$.

 For a Maxwell body of viscosity $\,\eta\,$, we have
 \ba
 -\;\Im\left[\,\bar{J}(\chi)\,\right]\;\equiv\;|\,\bar{J}(\chi)\,|\;\sin\delta(\chi)\;=\;\frac{1}{\eta\;\chi}\,\;,
 \label{}
 \ea
 wherefrom
 \ba
 \langle\,P\,\rangle_{\textstyle{_{\rm{tor}}}}\;=\;\frac{2\,\pi}{105}\;\frac{R^{\,7}\;\rho^2}{\eta}\;\chi^4\;{\cal{A}}^2
 \;\approx\;\frac{\rho^2\;\chi^4\;{\cal{A}}^2\;R^{\,7}}{17\;\eta}
 \,\;.
 \label{}
 \ea
 This expression is generic, in that it renders the contribution in the damped power, coming from a librational harmonic $\,\chi\,$ with a magnitude $\,{\cal{A}}\,$. This librational harmonic may be either free or forced.

 %  When longitudinal forced libration contains only the principal mode, we set $\,\chi\,=\,n\,$.
 In the presence of several libration modes with frequencies $\,\chi_j\,$ and magnitudes $\,{\cal{A}}_j\,$, we obtain:
 \ba
 \langle\,P\,\rangle_{\textstyle{_{\rm{tor}}}}\;=\;\frac{2\,\pi}{105}\;R^{\,7}\;\rho^2
 \;\sum_{j=1}^{\infty} \chi_j^5\;{\cal{A}}_j^2\;|\,\bar{J}(\chi_j)\,|\;\sin\delta(\chi_j)\,\;.
 \label{}
 \ea
 %   where the $\,j$-th harmonic is $\,\chi_j\,=\,j\,n\,$.
 For a Maxwell material, this becomes:
 \ba
 \langle\,P\,\rangle_{\textstyle{_{\rm{tor}}}}\;=
 \;\frac{2\,\pi}{105\,\eta}\;R^{\,7}\;\rho^2\;\sum_{j=1}^{\infty} \chi_j^4\;{\cal{A}}_j^2
 \;\approx\;
 \;\frac{R^{\,7}\;\rho^2}{17\,\eta}\;\sum_{j=1}^{\infty} \chi_j^4\;{\cal{A}}_j^2
 \,\;.
 \label{comment}
 \ea
 The number of terms to keep depends on how rapidly ${\cal{A}}(\chi_j)$  decreases with the increase of $j$.

 In Appendix \ref{bezobrazie}, we show that in the case of forced libration the above series can be approximated with its first term ($\,j=1\,$), provided the eccentricity is not too large:
 \ba
 \mbox{Forced libration, ~~$\;e\,<\,0.5\;\,:$}\qquad  \langle\,P\,\rangle_{\textstyle{_{\rm{tor}}}}\;\approx\;^{(z)}{\cal{A}}_1^{\,2}\;\frac{R^{\,7}\;\rho^2}{17\,\eta}\;n^4\,\;.
 \label{69}
 \ea
 On the right-hand side of this expression, we equipped the forced-libration magnitude with a superscript $\,{(z)}\,$, to emphasise that the expression for the magnitude depends not only on the harmonic (in our case, $\,j=1\,$) but also on the spin-orbit resonance the rotating body is in. The formulae for the magnitudes $\,^{(1:1)}{\cal{A}}_1\,$ and $\,^{(3:2)}{\cal{A}}_1\,$ are given by equations (\ref{10}) and (\ref{11}), correspondingly.

 \section{Dissipation due to the tidal deformation $\;\ubold_{\textstyle{_{\rm{tide}}}}\,$}

 Calculation of this input in dissipation differs from the calculation of the tidal dissipation rate in a non-librating rotator. The difference is two-fold. First, here we must include deformations due to libration. Second, we have to include the extra quadrupole component that comes from the (changing in time) centrifugal force.

 \subsection{Spectrum of the harmonics of the gravitational tides}

 As demonstrated in Frouard \& Efroimsky (2017\,a), in the presence of longitudinal libration at a frequency $\,\chi\,$,  the spectrum of the Fourier tidal modes is
 \ba
 \nonumber
 \beta_{lmpqs}\,=~(l\,-\,2\,p\,-\,m\,z\,')\,\stackrel{\bf\centerdot}{\omega}&+&(l\,-\,2\,p\,-\,m\,z\,+\,q)\;n\;-\;s\,\chi\\
 \label{beta}\\
 \nonumber
 &\approx&  (l\,-\,2\,p\,-\,m\,z\,+\,q)\;n\;-\;s\,\chi\,\;,
 \label{beta}
 \ea
 with both $\,q\,$ and $\,s\,$ running from $\,-\,\infty\,$ through $\,\infty\,$, and with $\,m\,$ and $\,p\,$ running from $\,0\,$ through $\,l\,$. $\,$Here $\;l\,,\,m\,,\,p\,,\,q\;$ are the integers showing up the Kaula expansion of the tidal torque (and in a similar expansion of the permanent-triaxiality-caused torque), while $\,n\,\equiv\,\,\stackrel{\bf\centerdot}{\cal M\,}\,$ is the anomalistic mean motion, and $\,\omega\,$ is the argument of the pericentre. The rational numbers $\,z\,$ and $\,z\,'\,$ mark the spin-orbit resonance the rotator is in, see Section \ref{definition}.$\,$\footnote{~Be mindful that in the above expression for $\,\beta_{lmpqs}\,$ the integers $\;l\,,\;m\,,\;p\,,\;q\;$ are arbitrary (within the permissible limits defined by $\,l\geq 2\,$, $\,0\leq m\leq l\,$, $\,0\leq p\leq l\,$), while in the expressions for $\,z\,$ and $\,z\,'\,$ in Section \ref{definition} they are fixed and correspond to a certain spin-orbit resonance in which the rotating body is trapped.}

 %  When the inclination is small, only the terms with $\,p=0\,$ should be kept in the expression for the permanent triaxiality-caused and tidal torques.
 In the quadrupole approximation, only the terms with $\,l=2\,$ are kept:
 \ba
 \beta_{2mpqs}\,=~(2\,-\,2\,p\,-\,m\,z\,+\,q)\;n\;-\;s\,\chi\,\;,
 \label{betaapprox}
 \ea
 the term with $\,\stackrel{\bf\centerdot}{\omega\,}$ being neglected.
 This spectrum includes the products of $\,n\,$ (and, likewise, the products of $\,\chi\,$) by all integer numbers, positive or negative. The spectrum also includes all semi-integers multiplied by $\,n\,$, in case the spin-orbit resonance $\,z\,$ is semi-integer.

 In the special case of forced libration, the main libration frequency coincides with the anomalistic mean motion: $\,\chi=\chi_1=n\,$, the other libration frequencies being $\,\chi_j=jn\,$, $\;j>1\,$.

  \subsection{A potential difficulty\label{difficulty}}

  While it is possible to calculate, separately, the damping rate $\,\langle P \rangle_{\textstyle{_{\rm{tide}}}}\,$ produced by the gravitational tides and the damping rate $\,\langle P \rangle_{\textstyle{_{\rm{cent}}}}\,$ produced by the quadrupole part of the centrifugal force, it is not easy to write down the combined dissipation rate in a general form. The problem is that the combined power is $\,${\it{not}}$\,$ equal to the sum of $\,\langle P \rangle_{\textstyle{_{\rm{tide}}}}\,$ and
 $\,\langle P \rangle_{\textstyle{_{\rm{cent}}}}\,$. To understand the reason for this, recall that the power exerted at a certain physical frequency is proportional to the squared magnitude of vibration at that frequency, multiplied by the sine of the phase lag between the reaction and action at that frequency. Therefore, if two processes have an overlap of their spectra, we should first sum up the vibrations at each common physical frequency (with phases taken into account)~---~and only then square the result, in order to find the power at that frequency.

 As we shall see in Section \ref{centrif}, longitudinal libration at a frequency $\,\chi\,$ causes dissipation associated with the forcing frequencies $\,\chi\,$ and  $\,2\,\chi\,$ in the material.$\,$\footnote{~This is not surprising, as the centrifugal force is quadratic in the angular velocity.} $\,$Each of these forcing frequencies has a counterpart in the spectrum (\ref{betaapprox}) of the gravitational tides. These are the tidal modes $\,\beta_{2mpqs}\,$ satisfying \footnote{~Recall that the physical forcing frequencies in the material of the rotating body are equal to the absolute values of the Fourier tidal modes, $\,|\,\beta_{lmpqs}\,|\,.\,$  Hence two Fourier modes of opposite signs correspond to the same physical frequency. For a detailed explanation, see Section 4.3 in Efroimsky \& Makarov 2013.}
 $\;\,|\,\beta_{2mpqs}\,|\,=\,\chi\,\;$ or $\;\,|\,\beta_{2mpqs}\,|\,=\,2\chi\,$.

 %  For example, the spectrum $\,\chi_j\,=\,j\,n\,$ of forced libration provides inputs into $\,\langle P \rangle_{\textstyle{_{\rm{cent}}}}\,$ at the doubled frequencies:
 %  \ba
 %  2\,\chi_j\,=\,2\,j\,n\quad\mbox{with}\quad\,j\,=\,1,\,2\,.\,.\,.\;
 %  \label{cental}
 %  \ea
 %  Each of these has counterparts in the spectrum (\ref{betaapprox}), i.e., the modes satisfying  $\;\,2\,\chi_j\,=\,|\,\beta_{2mpqs}\,|\,$,
 %  which is the same as $\;\,2\,j\,=\,|\,2\,-\,2\,p\,-\,m\,z\,+\,q\,-\,s\,|\,$.

 This way, the algorithm of finding the total power should be the following. In each particular spin-orbit resonance $\,z\,$, we must write down the terms of the gravitational perturbing tidal potential, that we wish to keep. We also must write down the terms of the quadrupole perturbing potential due to the centrifugal force, for each libration mode $\,\chi\,$. Thereafter, for each frequency $\,\chi\,$, we should calculate the appropriate disturbing potential by summing the centrifugal input at the forcing frequency $\,\chi\,$ with the gravitational disturbing tidal potential at all the modes satisfying $\;\chi\,=\,|\,\beta_{2mpqs}\,|\,$. Then we must calculate the resulting power at each such frequency, by squaring the so-obtained sum of the inputs.
 %  While at odd frequencies, only the gravitational tide will matter, at even both the gravitational and centrifugal will come into play.

 We are spared of these difficulties
 %  the necessity to perform this tedious procedure
 when one of the two effects~---~either the gravitational tides or the centrifugal tides~---~can be neglected as compared to the other.
 Below we shall see that this is definitely the case for the 1:1 spin-orbit resonance and, to a reasonable approximation, the case for the 3:2 resonance.

 \subsection{Dissipation due to the quadrupole part of the centrifugal force\label{centrif}}

 As we saw in Section \ref{centripetal}, the oscillating part of the centrifugal force can be split into two potential forces, one being purely radial, another quadrupole. Specifically, a harmonic $\,\chi\,$ in the longitudinal libration spectrum produces a quadrupole potential:
 \ba
 W_2^{(cent)}\;=\;
 \frac{2}{3}~{\cal{A}}\;\chi\;\dot{\theta}_{res}\,\erbold^{\,2}~P_2\left(\cos\varphi\right)~\cos \chi t
 \;+\;
 \frac{{\cal{A}}^2\;\chi^2\,\erbold^{\,2}}{6}~P_2\left(\cos\varphi\right)\;\cos 2 \chi t
 \label{}
 \ea
 or, in the complex notation:
 \ba
 W_2^{(cent)}\;=\;^{(1)}\bar{W}(\chi)\;+\;^{(2)}\bar{W}(2\chi)\,\;,
 \label{}
 \ea
 where
 \ba
 \nonumber
 ^{(1)}\bar{W}(\chi)&=&\frac{2}{3}\;{\cal{A}}\;\chi\;\dot{\theta}_{res}\;\erbold^{\,2}~P_2\left(\cos\varphi\right)\;e^{i\,\chi\,t}\,\;,\\
  \label{WW}\\
 ^{(2)}\bar{W}(2\chi)&=&\frac{1}{6}\;{\cal{A}}^2\;\chi^2\,\erbold^{\,2}~P_2\left(\cos\varphi\right)\;e^{i\,2\,\chi\,t}
         \,\;.
 \nonumber
 \ea
 The corresponding incremental potential produced by the tidal deformation is
 \ba
 U_2^{(cent)}\;=\;^{(1)}\bar{U}(\chi)\;+\;^{(2)}\bar{U}(2\chi)\,\;,
 \label{}
 \ea
 with
 \ba
 \nonumber
 ^{(1)}\bar{U}(\chi)&=&^{(1)}\bar{W}(\chi)\;\bar{k}_2(\chi)
     \;=\;\frac{2}{3}\;{\cal{A}}\;\chi\;\dot{\theta}_{res}\;\erbold^{\,2}~P_2\left(\cos\varphi\right)\;k_2(\chi)\;e^{-\,i\,\epsilon_2(\chi)}\;e^{i\,\chi\,t}\,\;,\\
 \label{U}\\
 ^{(2)}\bar{U}(2\chi)&=&^{(2)}\bar{W}(2\chi)\,\bar{k}_2(2\chi)
     \,=\,\frac{1}{6}\,{\cal{A}}^2\,\chi^2\,\erbold^{\,2}\,P_2\left(\cos\varphi\right)\,k_2(2\chi)\,e^{-\,i\,\epsilon_2(2\,\chi)}\;e^{i\,2\,\chi\,t}\,\;.
 \nonumber
 \ea
 Here the complex quadrupole Love numbers are written down as
 \ba
 \nonumber
 \bar{k}_2(\chi)&=&{k}_2(\chi)\;e^{-\,i\,\epsilon_2(\chi)}\,\;,\\
 \label{}\\
 \nonumber
 \bar{k}_2(2\chi)&=&{k}_2(2\chi)\;e^{-\,i\,\epsilon_2(2\chi)}\,\;,
 \ea
 where $\,\epsilon_2(\chi)\,$ and $\,{k}_2(\chi)\,=\,|\,\bar{k}_2(\chi)\,|\,$ are, correspondingly, the phase lag and the real dynamical Love number at the frequency $\,\chi\,$; while $\,\epsilon_2(2\chi)\,$ and $\,{k}_2(2\chi)\,=\,|\,\bar{k}_2(2\chi)\,|\,$ are those at the frequency $\,2\chi\,$.

 In a homogeneous near-spherical body, the power dissipated by a perturbation of the potential  can be written as a surface integral (e.g., Efroimsky \& Makarov 2014, eqn 61):
 \ba
 \langle \,P\,\rangle\,=\;\frac{1}{4\,\pi\,G\,R}\;\sum_{l\,=\,2}^{\infty}\,(2l+1) \int \langle \,W_l \;\dot{U}_l\,\rangle\;dS\;\,,
 \ea
 where angular brackets signify time averaging, overdot denotes time derivative, $\,G\,$ is the Newton gravity constant, $\,W_l\,$ and $\,U_l\,$ are the degree-$l\,$ perturbing and deformation-caused potentials, while $\,dS\,=\,2\,\pi\,R^{\,2}\,\sin\varphi\,d\varphi\,$ is a surface element (with $\,\varphi\,$ being the colatitude). As the centrifugal force produces a quadrupole perturbation, we calculate the resulting power as
 \bs
 \ba
 \nonumber
 \langle \,P\,\rangle_{\textstyle{_{\rm{cent}}}}&=&\frac{5}{4\,\pi\,G\,R}\; \int \langle \,W_2^{(cent)}\;\dot{U}_2^{(cent)}\,\rangle\;dS\\
 \label{}\\
 \nonumber
 &=&\frac{5}{4\,\pi\,G\,R}\; \int \langle \,^{(1)}W\;^{(1)}\dot{U}\,\rangle\;dS\;+\;\frac{5}{4\,\pi\,G\,R}\; \int \langle \,^{(2)}W\;^{(2)}\dot{U}\,\rangle\;dS
 \,\;.
 \ea
 In the complex notation, this reads as
 \ba
 \langle \,P\,\rangle_{\textstyle{_{\rm{cent}}}}\,=\;\frac{5}{4\,\pi\,G\,R}\;\frac{1}{2}\; \Re\int {^{(1)}\bar{W}}\;^{(1)}\dot{\bar{U}}^{\,*}\,dS
 \;+\;\frac{5}{4\,\pi\,G\,R}\;\frac{1}{2}\; \Re\int {^{(2)}\bar{W}}\;^{(2)}\dot{\bar{U}}^{\,*}\,dS
 \label{}
 \ea
 \es
 where $\,\Re\,$ signifies the real part, while asterisk stands for complex conjugation. Using formula (\ref{U}), we rewrite this as
 \bs
 \ba
 \nonumber
 \langle \,P\,\rangle_{\textstyle{_{\rm{cent}}}}&=&
 -\,\frac{5}{8\,\pi\,G\,R}\,\int\left\{\;\chi\,|\,^{(1)}\bar{W}(\chi)\,|^{\,2}\;\Im\left[\,\bar{k}_2(\chi)\,\right]
  \,+\, 2\chi\,|\,^{(2)}\bar{W}(2\chi)\,|^{\,2}\;\Im\left[\,\bar{k}_2(2\chi)\,\right]\;\right\}\,dS\\
 \label{}\\
 \nonumber
 &=&\frac{5}{8\,\pi\,G\,R}\;\int\left\{\;\chi\,|\,^{(1)}\bar{W}(\chi)\,|^{\,2}\;{k}_2(\chi)\;\sin\epsilon_2(\chi)
 \,+\,2\chi\,|\,^{(2)}\bar{W}(2\chi)\,|^{\,2}\;{k}_2(2\chi)\;\sin\epsilon_2(2\chi)\;\right\}\,dS
 \,\;,\;\qquad
 \ea
 \es
 where $\Im\,$ denotes the imaginary part, while the factor of $\,\chi\,$ in the first term and that of $\,2\chi\,$ in the second term pop up due to the differentiation of $\,^{(1)}{\bar{U}}\,$ and $\,^{(2)}{\bar{U}}\,$, correspondingly.

 Insertion of expression (\ref{WW}), with the subsequent integration, renders us
 \ba
 \langle \,P\,\rangle_{\textstyle{_{\rm{cent}}}}=
 \frac{1}{18}\;R^7\;\rho^2\;\chi^2\;\dot{\theta}^2_{res}\;{\cal{A}}^{\,2}\;\chi\;\frac{{k}_2(\chi)\;\sin\epsilon_2(\chi)}{G\;R^{\,2}\;\rho^2}
 \;+\;
 \frac{1}{72}\;R^7\;\rho^2\;\chi^4\;{\cal{A}}^4\;2\,\chi\;\frac{{k}_2(2\chi)\;\sin\epsilon_2(2\chi)}{G\;R^{\,2}\;\rho^2}\,\;,
 \label{ichi}
 \ea
 where we have singled multipliers $\;R^7\,\rho^2\,\chi^2\,\dot{\theta}^2_{res}\,{\cal{A}}^2\;$ and $\;R^7\,\rho^2\,\chi^4\,{\cal{A}}^4\;$ in order to make expression (\ref{ichi}) look similar to expression (\ref{power_radial}) for the power damped by the radial deformation.

 As explained in Appendix \ref{chi}, equation (\ref{manner}), for a near-spherical Maxwell body the remaining multiplier can be approximated, at not too low frequencies, with
 \ba
 \chi\;\frac{{k}_2(\chi)\;\sin\epsilon_2(\chi)}{G\;R^{\,2}\;\rho^2}\;=\;
 2\chi\;\frac{{k}_2(2\chi)\;\sin\epsilon_2(2\chi)}{G\;R^{\,2}\;\rho^2}\;\approx\;\frac{2}{3}\;\frac{1}{\eta}\,\;,
 \label{manner1}
 \ea
 $\,\eta\,$ and $\,\rho\,$ being the shear viscosity and mean density. This yields:
 \ba
 \nonumber
 \langle \,P\,\rangle_{\textstyle{_{\rm{cent}}}}&\approx&
 \frac{1}{27\;\eta}\;R^7\;\rho^2\;\chi^2\;\dot{\theta}_{res}^{\,2}\;{\cal{A}}^{\,2}
 \;+\;\frac{1}{108\;\eta}\;R^7\;\rho^2\;\chi^4\;{\cal{A}}^4
 ~\\
 \label{libela}\\
 &=&\frac{1}{27\;\eta}\;R^7\;\rho^2\;\chi^2\;\dot{\theta}_{res}^{\,2}\;{\cal{A}}^2\;
 \left[1\;+\;\frac{1}{4}\;\left(\frac{\chi}{\;\stackrel{\bf\centerdot\;\;\;\;}{\theta_{res}}\,}\right)^2\;{\cal{A}}^2\right]\,\;.
 \nonumber
 \ea
 Dependent upon the spin-orbit resonance the body is in, its resonant spin rate $\,\dot{\theta}_{res}\,$ is equal either to $\,n\,$ or to $\,n\,$ multiplied by some integer or semiinteger.

 In the case of forced libration, the libration frequency $\,\chi\,$ is equal either to $\,n\,$, if we are talking about dissipation due to the main mode, or to $\,n\,$ multiplied by an integer, if we are interested in dissipation due to a higher mode. The magnitude of forced libration is usually small~---~about $\,0.103$ rad for Epimetheus, and much less for all the other known moons.
 %    Even in hypothetical settings it will hardly exceed $\,0.5$ rad.
 So we can safely drop the second term in the square brackets, i.e., neglect the power generated at the double frequency $2\chi$.

 In the case of free libration, the libration frequency $\,\chi\,$ is lower (usually, much lower) than $\,n\,$, see equation (\ref{publish}) below. Also recall that our formalism is valid for weak libration, $\,|{\cal{A}}|\lesssim0.2$ rad, because stronger libration is highly nonlinear and requires a separate study. So in the case of small-magnitude free libration we also may omit the second term in the square brackets above.

 Thus, no matter whether the libration at a frequency $\,\chi\,$ is forced or free, almost all power is generated at this same frequency $\,\chi\,$:
 \ba
 \langle \,P\,\rangle_{\textstyle{_{\rm{cent}}}}\;\approx\;\frac{1}{27\;\eta}\;R^7\;\rho^2\;\chi^2\;\dot{\theta}_{res}^{\,2}\;{\cal{A}}^2\;\,.
 \label{prior}
 \ea
 When the libration spectrum comprises several frequencies $\,\chi_j\,$, with magnitudes $\,{\cal{A}}_j\,$, we have:
 \ba
 \langle\,P\,\rangle_{\textstyle{_{\rm{cent}}}}\;\approx\;
 \frac{1}{27\,\eta}\;R^7\;\rho^2\;\dot{\theta}_{res}^{\,2}\;\sum_{j=1}^{\infty}\chi_j^2\;{\cal{A}}_j^2\,\;.
 \label{ni}
 \ea

 Consider the special case of forced libration. By equation (\ref{follows}), its spectrum is:
 \ba
 \chi_j\,=\,j\,n\;\,,\quad j \geq 1\;\,,\quad n\,\equiv\,\stackrel{\bf\centerdot}{{\cal{M}}\,}\;.
 \label{should}
 \ea
 These $\,\chi_j\,$ should then be plugged into series (\ref{ni}). However, the analysis in Section \ref{check} shows that for moderate eccentricities the series can be approximated with its first term:
 \ba
 \nonumber
 \mbox{Forced libration, ~~$\;e\,\lesssim\,0.5\;\,:$} \qquad \qquad \qquad \qquad \qquad \qquad \qquad \qquad  \qquad \qquad \qquad \qquad \qquad \qquad  \;\;\;  \\
 \langle\,P\,\rangle_{\textstyle{_{\rm{cent}}}}
 \approx
  \frac{\textstyle R^{\,7}\;\rho^2}{\textstyle 27\;\eta}\;\dot{\theta}_{\rm{res}}^{\,2}\;n^2\;{\cal{A}}_1^2
 =\left\{
 \begin{array}{ll}
  ^{(1:1)}{\cal{A}}_1^2\;\frac{\textstyle R^{\,7}\;\rho^2}{\textstyle 27\;\eta}\;n^4
  \quad,\quad & \mbox{in~the~1:1~resonance,}\\
 ~\\
 ~\\
 ^{(3:2)}{\cal{A}}_1^2\;\frac{\textstyle R^{\,7}\;\rho^2}{\textstyle 12\;\eta}\;n^4      \,\;,
 & \mbox{in~the~3:2~resonance.} \qquad\qquad
 \end{array}
 \right.
 \label{centrum}
 \ea
 On the right-hand side of the above equation, we equip the forced-libration magnitude $\,^{(z)}{\cal{A}}_j\,$ not only with a subscript but also with a superscript.
 Within the general convention established by equation (\ref{5}), the superscript specifies in which spin-orbit resonance the rotator is
 (in our case, $\,z =$1:1 \,or $\,z=$3:2). The subscript $\,j\,$ shows which mode of libration we are referring to (in our case, $\,j=1\,$, as we are considering the principal mode$\,n\,$). The expressions for the magnitudes $\,^{(1:1)}{\cal{A}}_1\,$ and $\,^{(3:2)}{\cal{A}}_1\,$ are given by equations (\ref{10}) and (\ref{11}).

 \subsection{Power exerted by the gravitational tides\label{grav}}

 \subsubsection{Forced libration}

 Tidal dissipation in the presence of libration was addressed, in various approximations, by Wisdom (2004), Tiscareno et al. (2009), Correia et al. (2014), and Caudal (2017). A complete solution for a homogeneous near-spherical body performing forced libration in longitude about an arbitrary spin-orbit resonance was written down by Frouard \& Efroimsky (2017a). This solution reads as
\begin{equation}
\begin{split}
\langle P \rangle_{\rm{tide}}  &
=
 \frac{G M^{*\,2}}{a} \sum_{l=2}^{\infty}    \bigg( \frac{R}{a} \bigg)^{2l+1}   \sum_{q^{\prime}=-\infty}^{\infty} \sum_{m=0}^l \sum_{p=0}^l \sum_{q=-\infty}^{\infty} \sum_{s=-\infty}^{\infty}
  \frac{(l-m)!}{(l+m)!} (2-\delta_{0m}) \, \\
  ~\\
&     \,\;\;  F_{lmp}^{\,2} (i)  \, G_{lpq^{\prime}}(e) \, G_{lpq}(e) \, J_{(q^{\prime}-q+s)} (m {\cal{A}}_1) J_s (m {\cal{A}}_1) \,  \beta_{lmpqs} \,k_l(\beta_{lmpqs}) \sin\epsilon_l(\beta_{lmpqs})\;\,.
\label{long}
\end{split}
\end{equation}
 Here $\,\beta_{lmpqs}\,$, $\,\epsilon_l(\beta_{lmpqs})\,$, and $\,k_l(\beta_{lmpqs})\,$ are the forced-libration-caused tidal modes and the corresponding phase lags and Love numbers. The expression for the tidal mode $\,\beta_{lmpqs}\,$ is given above by formula (\ref{beta}). For $\,s=0\,$, the spectrum $\,\beta_{lmpqs}\,$ coincides with that of the tidal modes $\,\omega_{lmpq}\,$ emerging in the Darwin-Kaula theory of tides with no libration:
 \ba
 \beta_{lmpq0}\,=\;\omega_{lmpq}\,\;.
 \label{}
 \ea
  Under no libration ($\,{\cal A}_1 \rightarrow 0\,$), only the terms with $\,s=0\,$ and $\,q^{\,\prime} = q\,$ are left in the expression for $\;\langle P \rangle_{\textstyle{_{\rm{tide}}}}\;$, and we recover the result from Efroimsky \& Makarov (2014, eqn 65).
 %  $\,$\footnote{~To see what happens to expression (\ref{long}) for $\,{\cal A} \rightarrow 0\,$, recall the convention $\,0^0\,=\,1\,$.}

 In Appendix \ref{contrlibr}, footnote (\ref{asymptotic}), we provide an asymptotic expression for the Bessel functions, valid for a small argument $\,m{\cal{A}}_1\,$.

 %  Our goal here is to provide a very rough estimate for $\,\langle P \rangle_{\rm{tide}}\,$, wherefore is suffices to use the above expression (\ref{long}).
 %  The values of the Bessel functions in that expression are less than unity.
 %  So we shall assume that the extra input due to libration will not add more than an order of magnitude to $\,\langle P \rangle_{\rm{tide}}\,$,
 %  and shall take into account only the modes with $\,s=0\,$, i.e., those corresponding to $\,\beta_{lmpq0}\,$.
 In an $\,lmpq\,$ spin-orbit resonance, the tidal mode $\,\beta_{lmpq0}\,=\,\omega_{lmpq}\,$ becomes zero, and so becomes the $\,lmpq0\,$ input into the power.
 In the absence of libration, we would then look for the largest input from the other quadruples $\,lmpq\,$. In the presence of libration, however, we also must check if the terms $\,\beta_{lmpqs}\,$ with $\,s\neq 0\,$ contribute noticeably to damping.

 \subsubsection{Free libration\label{642}}

 The analytical theory of libration~---~including equation (\ref{long})~---~works insofar as the libration angle $\,\gamma\,$ is small enough to satisfy the condition (e.g., Frouard \& Efroimsky 2017\,a):
 \ba
 \cos(2 \gamma) \simeq 1\quad \mbox{and} \quad\sin(2 \gamma) \simeq 2 \gamma\,\;.
 \label{condi}
 \ea
 This is fulfilled for forced libration, because it magnitude usually amounts to fractions of a degree or, at most, to several degrees. The case of free libration is more complex, because right after the capture the libration magnitude is barely short of $\,2\pi\,$, and libration is very non-sinusoidal. As the magnitude $\,{\cal{A}}\,$ is damped down to about $\,10 - 12\,$ angular degrees, the condition (\ref{condi}) begins to hold, so the libration becomes roughly sinusoidal:
 \ba
 \gamma(t)\;=\;{\cal{A}}\;\sin\chi t\;\;,\qquad|\,{\cal{A}}\,|\,\lesssim\,12^\circ\approx\,0.2\;\mbox{rad}\,,
 \label{}
 \ea
 and expression (\ref{long}) can be used. The natural frequency $\,\chi\,=\;^{(z)}\chi\,$ of sinusoidal free libration in a spin-orbit resonance $\,z\,$ is given by expression (\ref{publish}) and is usually much lower than $\,n\,$, Epimetheus being an exception.

 The power damped by such a sinusoidal mode of libration is given by an expression similar but not identical to equation (\ref{long}):
 \footnote{~The expression (\ref{short}) is but equation (\ref{long}) with $\,q^{\,\prime}=q\,$. This simplification becomes affordable, because in the case of  free libration we average the power not only over the orbital period, but also over the libration period. (In the case of forced libration these two periods coincide.)}
 \begin{equation}
 \begin{split}
 \langle P \rangle_{\rm{tide}}  &
 =
 \frac{G M^{*\,2}}{a} \sum_{l=2}^{\infty}    \bigg( \frac{R}{a} \bigg)^{2l+1}   \sum_{m=0}^l \sum_{p=0}^l \sum_{q=-\infty}^{\infty} \sum_{s=-\infty}^{\infty}
  \frac{(l-m)!}{(l+m)!} (2-\delta_{0m}) \, \\
  ~\\
 &     \,\;\;  F_{lmp}^{\,2} (i) \, G^{\,2}_{lpq}(e)\,J^{\,2}_{s}(m {\cal{A}})\,\beta_{lmpqs} \,k_l(\beta_{lmpqs}) \sin\epsilon_l(\beta_{lmpqs})\;\,.
 \label{short}
 \end{split}
 \end{equation}

 \subsubsection{Example 1. Forced libration about the 1:1 spin-orbit resonance\label{6.4.1}}

 As shown in Appendix \ref{powergrav}, the quadrupole part of the power exerted in the synchronised spin state is
 \ba
 ^{(1:1)}\langle P\rangle_{\rm{tide}}&=&^{(1:1)}\langle P\rangle_{\rm{tide}}^{\rm(main)}\,+\;^{(1:1)}\langle P\rangle_{\rm{tide}}^{\rm(forced)}\,+\;^{(1:1)}\langle P\rangle_{\rm{tide}}^{\rm(obliquity)}
 \nonumber\\
 \label{101}\\
  &=&\frac{G~{M^{\,*}}^{\,{2}}~R^{\,5}}{a^6}\,n\,k_2(n)\;\sin\epsilon_2(n)\;
 \left[\,\frac{21}{2}\;e^2\,-\;6\;{\cal{A}}_1\;{e}\;+\;\frac{3}{2}\;{\cal{A}}^2_1\;+\;\frac{3}{2}\;\sin^2 i\,\right]\;,\,\;\qquad\;
 \nonumber
  \ea
 where $\,{\cal{A}}_1\,$ is a shortened notation for the magnitude $\,^{(1:1)}{\cal{A}}_1\,$ of the forced-libration principal mode in the 1:1 spin-orbit resonance.
 This magnitude is given by expression (\ref{10}).

 If the frequency $\,n\,$ is not too low, a near-spherical Maxwell body satisfies the relation
 (\ref{manner}) derived in Appendix \ref{chi}. Employment thereof makes the power look as~\footnote{~Approximating the osculating mean motion $\,\sqrt{G(M^*+M)/a^3\,}\approx\sqrt{GM^*/a^3\,}\,$ with the anomalistic mean motion $\,n\equiv\,\stackrel{\bf{\centerdot}}{\cal{M}\,}$,
 we rewrite the common factor in equation (\ref{101}) as
 \ba
 \nonumber
 \frac{G~{M^{\,*}}^{\,{2}}~R^{\,5}}{a^6}\,n\,e^2\;k_2(n)\;\sin\epsilon_2(n)~=\;R^{\,7}\,\rho^2\,n^4\;e^2\;n\;\frac{k_2(n)\;\sin\epsilon_2(n)}{R^{\,2}
 \;\rho^2\;G}\,\;.
 \ea
 Then, combining the right-hand side of the above equality with the formula (\ref{manner}), and inserting the outcome into the equation (\ref{101}), we arrive at (\ref{103}).
  \label{foo}}
 \ba
 ^{(1:1)}\langle P\rangle_{\rm{tide}}\,=\,\frac{7}{\eta}\;R^{\,7}\,\rho^2\,n^4\,e^2
 \left[\,1\,-\;\frac{4}{7}\;\frac{{\cal{A}}_1}{e}\;+\;\frac{1}{7}\;\frac{{\cal{A}}^2_1}{e^2}\;+\;\frac{1}{7}\;\frac{\sin^2 i}{e^2}\,\right]\,\;.\;\qquad\;
 \label{103}
 \label{gra}
 \ea
 Recall that, according to equation (\ref{10}), the magnitude $\,{\cal{A}}_1\,$ of the principal mode of forced libration in the 1:1 spin-orbit resonance is negative. So all summands in the above expression are positive definite.

 In expression (\ref{gra}), both the principal part and the leading libration-caused input are associated with the forcing frequency $\,n\,$ in the material of the body. The quadrupole centrifugal input is associated with the forcing frequency $\,n\,$ also (see our comment prior to equation \ref{prior}).  Thence, as explained in Section \ref{difficulty},  the total power exerted by tides in the 1:1 resonance cannot be obtained simply by summing $\,^{(1:1)}\langle P\rangle_{\rm{tide}}\,$ and $\,\langle P\rangle_{\rm{cent}}\,$. Luckily, this is not a big deal because the tidal power due to the quadrupole part of the centrifugal force, (\ref{centrum}), is much lower than the power (\ref{gra}) due to the gravitational tides. So the expression (\ref{gra}) alone serves as a fair approximation to the overall tidal power.

 {\small
  \begin{table}[htbp]
 \begin{center}
 \begin{tabular}{lcccccc}
 \hline
 \hline
     parameter                 &   notation  &  Moon &     Phobos     &          Mimas          &          Enceladus        &        Epimetheus         \vspace{1mm}\\
 \hline
 \hline
 eccentricity                  &     $e$      & 0.055  &    0.015     &        $0.0202$         &          $0.0045$         &          $0.009$          \vspace{1mm}\\
 \hline
 equatorial\\
 obliquity\\
 to orbit                  &     $i$  &0.1166&    $\sim 10^{-5}$     &
 %  (5.8 --- 8.7)
 $
 % \times
 \sim 10^{-4}$         &          $\sim 10^{-6}$         &          $< 0.017$          \vspace{1mm}\\
 \hline
 forced\\
 libration\\
 magnitude\\
  (in rad) &${\cal{A}}_1$ &$-\;8.144\times 10^{-5}$&  $-\;0.021$   &        $-\;0.0146$       &          $-\;0.0021$      &          $-\;0.103$       \vspace{1mm}\\
 \hline
 libration\\
 -caused\\
 input in\\
   tidal heating
   &  $\,-\,\frac{\textstyle 4}{\textstyle 7}\,\frac{\textstyle {\cal{A}}_1}{\textstyle e}\,+\,\frac{\textstyle 1}{\textstyle 7}\,\frac{\textstyle {\cal{A}}^2_1}{\textstyle e^2}\,$   & $0.85\times 10^{-3}$ &   1.08   &        $0.49$           &          $0.30$         &         $25.25$          \vspace{1mm}\\
\hline
  obliquity\\
  -caused\\
  input in\\
  tidal heating
  &   $\frac{\textstyle \sin^2i}{\textstyle 7\;e^2}$        &  1.85   &    $\sim 10^{-3}$      &
  % (1.64 --- 2.46)
  $
  % \times
  \sim 10^{-2}$         &          $\sim 10^{-3}$         &          $<32$          \vspace{1mm}\\
  \hline
 \hline
 \end{tabular}
 \end{center}
 \caption{\small{.  The libration- and obliquity-generated inputs into the tidal dissipation, as compared to the main input (see equation \ref{gra}).
 Here $\,{\cal{A}}_1\,$ is a shortened notation for the main-mode forced-libration magnitude $\,^{(1:1)}{\cal{A}}_1\,$ in the 1:1 spin-orbit resonance. It is given in radians. The value of $\,{\cal{A}}_1\,$ for the Moon is borrowed from the second line of Table 5 in the Supplementary Electronic Materials to Rambaux \& Williams (2011). The value of $\,{\cal{A}}_1\,$ for Mimas is taken from Tajeddine et al. (2014). The value of $\,{\cal{A}}_1\,$ for Enceladus is from Thomas et al. (2016).  The values of $\,{\cal{A}}_1\,$ for Phobos and Epimetheus are from Rambaux (2014).
 The obliquities are those with respect to orbit. Their values are taken from Rambaux et al. (2012), for Phobos; from Noyelles et al. (2011), for Mimas; and from Baland et al. (2016), for Enceladus. The upper limit for Epimetheus' obliquity, $\;\sim 1^\circ=0.017$ rad$\,$, $\,$is borrowed from Tiscareno et al. (2009).}}
 \label{table2}
 \end{table}
 }

 From Table \ref{table2}, we see that the forced libration in longitude adds virtually nothing to the tidal friction in the Moon, while the Moon's equatorial obliquity on the orbit plane almost triples the tidally dissipated power. For Phobos the situation is opposite: while the obliquity is unimportant, the physical libration more than doubles the tidal power. In Mimas and Enceladus the obliquity produces almost no additional heat either, while the physical libration contributes noticeable portions.
 As demonstrated in Efroimsky (2018), the tidal power produced currently in Enceladus is sufficient to support the observed plume activity, provided the Enceladean mean viscosity is set equal to the viscosity of ice near melting point.

 The case of Epimetheus is most intriguing. Its latitudinal libration does not exceed $\,1^\circ\,$, a number we may use as the upper limit for the obliquity. This tells us that the obliquity-caused heat cannot exceed the main input by more than about 32 times. More definite is our evaluation of the input generated by the longitudinal physical libration: it is more than 25 times larger than the ``main'' input, one unrelated to libration or obliquity. So in application to this satellite the term ``main'' definitely needs quotation marks.

 As follows from equation (\ref{10}), the scatter in the libration contribution to tidal heating in different moons is explainable by the difference in their permanent  triaxiality.\,\footnote{~We would remind that the permanent triaxiality $\,(B=A)/C\,$ is essentially a \,{\it{dynamical}}\, triaxiality. This means that its value is
  defined not only by irregularity of shape but also by inhomogeneity of structure~---~and at times the inhomogeneity is of a greater importance. E.g., while the shape of Epimetheus is not as irregular as that of (say) Phobos, the Epimethean strong libration and, therefore, large dynamical triaxiality can be put down to the rock fraction frozen into this moon in an asymmetric manner.}

 \subsubsection{Example 2.\\ Superimposed free and forced libration about the 1:1 spin-orbit resonance\label{6.4.2}}

 Let us assume that the amplitude of free libration is not very large: $\,|{\cal{A}}|\,\lesssim\,12^\circ\approx\,0.2\,$ rad. Under this limitation, libration can be assumed sinusoidal, as was  explained in Section \ref{642}. The natural frequency $\,\chi\,=\,^{(z)}\chi\,$ of such libration in a spin-orbit resonance $\,z\,$ is given by expression (\ref{publish}). In the 1:1 spin-orbit resonance (i.e., for $\,z=1\,$), this frequency becomes
 \ba
 \chi
 %  \;=\;^{(z=1)}\chi
 \;\approx\;n\;\sqrt{\,3\;\frac{B-A}{C}\,}\,\;,
 \label{ravnina}
 \ea
 where for brevity we denote $\,^{(z=1)}\chi\,$ simply with $\,\chi\,$.

 Then employment of expression (\ref{short}) entails:
 \ba
 \nonumber
  ^{(1:1)}\langle P\rangle_{\rm{tide}}\,=\,^{(1:1)}\langle P\rangle_{\rm{tide}}^{\rm(main)}\,+\,^{(1:1)}\langle P\rangle_{\rm{tide}}^{\rm(free)}\,+\;^{(1:1)}\langle P\rangle_{\rm{tide}}^{\rm(obliquity)}\,=\qquad\;\qquad\;\qquad\;\qquad\;\qquad\;\;\;
  ~\\
  \label{a}\\
  \nonumber
  \frac{G~{M^{\,*}}^{\,{2}}~R^{\,5}}{a^6}\,n\,k_2(n)\;\sin\epsilon_2(n)
  \;\left[\,\frac{21}{2}\;e^2\,+\;2\,{\cal{A}}^2\;\frac{\chi\,k_2(\chi)\;\sin\epsilon_2(\chi)}{n\,k_2(n)\;\sin\epsilon_2(n)}\;
  \;+\;\frac{3}{2}\;\sin^2 i\,\right]\;.\,\;\qquad\;
  \ea
  Derivation of this expression repeats almost verbatim the derivation carried out in Appendix \ref{powergrav} for forced libration. The only difference is that here we obtain no term linear in $\,{\cal{A}}\,$, because such terms are averaged out in the expression (\ref{short}).

  Just like the dissipation due to forced libration, so the dissipation due to free libration can be leading, and can considerably (by orders of magnitude) enhance tidal heating.

 In realistic situations, a triaxial body will experience free libration superimposed with the forced libration. Together, the equations (\ref{101}) and (\ref{a}) will then give:
  \footnote{~Justification of expression (\ref{b}) is nontrivial. In the presence of both a forced libration with a magnitude $\,{\cal{A}}_1\,$ and a free
 libration with a magnitude $\,{\cal{A}}\,$, derivation of the total tidal power produces a sum of terms containing all the products $\;J_{q^{\,\prime}-q+s_1}(m {\cal{A}}(n)\,)\,J_{s_1}(m {\cal{A}}_1\,)\,J^2_{s_2}(m {\cal{A}}\,)\;$. Besides, in this situation the tidal modes are parameterised not with five but with six indices: $\,\beta_{lmpqs_1s_2}\,$. If however we intend to keep only the terms which are, at most, quadratic in the libration magnitudes, then the cross terms may be neglected and we shall be left only with the terms containing $\;J_{q^{\,\prime}-q+s}(m {\cal{A}}_1\,)\,J_{s}(m {\cal{A}}_1\,)\;$ and those containing $\;J^2_{s}(m {\cal{A}}\,)\;$. In this approximation, the total power will consist of two separate groups of terms. One group will comprise the terms due to the free libration, with the five-index modes $\,\beta_{lmpqs}\,$ given by (\ref{beta}), where $\,\chi\,$ is the free libration frequency. Another group will comprise the terms due to the forced libration, with the five-index modes $\,\beta_{lmpqs}\,$
 given by (\ref{beta}), where $\,\chi\,$ is set equal to $\,n\,$. Hence, in the said approximation, the expression (\ref{b}) can be used (Frouard \& Efroimsky 2017\,a).
 \label{justification}}
 \ba
 \nonumber
 ^{(1:1)}\langle P\rangle_{\rm{tide}}\,=\,^{(1:1)}\langle P\rangle_{\rm{tide}}^{\rm(main)}\,+\,^{(1:1)}\langle P\rangle_{\rm{tide}}^{\rm(forced)}\,+\,^{(1:1)}\langle P\rangle_{\rm{tide}}^{\rm(free)}\,+\;^{(1:1)}\langle P\rangle_{\rm{tide}}^{\rm(obliquity)}\,=\qquad\;\quad\;\quad\quad
  ~\\
  \label{b}\\
  \nonumber
 \frac{G{M^{\,*}}^{\,{2}}\,R^{\,5}}{a^6}\,n\,k_2(n)\;\sin\epsilon_2(n)\,
 \left[\frac{21}{2}\,e^2\,-\,6\,{\cal{A}}_1\,{e}\,+\,\frac{3}{2}\,{\cal{A}}^2_1\,+\,\frac{3}{2}\,{\cal{A}}^2\,\frac{\chi\,k_2(\chi)\;\sin\epsilon_2(\chi)}{n\,k_2(n)\;\sin\epsilon_2(n)}
 \;+\;\frac{3}{2}\;\sin^2 i\right]\;.
   \ea

 If we set the factor $\,k_2\,\sin\epsilon_2\,$ frequency-independent, the free-libration-produced term in the square brackets becomes equal to $\;\frac{\textstyle 3}{\textstyle 2}\,{\cal{A}}^2\,\frac{\textstyle \chi}{\textstyle n}\;$. Then, in the special case of $\,\chi/n = 1/3\,$, our expression (\ref{b}) will coincide with equation (45) from Wisdom (2004).$\,$\footnote{~Wisdom (2004) addressed a situation where the main mode of the forced libration with the frequency $\,n\,$ and magnitude $\,F\,$ was superimposed with additional libration, which originated due to a secondary resonance and which had a lower frequency $\,\chi=n/3\,$ and a magnitude $\,S\,$.
 Mathematically, it looked like free libration. Wisdom's notation for the magnitudes relates to our notation as $\,F\,=\,-\,{\cal{A}}_1\,$ and $\,S\,=\,{\cal{A}}\,$.}

 If however the body is Maxwell and the frequencies are not too low (see Appendix \ref{chi}), then the free-libration-produced term
 becomes simply $\,\frac{\textstyle 3}{\textstyle 2}\,{\cal{A}}^2\,$. Transforming the overall factor as explained in Footnote \ref{foo}, we write the total power as
 \ba
 \nonumber
 ^{(1:1)}\langle P\rangle_{\rm{tide}}\,=\,\qquad\;\qquad\;\qquad\;\qquad\;\;\;\qquad\,\qquad\;\qquad\;\qquad\;\qquad\;\;\;\qquad
  ~\\
  \label{c}\\
  \nonumber
 \frac{G~{M^{\,*}}^{\,{2}}\,R^{\,5}}{a^6}\,n\,k_2(n)\;\sin\epsilon_2(n)\;
 \left[\,\frac{21}{2}\,e^2\,-\,6\,{\cal{A}}_1\,{e}\,+\,\frac{3}{2}\,{\cal{A}}^2_1\,+\,\frac{3}{2}\,{\cal{A}}^2\,
 +\;\frac{3}{2}\;\sin^2 i\,\right]\;\,.\;\qquad\;\qquad
  \ea
 For a Maxwell body, the overall factor can be transformed as in (\ref{gra}). So we eventually obtain
  \ba
 ^{(1:1)}\langle P\rangle_{\rm{tide}}\,=\,\frac{7}{\eta}\;R^{\,7}\,\rho^2\,n^4\,e^2
 \left[\,1\,-\;\frac{4}{7}\;\frac{{\cal{A}}_1}{e}\;+\;\frac{1}{7}\;\frac{{\cal{A}}^2_1}{e^2}
 \;+\;\frac{1}{7}\;\frac{{\cal{A}}^2}{e^2}
 \;+\;\frac{1}{7}\;\frac{\sin^2 i}{e^2}\,\right]\,\;.\;\qquad\;
 \label{d}
 \ea

 When free libration has a large magnitude and is strongly nonsinusoidal, we can expand it over the Fourier modes $\,\chi_i\,$, each of these contributing its
 $\;\frac{\textstyle 1}{\textstyle 7}\;\frac{\textstyle{\cal{A}}^2_i}{\textstyle e^2}\;$ input into the power. This sum, however, will be truncated at the frequencies as low as the peak of the frequency-dependence of $\,k_2\,\sin\epsilon_2\,$. Below that peak, the intensity of the tidal friction quickly falls to zero, see Appendix \ref{chi}.

 \subsubsection{Example 3. Forced libration about the 3:2 spin-orbit resonance\label{6.4.3}}

 We keep the assumptions adopted in Section \ref{6.4.1}, except that now the tidally perturbed body is in the spin-orbit resonance $\,z\,=\,3/2\,$. The calculation in Appendix \ref{ken} renders the dissipated power within an arbitrary tidal model, i.e., for an arbitrary form of the frequency-dependence of $\,\beta\,k_l(\beta)\,\epsilon_l(\beta)\,$. In the special case of a Maxwell body at not too low frequencies, relation (\ref{manner}) works and makes the final answer
 much less cumbersome than in the general case. That answer is given by the expression (\ref{altogether}) from Appendix \ref{ken}:
 \ba
  ^{(3:2)}P_{\rm{tide}}\,=\,^{(3:2)}P_{\rm{tide}}^{\rm(main)}\,+\,^{(3:2)}P_{\rm{tide}}^{\rm({obliquity})}\,+\,^{(3:2)}P_{\rm{tide}}^{\rm(forced)}\,=\;\qquad\;\qquad\,\qquad\qquad\qquad\qquad\qquad\qquad
 \nonumber ~\\
 \label{gro}\\
 %  \ea
 %  \ba
     \frac{1}{2\eta}\,R^7\,\rho^2\,n^4\,\left[\left(1-\frac{13}{4}\,e^2\right)\,+\,\left(1\,+\,
 \frac{61}{4}\,e^2\right)\,\sin^2 i
  \,+\,
    7\,e\,{\cal{A}}_1\,\cos^2i\,-\,\left(1-\frac{\textstyle 193}{\textstyle 4}e^2\right) {\cal{A}}_1^2\,\cos^2i
   \right]
  \,\;,
  \nonumber
  \ea
  where $\,{\cal{A}}_1\,$ is a shortened notation for the main-mode forced-libration magnitude $\,^{(3:2)}{\cal{A}}_1\,$ in the 3:2 spin-orbit resonance. It is related to the permanent triaxiality via the equation (\ref{11}).

  Mind that power (\ref{gro}) comprises terms associated with several different physical frequencies, see Appendix \ref{ken}. Some of those terms are associated with the physical forcing frequency $\,n\,$. This means that our derivation of the above expression for power falls under the auspices of the difficulty explained in Section \ref{difficulty}. Fortunately, at small libration magnitudes this difficulty can be ignored because, as we shall see shortly, the centrifugal input is much smaller than all the terms kept in (\ref{gro}).

  An interesting fact revealed by expression (\ref{gro}) is that in the 3:2 spin-orbit state the forced libration does not necessarily add to tidal damping. Forced libration serves to reduce the damped power for $\,|{\cal{A}}_1|\,\gtrsim \,7\,e\,$, provided the eccentricity is small.

 \subsubsection{Example 4.\\ Superimposed free and forced libration about the 3:2 spin-orbit resonance\label{6.4.4}}

 We again assume that the magnitude of the free libration obeys: $\,|{\cal{A}}|\,\lesssim\,12^\circ\approx\,0.2\,$ rad, so the libration can be regarded sinusoidal, see Section \ref{642}.  According to expression (\ref{publish}), the frequency $\,\chi\,=\,^{(z)}\chi\,$ of sinusoidal free libration about the resonance $\,z=3/2\,$ is equal to
 \ba
 \chi\;\approx\;n\;\sqrt{\,\frac{21}{2}\;e\;\frac{B-A}{C}\,}\,\;,
 \label{dolina}
 \ea
 where we denote $\,^{(3:2)}\chi\,$ with $\,\chi\,$, to simplify notation.

 Like in the preceding subsection, we shall assume that the body is Maxwell and that the frequencies are not exceedingly low. So the expression (\ref{manner}) can be employed.
 Then, owing to the explanation provided in Footnote \ref{justification}, the total tidal power will look as
    \ba
 \nonumber
   ^{(3:2)}P_{\rm{tide}}\,=\,^{(3:2)}P_{\rm{tide}}^{\rm(main)}\,+\,^{(3:2)}P_{\rm{tide}}^{\rm({obliquity})}\,+\,^{(3:2)}P_{\rm{tide}}^{\rm(forced)}\,+\,^{(3:2)}P_{\rm{tide}}^{\rm(free)}\,=\;\qquad\;\qquad\,\qquad\qquad\qquad\qquad\\
 \label{grot}\\
 \nonumber
  \frac{1}{2\eta}\,R^7\,\rho^2\,n^4\,\left[\left(1-\frac{13}{4}\;e^2\right)\,+\,\left(1+\frac{61}{4}\,e^2\right)\,\sin^2 i
   \,+\,
    7\,e\,{\cal{A}}_1\,\cos^2i\,-\,\left(1-\frac{\textstyle 193}{\textstyle 4}e^2\right) {\cal{A}}_1^2\,\cos^2i
    \right.  \qquad\;
  ~\\
  \nonumber\\
   \left.
  +\,\left(1\,+\,\frac{39}{2}
 \,e^2\right)\,{\cal{A}}^2\,\cos^2i\,
   \right]
  \,\;,\qquad
  \nonumber
  \ea
 where the term with $\,{\cal{A}}\,$ is due to the free libration. This input is derived in Appendix \ref{pizda}, equation (\ref{zaebalo}).
  Recall that we limited the value of $\,|\,{\cal{A}}\,|\,$ by $\,0.2\,$ rad, to ensure that libration is sinusoidal. The case of large-magnitude free libration needs a separate study.

 \section{Comparison of the damping rates}

 We have derived expressions for dissipation rate in a homogeneous near-spherical body performing small-amplitude libration in longitude. Below is a list of the inputs into the power, written for a Maxwell body.

 To each of the formulae below, we provide a comment explaining for what spin-orbit resonance the formula is intended. Therefore, to save type, we keep using the shortened notation: simply $\,\chi\,$ for $\,^{(z)}\chi\,$, and $\,{\cal{A}}_1\,$ for $\,^{(z)}{\cal{A}}_1\,$.

 In the 1:1 spin-orbit resonance, the free-libration frequency $\,\chi\,=\,^{(1:1)}\chi\,$ is given by expression (\ref{ravnina}), while the magnitude of the main-mode forced libration,
 $\,{\cal{A}}_1\,=\,^{(1:1)}{\cal{A}}_1\,$ is rendered by equation (\ref{10}).

 In the 3:2 resonance, the frequency $\,\chi\,=\,^{(3:2)}\chi\,$ is given by expression (\ref{dolina}), while the forced-libration magnitude
 $\,{\cal{A}}_1\,=\,^{(3:2)}{\cal{A}}_1\,$ is furnished by (\ref{11}).

 As ever, $\,{\cal{A}}\,$ will denote the magnitude of free libration.

 \subsection{Inventory of formulae}

 When neither the forced libration magnitude $\,{\cal{A}}_1\,$ nor the free libration magnitude $\,{\cal{A}}
 %  _{\textstyle{_{\rm{\,free}}}}
 \,$ exceed $\,\sim 12^\circ\approx\,0.2\,$ rad, the power generated by the gravitational tides is
 \ba
 \langle\,P\,\rangle_{\textstyle{_{\rm{tide}}}}=
 \left\{
 \begin{array}{ll}
  \frac{\textstyle 7}{\textstyle \eta}\,R^{\,7}\,\rho^2\,n^4\,
 \left[e^2\,-\,\frac{\textstyle 4}{\textstyle 7}\;\,
 {\cal{A}}_1\,e\,+\,\frac{\textstyle 1}{\textstyle 7}\,\;
  {\cal{A}}^2_1
   \right. ~\\ ~\\ \qquad\qquad\;\qquad \left.
 \,+\,\frac{\textstyle 1}{\textstyle 7}\, {\cal{A}}^2
  \,+\,\frac{\textstyle 1}{\textstyle 7}\, \sin^2 i\,\right]\,\;,            & \mbox{in~the~1:1~resonance,}\\
 ~\\
 ~\\
  \frac{\textstyle 1}{\textstyle 2\eta}\;R^7\,\rho^2\,n^4\;\left[\,\left(1\;-\;\frac{\textstyle 13}{\textstyle 4}\;e^2\right)\;+\;\left(1\;+\;
 \frac{\textstyle 61}{\textstyle 4}\,e^2\right)\;\sin^2 i
  % \,\right]\;\beta\,k_2(\beta)\,\sin\epsilon_2(\beta)
 \right.
 ~\\
 ~\\
  \qquad\;\;
 \;+\;
    7\,e\,\;
  %  ^{\textstyle{^{{(3:2)}}}}
    {\cal{A}}_1\,\cos^2i\,-\,\left(1\,-\,\frac{\textstyle 193}{\textstyle 4}\,e^2\right)\;\,
  %  ^{\textstyle{^{{(3:2)}}}}
    {\cal{A}}_1^2\,\cos^2i
  ~\\
  ~\\
    \left.
  \qquad\,\quad
  +\,\left(1\,+\,\frac{\textstyle 39}{\textstyle 2}
 \,e^2\right)\,{\cal{A}}^2
 %  _{\textstyle{_{\rm{\,free}}}}
 \,\cos^2i\,
   \right]\,\;,
 & \mbox{in~the~3:2~resonance}. \qquad
 \end{array}
 \right.
 \label{pow}
 \ea

 To write down the dissipation rate produced by the quadrupole part of the centrifugal force, we use the formulae (\ref{ni}) and (\ref{centrum}),
 and recall that the resonant spin rate $\,\dot{\theta}_{res}\,$ is equal to $\,n\,$, in the 1:1 spin-orbit state, and to $\,3n/2\,$ in the 3:2 resonance. This yields:
 \ba
 \nonumber
 \langle\,P\,\rangle_{\textstyle{_{\rm{cent}}}}&=&\langle\,P\,\rangle_{\textstyle{_{\rm{cent}}}}^{\rm(forced)}\,+\,\langle\,P\,\rangle_{\textstyle{_{\rm{cent}}}}^{\rm(free)}
  ~\\
    %    \nonumber\\
    %    \nonumber
    %    &\approx&\frac{1}{27\,\eta}\;R^7\;\rho^2\;\dot{\theta}^{\,2}_{res}\;n^2\;{\cal{A}}(n)^2\,+\,
    %    \frac{1}{27\,\eta}\;R^7\;\rho^2\;\dot{\theta}^{\,2}_{res}\;\chi^2\;{\cal{A}}(\chi)^2
    %    ~\\
    %    \nonumber\\
 \nonumber\\
  &=&\left\{
 \begin{array}{ll}
  %  ^{\textstyle{^{{(1:1)}}}}
  {\cal{A}}_1^2\;\,\frac{\textstyle \,R^{\,7}\,\rho^2\,n^4\,}{\textstyle 27\;\eta}\;+\,{\cal{A}}^2\,\frac{\textstyle \,R^7\,\rho^2\,n^2\;\chi^2\,}{\textstyle 27\;\eta} \quad,\qquad & \mbox{~in~the~1:1~resonance,}\\
 ~\\
 ~\\
  %  ^{\textstyle{^{{(3:2)}}}}
  {\cal{A}}_1^2\;\,\frac{\textstyle \,R^{\,7}\,\rho^2\,n^4\,}{\textstyle 12\;\eta}\;+\,{\cal{A}}^2\,\frac{\textstyle \,R^7\,\rho^2\,n^2\;\chi^2\,}{\textstyle 12\;\eta}\;\;\,\;,
 & \mbox{~in~the~3:2~resonance,} \qquad\qquad
 \end{array}
 \right.
 \label{power_cent}
 \ea
  The powers produced by the radial and toroidal deformations are given, correspondingly, by
 \ba
 \nonumber
 \langle\,P\,\rangle_{\textstyle{_{\rm{rad}}}}&=&\langle\,P\,\rangle_{\textstyle{_{\rm{rad}}}}^{\rm(forced)}\,+\,\langle\,P\,\rangle_{\textstyle{_{\rm{rad}}}}^{\rm(free)}\\
 \nonumber\\
  \label{power_rad}
  &=&
   \left\{
 \begin{array}{ll}
  %  ^{(1:1)}
  {\cal{A}}_1^2\;\frac{\,\textstyle R^{\,7}\,\rho^2\;n^4\,}{\textstyle 4.7\;\zeta}\;+\,{\cal{A}}^2\;\frac{\textstyle R^{\,7}\,\rho^2\,n^2\,\chi^2}{\textstyle 4.7\;\zeta}
  \quad,\quad & \mbox{in~the~1:1~resonance,}\\
 ~\\
 ~\\
 %  ^{(3:2)}
 {\cal{A}}_1^2\;\frac{\,\textstyle R^{\,7}\,\rho^2\;n^4\,}{\textstyle 2.1\;\zeta}\;+\,{\cal{A}}^2\;\frac{\textstyle R^{\,7}\,\rho^2\,n^2\,\chi^2}{\textstyle 2.1\;\zeta}      \,\;,
 & \mbox{in~the~3:2~resonance,} \qquad\qquad
 \end{array}
 \right.
  \,\;
 \ea
 \ba
 \nonumber
 \langle\,P\,\rangle_{\textstyle{_{\rm{tor}}}}&=&\langle\,P\,\rangle_{\textstyle{_{\rm{tor}}}}^{\rm(forced)}\,+\,\langle\,P\,\rangle_{\textstyle{_{\rm{tor}}}}^{\rm(free)}\\
 \nonumber\\
 \label{power_tor}
 &=&
 \left\{
 \begin{array}{ll}
  %  ^{(1:1)}
  {\cal{A}}_1^2\;\frac{\,\textstyle R^{\,7}\;\rho^2\;n^4\,}{\textstyle 17\;\eta}\;+\;{\cal{A}}^2\,\frac{\,\textstyle R^{\,7}\;\rho^2\;\chi^4\,}{\textstyle 17\;\eta}
  \quad,\quad & \mbox{in~the~1:1~resonance,}\\
 ~\\
 ~\\
 %  ^{(3:2)}
 {\cal{A}}_1^2\;\frac{\,\textstyle R^{\,7}\;\rho^2\;n^4\,}{\textstyle 17\;\eta}\;+\;{\cal{A}}^2\,\frac{\,\textstyle R^{\,7}\;\rho^2\;\chi^4\,}{\textstyle 17\;\eta}      \,\;,
 & \mbox{in~the~3:2~resonance.} \qquad\qquad
 \end{array}
 \right.
 \,\;
 \ea
 where $\,\eta\,$ and $\,\zeta\,$ are the shear and bulk viscosities (the rheology being set Maxwell).

 In expressions (\ref{power_cent} - \ref{power_tor}) we used the afore-proven fact that, for not too high eccentricities, an overwhelming share of heat due to the forced libration is produced at the principal frequency $\,n\,$. We also assumed that most heat due to the free libration is produced at the frequency $\,\chi\,$ which is rendered by expressions (\ref{ravnina}) and (\ref{dolina}) in the 1:1 and 3:2 spin-orbit resonances, accordingly. From these expressions, we observe that $\,\chi\,$ is equal to $\,n\,$ multiplied by a usually small factor: $\;\sqrt{3\,\frac{\textstyle B-A}{\textstyle C}\,}\;$ in the 1:1 resonance, and $\;\sqrt{\frac{\textstyle 21}{\textstyle 2}\,e\,\frac{\textstyle B-A}{\textstyle C}\,}\;$ in the 3:2 resonance. This tells us that in most situations the contribution from free libration is much lower than that from forced libration, especially in the 3:2 spin-orbit resonance.$\,$\footnote{~Among synchronised moons, Epimetheus may, arguably, be a rare exception of this rule.} Thus we end up simply with

 \ba
 \langle\,P\,\rangle_{\textstyle{_{\rm{cent}}}}\,\approx\,\langle\,P\,\rangle_{\textstyle{_{\rm{cent}}}}^{\rm(forced)}
  =\;\left\{
 \begin{array}{ll}
  %  ^{\textstyle{^{{(1:1)}}}}
  {\cal{A}}_1^2\;\frac{\textstyle \,R^{\,7}\,\rho^2\,n^4\,}{\textstyle 27\;\eta}\quad,\quad & \mbox{~in~the~1:1~resonance,}\\
 ~\\
 ~\\
  %  ^{\textstyle{^{{(3:2)}}}}
  {\cal{A}}_1^2\;\frac{\textstyle \,R^{\,7}\,\rho^2\,n^4\,}{\textstyle 12\;\eta}\;\,\;,
 & \mbox{~in~the~3:2~resonance.}
 \end{array}
 \right.
 \,\;
 \label{power_cent1}
 \ea
 \ba
 \langle\,P\,\rangle_{\textstyle{_{\rm{rad}}}}\,\approx\,\langle\,P\,\rangle_{\textstyle{_{\rm{rad}}}}^{\rm(forced)}\,=\,
 \left\{
 \begin{array}{ll}
  %  ^{(1:1)}
  {\cal{A}}_1^2\;\frac{\,\textstyle R^{\,7}\;\rho^2\;n^4\,}{\textstyle 4.7\;\zeta}
  \quad,\quad & \mbox{in~the~1:1~resonance,}\\
 ~\\
 ~\\
 %  ^{(3:2)}
 {\cal{A}}_1^2\;\frac{\,\textstyle R^{\,7}\;\rho^2\;n^4\,}{\textstyle 2.1\;\zeta}      \,\;,
 & \mbox{in~the~3:2~resonance.} \qquad\qquad
 \end{array}
 \right.\,\;
 \label{power_rad1}
 \ea
 \ba
 \langle\,P\,\rangle_{\textstyle{_{\rm{tor}}}}\;\approx\;\langle\,P\,\rangle_{\textstyle{_{\rm{tor}}}}^{\rm(forced)}\;=\;
 \left\{
 \begin{array}{ll}
  %  ^{(1:1)}
  {\cal{A}}_1^2\;\frac{\,\textstyle R^{\,7}\;\rho^2\;n^4\,}{\textstyle 17\;\eta}
  \quad,\quad & \mbox{in~the~1:1~resonance,}\\
 ~\\
 ~\\
  %  ^{(3:2)}
 {\cal{A}}_1^2\;\frac{\,\textstyle R^{\,7}\;\rho^2\;n^4\,}{\textstyle 17\;\eta}      \,\;,
 & \mbox{in~the~3:2~resonance.} \qquad\qquad
 \end{array}
 \right.\,\;
 \label{power_tor1}
 \ea

  Recall that in different spin-orbit resonances the forced-libration magnitude $\,{\cal{A}}_1\,$ looks differently. It is given by expression (\ref{10}) in the 1:1 resonance, and by (\ref{11}) in the 3:2 resonance. Both of those expressions are valid for $\,\chi\ll n\,$. Both must be amended with a factor of $\,1/[1\,-\,(\chi/n)^2]\,$ otherwise.

 \subsection{Comparison}

 \subsubsection{Power inputs from different types of deformation}

 From formulae (\ref{pow}) and (\ref{power_cent1} - \ref{power_tor1}) it ensues that
 \bs
 \ba
 %    \;\,\left\{ \begin{array}{ll}
  \langle\,P\,\rangle_{\textstyle{_{\rm{rad}}}}\,\ll\,
 \langle\,P\,\rangle_{\textstyle{_{\rm{cent}}}}\,\lesssim\,\langle\,P\,\rangle_{\textstyle{_{\rm{tor}}}}
  \,\ll\, \langle\,P\,\rangle_{\textstyle{_{\rm{tide}}}}
  ~~~            & \mbox{in~the~1:1~resonance}\;\,,
  \label{result_forced_1:1}\\
 \nonumber\\
  \langle\,P\,\rangle_{\textstyle{_{\rm{rad}}}}\,\ll\,
 \langle\,P\,\rangle_{\textstyle{_{\rm{tor}}}}\,\lesssim\,\langle\,P\,\rangle_{\textstyle{_{\rm{cent}}}}
 \,\lesssim\, \langle\,P\,\rangle_{\textstyle{_{\rm{tide}}}}~~~            & \mbox{in~the~3:2~resonance}\;\,.
 \label{result_forced_3:2}
 %   \end{array} \right.
 \ea
 \es

 Designating the input $\,\langle\,P\,\rangle_{\textstyle{_{\rm{rad}}}}\,$ as the smallest one, we assumed that $\,\eta\ll\zeta\,$, an inequality fulfilled for solids. Whether this inequality is valid also for semimolten materials~---~depends on the porosity and percentage of partial melt. For details, see Footnote \ref{Kalous} in Appendix \ref{AppendixRad}.

 When comparing $\,\langle\,P\,\rangle_{\textstyle{_{\rm{cent}}}}\,$  with $\,\langle\,P\,\rangle_{\textstyle{_{\rm{tide}}}}\,$ in the 3:2 resonance case, we wrote ``$\,\lesssim\,$'', because in that resonance the centrifugal tidal deformation can provide a power input of more than $\,\sim 1/6\,$ of the power exerted by the gravitational tides.

 Our calculations are intended for weak libration: $\,|{\cal{A}}_1|\;,\;|{\cal{A}}|\lesssim 12^{\,\circ}\approx 0.2$ rad. This is a situation where the theory of libration can be linearised, the free and forced components of libration can be separated from one another, and the Fourier expansion is straightforward (see, e.g., Frouard \& Efroimsky 2017a and references therein).

 Generalisation of our theory to strong libration will be trivial for the radial and toroidal inputs, will demand some attention for the centrifugal input, and will require a formidable effort for calculation of the tidal power.\,\footnote{~Our expressions (\ref{power_rad1} - \ref{power_tor1}) are valid for an arbitrarily large magnitude of libration, forced or free. Both expressions give contributions due to the main librational harmonic $\,n\,$ with a magnitude $\,{\cal{A}}_1\,$. Under strong libration, when the linear theory of libration is no longer valid but the total libration still can be expanded into a Fourier series, these expressions remain in force for every Fourier harmonic, separately.

 While the expression (\ref{power_cent1}) for the centrifugal input is correct for weak libration, in the case of strong libration an extra term must be taken into account, see equation (\ref{libela}). The so-amended expression will become valid for any harmonic entering the libration spectrum.

 Our expressions for the forced-libration-caused and free-libration-caused parts of the tidal power are correct only for weak libration. For larger magnitudes, separation of libration into a free and forced part cannot be performed, the procedure of time averaging is no longer straightforward, while the Bessel functions in the expression for the tidal power can no longer be approximated with their asymptotic expressions like in Footnote \ref{asymptotic}. It is not apparently clear if the analytical calculation of the tidal power can at all be generalised to the strong-libration case.\label{above}}

 \subsubsection{Components of the power exerted by the gravitational tides}

 Equation (\ref{pow}) tells us that in the 1:1 spin-orbit resonance, dependent upon the values of the eccentricity and the libration magnitude, both the forced- and free-libration-caused tidal inputs may be smaller, or comparable, or larger than the ``main'' tidal contribution to heating.
 The tidal power damped due to the forced libration about the 1:1 spin-orbit resonance was illustrated with examples in Table \ref{table2}. There we saw that, under synchronicity, the forced-libration-caused contribution to the power can be leading and even overwhelmingly leading (as was predicted by Makarov \& Efroimsky 2014).

 The case of forced libration about the 3:2 spin-orbit resonance lacks this intrigue. As can be seen from equation (\ref{pow}), the main input is always leading in that case, with libration providing only a small correction to power.

 Also recall that, by formulae (\ref{10}) and (\ref{11}), the forced-libration $\,{\cal{A}}_1\,$ is negative in the 1:1 resonance and is positive in the 3:2 resonance. Then we observe from equation (\ref{pow}) that the forced-libration-caused input into the tidal power is always positive in the 1:1 case. In the 3:2 case, however, the sign of this contribution depends upon the values of $\,{\cal{A}}_1\,$ and $\,e\,$.

 \section{Conclusions}

 We have demonstrated that under weak libration in longitude, forced or free, most dissipation in a homogeneous rotator is due to the gravitational tides (including the additional tides generated by libration). The other three sources of dissipation~---~which are the alternating parts of the centripetal, toroidal and purely radial deformations~---~are less important when libration is weak. Whether this is so for large-magnitude libration requires a separate study, because some parts of our calculation are valid only for libration not exceeding $\sim 0.2$ rad, see Footnote \ref{above}.

 In the 3:2 spin-orbit resonance, the input of forced libration in the gravitational tides' power is relatively small. Likewise, small is the input of free libration, provided libration is weak.

 In the 1:1 case, the inputs of both forced and free libration in the gravitational tides' power can be considerable and even leading~---~and even overwhelmingly leading. Independent of the rheology, forced libration in longitude is responsible for 52\% of the tidally dissipated power in Phobos, 33\% in Mimas, 23\% in Enceladus, and 96\% in Epimetheus.

 The example of Epimetheus demonstrates that, as was hypothesised in Makarov \& Efroimsky (2014), tidal heating due to libration might have contributed to the early heating of some large moons (and, perhaps, of some very close-in terrestrial exoplanets). One possibility may be the following: (1) a synchronised moon gets chipped by a collision, but remains synchronised; (2) the resulting change of shape produces a much higher permanent triaxiality and, thereby, a much higher amplitude of the forced libration; (3) the intensified libration leads to a one to two orders of magnitude more intense tidal heating; (4) this additional heating softens up the body, so its permanent triaxiality decreases to a value for which tidal heating becomes less intense, and the body freezes. This scenario may be of use to explain why some of the Saturnian large icy moons (Enceladus, Iapetus) warmed up in the past, while others (Mimas) remained cold.

 \section*{Acknowledgments}

 The author is grateful to Julien Frouard, Valery Makarov, and Tim Van Hoolst for their kind consultation.
 This research has made use of NASA's Astrophysics Data System\\

 \pagebreak

 \appendix
 \begin{center}
  {\underline{\Large{\bf{Appendix}}}}
 \end{center}

 \section{Comparison of the tidal and permanent-triaxiality-caused torques\label{comparison}}

 Our treatment pertains to the situations where
 \ba
 |\,\vec{\cal{T}}^{\rm{^{\,(TIDE)}}}\,|\,\ll\;|\,\vec{\cal{T}}^{\rm{^{\,(TRI)}}}\,|\;\,.
 \label{inequa}
 \ea
 Although this inequality is not guaranteed to be obeyed \,{\it{a priori}}, in realistic settings it is satisfied, as ensues from the estimates below.

 \subsection{Scheme of the estimate}

 A homogeneous ellipsoid of mass $\,M\,$ and the principal semiaxes $\,a\,>\,b\,>\,c\,$ has the moments of inertia: $\,A\,=\,M\,(b^2+c^2)/5\,$, $\,B\,=\,M\,(c^2+a^2)/5\,$, $\,C\,=\,M\,(a^2+b^2)/5\,$. Hence its triaxiality is $\,(B-A)/C\,=\,(a^2-b^2)/(a^2+b^2)\,$. When the shape is not very different from oblate, we have $\,a\approx R+u\,$ and $\,b\approx R-u\,$, where $\,R\,$ is the mean equatorial radius. Thence,
 \ba
 \frac{B-A}{C}\;\approx\;\frac{4\,R\,u}{2\,R^2}\;=\;2\;\frac{u}{R}\;\,.
 \label{}
 \ea
 Interpreting $\,u\,$ as the tidal elevation, we shall use the above formula as a rough estimate of the tidal input in the triaxiality.

 In the quadrupole approximation, the tidal elevation is
 \ba
 u\;=\;h_2\;\frac{W_2}{\mbox{g}}\,\;,
 \label{}
 \ea
 where \,g\, is the near-surface gravitational acceleration, and $\,W_2\,$ is the quadrupole part of the tide-raising potential. (We, of course, imply the Fourier components of both $\,u\,$ and $\,W_2\,$.)

 For a near-spherical body of mass $\,M\,$ and radius $\,R\,$, we have:
 \ba
 \mbox{g}\,=\,\frac{GM}{R^2}\,\;.
 \label{}
 \ea
 The semidiurnal component of the perturbing potential is equal to $\,G_{200}(e)\,G\,M^{\,*}\,R^2/a^3\,\approx\,G\,M^{\,*}\,R^2/a^3\,$, where $\,M^{\,*}\,$ is the mass of the perturber and $\,a\,$ is the major semiaxis:
 \ba
 W_2\,\approx\;\frac{G\,M^{\,*}}{a}\;\left(\frac{R}{a}\right)^2\,\;.
 \label{}
 \ea
 Gleaning these formulae together, we end up with
 \ba
 \frac{u}{R}\;\approx\;h_2\;\frac{R^{\,3}\,n^2}{G\,M}\,\;.
 \label{}
 \ea
 Thence
  \ba
 \left(\frac{B-A}{C}\right)_{\rm{tidal}}\;\approx\;2\;h_2\;\frac{R^{\,3}\,n^2}{G\,M}\,\;,
 \label{luna}
 \ea
 where we approximated the mean motion with its Keplerian value: $\,n^2\,\approx\,G\,(M^{\,*}+M)/a^3\,\approx\,G\,M^{\,*}/a^3\,$.

 \subsection{The Moon}

 For the Moon, the needed orbital and physical parameters are: $\;n\,\approx\,2.665\,\times\,10^{-6}$ s$^{-1}\,$, $\;M\,=\,7.342\,\times\,10^{22}$ kg\,, $\;R\,=\,1.737\,\times\,10^6\,$ m\,, $\;G\,=\,6.674\,\times\,10^{-11}$ m$^3$ kg$^{-1}$ s$^{-2}\,$. As was determined by the GRAIL mission
 (Williams et al. 2014), the quadrupole lunar Love number is $\,h_2\,=\,0.045\,$. Insertion of all these values in the above formula furnishes:
 \ba
 \frac{u}{R}\;\approx\;3.4\,\times\,10^{-7}\,\;,
 \label{}
 \ea
 which is only several times larger than the actually measured deformation.\,\footnote{~The observed elevation of about $\,0.1$ m\, entails $\,u/R\,\approx\,0.56\,\times\,10^{-7}\,$.} The difference may be put down to our neglect of the layered structure of the Moon. For our crude estimate, this difference is unimportant.

 So the tidally induced triaxiality of the Moon is about
 \ba
 \left(\frac{B-A}{C}\right)_{\rm tidal}\;\approx\;2\;\frac{u}{R}\;\approx\;10^{-6}\;\,.
 \label{121}
 \ea

 At the same time, its permanent triaxiality is
 \ba
 \left(\frac{B-A}{C}\right)_{\rm permanent}\;\approx\;\frac{1}{2}\;f\;\,,
 \label{}
 \ea
 where the lunar equatorial flattening $\,f\;\equiv\;(a\,-\,b)/a\,$ is of the order of $\,1.2\,\times\,10^{-3}\,$ (see the Moon Fact Sheet provided by NASA).

 Comparison of the former and the latter formulae supports the validity of inequality (\ref{inequa}).

 \subsection{Epimetheus}

 As derived from the measured forced libration, the permanent-figure triaxiality of Epimetheus is much higher than might have been expected from Epimetheus' shape (Tiscareno et al. 2009):
 \ba
 \left(\frac{B-A}{C}\right)_{\rm permanent}\;=\;0.296\;\,.
 \label{perm}
 \ea
 The only explanation to such a strong triaxiality is an nonuniform distribution of the rock fraction over the volume.

 To calculate the tidally-induced triaxiality, we must insert in formula (\ref{luna}) the following parameters: $\;n\,\approx\,1.05\,\times\,10^{-4}$ s$^{-1}\,$, $\;M\,=\,0.53\,\times\,10^{18}$ kg\,, $\;R\,=\,0.52\,\times\,10^5\,$ m\,, $\;G\,=\,6.674\,\times\,10^{-11}$ m$^3$ kg$^{-1}$ s$^{-2}\,$.
 We also need to estimate the Epimethean Love number $\,h_2\,$. Epimetheus being rigid, it will not be very wrong to assume that $\,h_2\,\approx\,5k_2/3\,$ and to approximate $\,k_2\,$ with its static counterpart~\footnote{~To examine the validity of the latter approximation, we have to start out with the exact expression for the dynamical Love number, see equation (49b) in Efroimsky (2015). In that equation, the tidal frequency $\,\chi\,$ is equal to the mean motion (provided the moon is synchronised), while the Maxwell time is the ratio of the viscosity and the rigidity (which are, for cold ices, $\,\gtrsim 10^{17}$ Pa s \,and $\,\sim 10^{10}$ Pa, correspondingly). Insertion of these values in the said formula shows that, indeed, the absolute value of the complex Love number is, in this situation, very close to the static Love number.}
 \ba
 k^{\textstyle{^{(static)}}}_{2}\,=\;\frac{3}{2}\;\,\frac{1}{1\;+\;A^{\textstyle{^{(static)}}}_{2}}~~~,
 \label{Mga}
 \ea
 where the quantity
 \ba
 A^{\textstyle{^{(static)}}}_2=\,\frac{\textstyle 57}{\textstyle 8\,\pi}~\frac{\textstyle \mu(\infty)}{\textstyle
 G\,\rho^2\,R^2}\,\;,
 \label{}
 \ea
 sometimes denoted as $\,\tilde{\mu}\,$, is a dimensionless measure of strength-dominated versus gravity-dominated behaviour. The notation $\,\mu(\infty)\,$ stands for the relaxed shear modulus, which is of the order of $\,10^{10}$ Pa\, for cold ices.

 As can be seen from Table 1 in Efroimsky (2012\,b), the quantity $\,A^{\textstyle{^{(static)}}}_2\,$ is lower than unity for large super-Earths, is barely higher than unity for the solid Earth, and is much higher than unity for smaller terrestrial bodies. Simply speaking, for large bodies self-gravitation is more important than rheology, while for small bodies rheology ``beats'' gravity. Hence, for small bodies,
 \ba
 k^{\textstyle{^{(static)}}}_{2}\,\approx\;\frac{3}{2}\;\,\frac{1}{A^{\textstyle{^{(static)}}}_2}\;=\;\frac{\textstyle 4\,\pi}{19}\;\frac{\textstyle G\,\rho^2\,R^2}{\textstyle \mu(\infty)}\,\;.
 \label{k}
 \ea
 Epimetheus' mean density being $\,640$ kg m$^{-3}\,$, \,we obtain for this satellite:
 \ba
 k^{\textstyle{^{(static)}}}_{2}\,\approx\;4.9\;\times\;10^{-6}\;\;,\quad h^{\textstyle{^{(static)}}}_{2}\,\approx\,\frac{5}{3}\;k^{\textstyle{^{(static)}}}_{2}
 \;\approx\;8.1\;\times\;10^{-6}\,\;,
 \label{}
 \ea
 which is more than a thousand times lower than the appropriate values for the Moon. This is not surprising, owing to Epimetheus' weak self-gravitation.

 Insertion of expression (\ref{k}) in equation (\ref{luna}) gives us:
 \ba
 \left(\frac{B-A}{C}\right)_{\rm{tidal}}\;\approx\;2\;h_2\;\frac{R^{\,3}\,n^2}{G\,M}\;\approx\;\frac{40\,\pi}{57}\;\frac{\rho^2\,R^5\,n^2}{\mu(\infty)\;M}
\;=\;\frac{15}{38\,\pi}\;\frac{M\;n^2}{R\;\mu(\infty)}\;\approx\;1.3\,\times\,10^{-6}\,\;,
 \label{128}
 \ea
 which is much smaller than the Epimethean permanent triaxiality given by equation (\ref{perm}).

 \subsection{Enceladus and other small bodies}

 For an independent check of the expression (\ref{128}), we can apply this expression to the Moon. For the mean lunar rigidity of $\,4\times 10^{10}$ Pa (see Eckhardt 1993), this expression renders the value of $\,10^{-6}\,$ coinciding with that given by the expression (\ref{121}). So, for the purpose of estimation, the expression (\ref{128}) can be employed for the bodies of the Moon size and smaller.

 To make such an estimate for Enceladus, we set its rigidity equal to $10^{10}$ Pa and arrive at:
 \ba
 \left(\frac{B-A}{C}\right)_{\rm{tidal}}\;\approx\;\frac{15}{38\,\pi}\;\frac{M\;n^2}{R\;\mu(\infty)}\;\approx\;3.9\,\times\,10^{-5}\,\;.
 \label{}
 \ea
 This is much smaller than Enceladus' permanent triaxiality.\,\footnote{~If we model Enceladus with a homogeneous ellipsoid with the principal radii equal to $\,a = 256.3$ km, $\,b = 247.3$ km, and $\,c = 244.6$ km (Thomas et al. 2016), we obtain the following value for the permanent triaxiality: $\,(B-A)/C = 0.036\,$.
 The easiest way to account for the layered structure would be to assume that the shell is decoupled from the core and is librating (more or less)
 independently from the interior. In this case, the triaxiality should be calculated as $\,(B-A)/C_s\,$, where $\,C_s\,$ is the moment of inertia of the shell alone. Dependent on a particular model of the interior, this will yield a result larger than $\,0.036\,$ by a factor of 2 to 3. The lower value of this factor is more realistic owing to the friction between layers.

 We may as well insert in expression (\ref{10}) the measured value of $\;^{(1:1)}{\cal{A}}_1\,=\,-\,0.12^{\circ}\,=\,-\,0.0021$ rad. This gives a permanent triaxiality of $\,0.078\,$ which is about 2.2 times larger than $\,0.036\,$.
 }

 For other moons, expression (\ref{128}) will give comparable values, give or take an order of two of magnitude. Such values will not be sufficient for the tidal triaxiality to compete with the permanent triaxiality.

 \section{The equation of motion for a small parcel of material,
 in the centre-of-mass frame\label{A}}

 Consider a body of mass $\,M\,$ spinning at a rate $\,\omegabold\,$ and also performing some orbital motion.
 % (for example, is orbiting, with its partner of mass $\,M_{sec}\,$, around their mutual centre of mass).
 In an inertial frame, the centre of mass of the body is located at $\,\xbold_{_{CM}}\,$, while a small parcel of its material resides at $\,\xbold\,$.
 Relative to the centre of mass of the body, the parcel is located at $\,\erbold\,=\,\xbold\,-\,\xbold_{_{CM}}\,$. If the body is deformable and its shape oscillates, we can decompose $\,\erbold\,$ into its average value $\,\erbold_0\,$ and an instantaneous displacement $\,\ubold~$:
 \ba
 \left.
 \begin{array}{c}
 {\mbox{In an inertial frame: \qquad\qquad}}\xbold\;=\;\xbold_{_{CM}}\,+\;\erbold~~\\
 ~\\
 {\mbox{In the centre of mass frame:~~~}}\rbold\;=\;\rbold_{0}\,+\;\ubold~~~~
 \end{array}
 \right\}~\quad\Longrightarrow\quad\quad\xbold\;=\;\xbold_{_{CM}}\,+\;\erbold_0\,+\;\ubold~.\,\;\;\;
 \label{oko}
 \ea
 Denote with $\,D/Dt\,$ the time-derivative in the inertial frame. The symbol $d/dt$ and its synonym, overdot,
 will be reserved for the time-derivative in the body frame, whence $\,{d\rbold_0}/{dt}=0\,$ and
 \ba
 \frac{d\rbold}{dt}\,=\,\frac{d\ubold}{dt}\,\;.
 \label{glinka}
 \ea
 From $\,\frac{\textstyle D\rbold}{\textstyle Dt}\;=\;\frac{\textstyle d\rbold}{\textstyle dt}~+~\omegabold\,\times\,\rbold\;$, follows the well-known relation
 \ba
 %  \frac{D\rbold}{Dt}\;=\;\frac{d\rbold}{dt}~+~\omegabold\,\times\,\rbold\quad~\quad\mbox{and}\quad~\quad
 \frac{D^2\rbold}{Dt^2}\;=\;\frac{d^2\rbold}{dt^2}\;+\;2\;\omegabold\,\times\,\frac{d\rbold}{dt}\;+\;
 \omegabold\,\times\,\left(\omegabold\,\times\,\rbold\right)\;+\;\dotomegabold\,\times\,\rbold~~.\quad\quad
 \label{oj}
 \ea
 Double differentiation of formula (\ref{oko}), with equalities (\ref{glinka} - \ref{oj}) taken into account, yields:
 \ba
 \nonumber
 \frac{D^2\xbold}{Dt^2}&=&\frac{D^2\xbold_{_{CM}}}{Dt^2}\;+\;\frac{D^2\rbold}{Dt^2}
 ~\\
 \label{ojo}\\
 %  &=&\frac{D^2\xbold_{_{CM}}}{Dt^2}\;+\;\frac{d^2\rbold}{dt^2}\;+\;2\;\omegabold\,\times\,\frac{d\rbold}{dt}\;+\;
 %  \omegabold\,\times\,\left(\omegabold\,\times\,\rbold\right)\;+\;\dotomegabold\,\times\,\rbold\\
 %  \nonumber\\
 &=&\frac{D^2\xbold_{_{CM}}}{Dt^2}\;+\;\frac{d^2\ubold}{dt^2}\;+\;2\;\omegabold\,\times\,\frac{d\ubold}{dt}\;+\;
 \omegabold\,\times\,\left(\omegabold\,\times\,\rbold\right)\;+\;\dotomegabold\,\times\,\rbold~~.\quad\quad\quad
 \nonumber
 \ea
 The equation of motion for a small parcel of material is
 \ba
 \rho\;\frac{D^2\xbold}{Dt^2}\;=\;\nabla{\mathbb{S}}\;
 %  -\;\nabla p
 \;+\;\Fbold_{self}\;+\;\Fbold_{ext}~~~,
 \label{joj}
 \ea
 where $\,\rho\,$ is the density, $\,{\mathbb{S}}\,$ is the stress,
 %  $\,p\,$ is the pressure,
 $\,\Fbold_{ext}\,$ is the exterior gravity force {\it{per unit volume}}, while $\Fbold_{self}$ is the gravity force
 % (also {\it{per unit volume}})
 wherewith the rest of the body acts on the selected parcel. Insertion of
 % (\ref{ojo}) in (\ref{joj}) furnishes:
 the former equation into the latter gives:
 \ba
 \rho\,\left[\,\frac{D^2\xbold_{_{CM}}}{Dt^2}\,+\,\ddotubold\,+\,2\,\omegabold\,\times\,\dotubold\,+\,\omegabold\,\times\,
 \left(\omegabold\,\times\,\rbold\right)\,+\,\dotomegabold\,\times\,\rbold\,\right]\,=\,\nabla{\mathbb{S}}
 %  \,-\,\nabla p
 \,+\,\Fbold_{self}\,+\,\Fbold_{ext}~~.\quad\quad
 \label{zy}
 \ea
 At the same time, for the body as a whole, we have, in an inertial frame:
 \ba
 M\;\frac{D^2\xbold_{_{CM}}}{Dt^2}\;=\;\int_{V}\Fbold_{ext} \;d^3\rbold~~,
 \label{zx}
 \ea
 the integral being taken over the volume $\,V\,$ of the body.
 (Mind that $\,\Fbold_{ext}\,$ is a force per unit volume.)
 Insertion of equation (\ref{zx}) into (\ref{zy}) entails:
 \ba
 \rho\left[\,\ddotubold+\,2\,\omegabold\times\dotubold+\,\omegabold\times
 \left(\omegabold\times\rbold\right)+\dotomegabold\times\rbold\,\right]\,=\,\nabla{\mathbb{S}}
 %  \,-\,\nabla p
 \,+\,
 \Fbold_{self}\,+\,\Fbold_{ext}\,-\,\frac{\rho}{\,M\,}\int_{V}\Fbold_{ext} \;d^3\rbold~~.\quad~
 \label{eqm}
 \ea
 Now recall that
 \ba
 \Fbold_{ext}\,-\,\frac{\rho~}{\,M\,}\int_{V}\Fbold_{ext}~d^3\rbold~=~-\;\rho~\sum_{l=2}^\infty\nabla W_l~~,
 \label{}
 \ea
 where $\;W\,=\,\sum_{l=2}^{\infty}W_l\;\,$ is the perturbing potential due to an external body.

 To each disturbing term of the exterior potential, $\,W_l\,$, corresponds a term $\,U_l\,$ of the self-potential, the self-force thus being expanded into
 \ba
 \Fbold_{self}\,=\,-\,\rho\,\sum_{l=2}^\infty\nabla U_l\,-~\rho~\nabla V
 % ^{{\rm{^{(N.-S.)}}}}
 -\,\rho\,\nabla U_{in}\,\;.
 \label{}
 \ea
 where $\,V
 % ^{{\rm{^{(N.-S.)}}}}
 \,$ is the constant in time (average) potential
 %  due to the nonsphericity
 of the body.  The constant term $\;-~\rho~\nabla V
 % ^{{\rm{^{(N.-S.)}}}}
 \,$ is compensated by the constant in time part of the stress. This way, when inserting the above expression for $\,\Fbold_{self}\,$ into equation of motion, we may omit both the term $\;-~\rho~\nabla V
 % ^{{\rm{^{(N.-S.)}}}}
 \,$ and the constant stress compensating it, thus implying that $\,\nabla{\mathbb{S}}\,$ comprises only the oscillating components of the stress.

 The potential $\,U_{in}\,$ is due to the changes of shape generated by the inertial forces. It comprises an input $\,U_{\smallfirst}\,$ generated by shape variation owing to the Coriolis force,
 and an input $\,U_{\smallsecond}\,$ produced by the change of shape due to the centrifugal force:  \footnote{~The toroidal force renders no change of shape and no extra potential, when the body has a rotational symmetry. In our case, we endow the body with some permanent triaxiality. This triaxiality, however, is very small, wherefore the change of shape due the toroidal force is of a higher order of smallness, as compared to the changes due to the Coriolis or centrifugal forces.}
 \ba
 U_{in}\,=\;U_{\smallfirst}\,+\;U_{\smallsecond}\,\;.
 \label{}
 \ea
 The centrifugal force $\;-\,\omegabold\times\left(\omegabold\times\rbold\right)\;$ and therefore also the additional potential $\,U_{\smallsecond}\,$ comprise both constant and oscillating parts. Just as $\,V
 % ^{{\rm{^{(N.-S.)}}}}
 \,$, their constant parts are compensated by the constant part of the stress. So these mutually compensating inputs may be omitted, and we should be interested only in the oscillating component $\;-\,\omegabold\times\left(\omegabold\times\rbold\right)^{(oscill)}\;$ of the centrifugal force. Also, by $\,U_{\smallsecond}\,$ we hereafter imply only the oscillating part of that potential.

 With all these observations and conventions, equation of motion (\ref{eqm}) becomes
 \ba
 \nonumber
 \rho\,\ddotubold&=&\nabla{\mathbb{S}}
 %  \,-\,\nabla p
 \,-\,\rho\,\sum_{l=2}^\infty\nabla(U_l\,+\;W_l)~-~\rho\,\dotomegabold\times\rbold\\
 \label{eqq}
 &-&2\,\rho~\omegabold\times\dotubold\;-\;\rho\,\nabla U_{\smallfirst}
 ~-~\rho~\omegabold\times\left(\omegabold\times\rbold\right)^{(oscill)}\;-\;\rho\,\nabla U_{\smallsecond}~~.\quad~
 \ea

 \section{Development of the centrifugal force\label{cen}}

 Employing the vector formula
 \ba
 \omegabold\times\left(\omegabold\times\rbold\right)~=~\omegabold~(\omegabold\cdot\erbold)~-~\erbold~\omegabold^{\,2} =
 ~\nabla\left[\,\frac{1}{2}\,\left(\omegabold\cdot\erbold\right)^2\,-~\frac{1}{2}\,\omegabold^{\,2}\,\erbold^{\,2}\,\right]\,\;,
 \label{}
 \ea
 and introducing the colatitude $\,\varphi\,$ through
 $~
 %  \ba
 \cos\varphi~=~\frac{\omegabold}{\,|\omegabold|\,}\cdot\frac{\erbold}{\,|\erbold|\,}
 %   \nonumber\ea
 \;$, $\;$we split the centrifugal force into a second-harmonic part and a purely radial part: \footnote{~Recall that $~P_2(\cos\varphi)\,=~\frac{\textstyle 1}{\textstyle 2}\left(3\,\cos^2\varphi\,-~1\right)~$, wherefrom
  \ba
 \omegabold\times\left(\omegabold\times\rbold\right)~=~\nabla\left[~~\frac{1}{2}\,\omegabold^{\,2}\,\erbold^{\,2}\,\left(\,\cos^2\varphi~-~1\,\right)~\right]
 ~=~\nabla\left[~~\frac{1}{3}~\omegabold^{\,2}\,\erbold^{\,2}
 \,\left[\,P_2\left(\cos\varphi\right)~-~1~\right]~~\right]~~.~~
 \nonumber
 \ea
 }
 \ba
 -~\rho~\omegabold\times\left(\omegabold\times\rbold\right)~=~-~\nabla\left[~\frac{\rho}{3}~\omegabold^{\,2}\,\erbold^{\,2}
 \,P_2\left(\cos\varphi\right)~\right]
 ~+~\nabla\;\left[~\frac{\rho}{3}~\omegabold^{\,2}\,\erbold^{\,2}~\right]~~~,
  \ea
 where we assumed the body homogeneous. The second-harmonic part can be incorporated into
 % the term $\,W_2$ of
 the external potential. The ensuing deformation will be defined by the degree-2 Love number.

 Above we agreed that our equation of motion should incorporate only oscillating parts of the body forces. The constant parts of the body forces are compensated by the constant part of the stress, and thus are irrelevant in our study. So we should insert in the equation of motion only the oscillating part of the above expression,
 \ba
 -\;\rho~\omegabold\times\left(\omegabold\times\rbold\right)^{(oscill)}=   \qquad\qquad\qquad\qquad\qquad\qquad\qquad\qquad\qquad\qquad\qquad\qquad\qquad\qquad
 \label{}\\
 \nonumber\\
 \nonumber
 -\;\nabla\left[~\frac{\rho}{3}~\left(\omegabold^{\,2}\,-\;\langle\,\omegabold^{\,2}\,\rangle
 \right)
 \,\erbold^{\,2}
 \,P_2\left(\cos\varphi\right)~\right]
 ~+~\nabla\;\left[~\frac{\rho}{3}~\left(\omegabold^{\,2}\,-\;\langle\,\omegabold^{\,2}\,\rangle
 \right)\,\erbold^{\,2}~\right]~~~,
  \ea
  where angular brackets denote time averaging.

 Assume libration to be purely longitudinal, and single out one Fourier mode $\,\chi\,$ out of it:
 %  ~\footnote{~For example, forced libration in longitude often can be approximated with its principal mode:
 %   \ba
 %   \gamma\;=\;{\cal{A}}\;\sin{\cal{M}}\;=\;{\cal{A}}\;\sin n t\,\;,
 %   \nonumber
 %   \ea
 %   $n\equiv\,\stackrel{\bf{\centerdot}}{\cal{M}\,}$ being the anomalistic mean motion.
 %  }
  \ba
  \gamma\;=\;{\cal{A}}\;\sin \chi t\,\;.
  \label{zulu}
  \ea
  Thence, for the angular velocity $\,\omegabold\,$, we obtain:
 \ba
 \omegabold\;\equiv\;\hat{\bf{e}}_z\stackrel{\bf\centerdot}{\theta\,}\,=\,
 \hat{\bf{e}}_z\,(\,\stackrel{\bf\centerdot}{\theta\,}_{res}\,+\,\stackrel{\bf\centerdot}{\gamma\,})\,
 =\;\hat{\bf{e}}_z\,[\,\stackrel{\bf\centerdot}{\theta\,}_{res}+\,\chi\;{\cal{A}}\;\cos \chi t\,]\,\;,
 \label{shaka}
 \ea
 \ba
 \omegabold^{\,2}\,-\;\langle\,\omegabold^{\,2}\,\rangle\;=\;\frac{1}{2}\;\chi^2\,{\cal{A}}^2\;\cos 2 \chi t
 \;+\;2\;{\cal{A}}\;\chi\;\dot{\theta}_{res}\;\cos\chi t
 \,\;
 \label{}
 \ea
 and, consequently:
 \ba
 \nonumber
 &\,&-\;\rho~\omegabold\times\left(\omegabold\times\rbold\right)^{(oscill)}=
  \qquad\qquad\qquad\qquad\qquad\qquad\qquad\qquad\qquad\qquad\qquad\qquad\qquad\qquad
  \label{}\\
  \nonumber\\
  \nonumber
 &\,&\qquad-\;\rho\,\nabla\left[~\frac{1}{6}\;\chi^2\,{\cal{A}}^2\,\erbold^{\,2}~P_2\left(\cos\varphi\right)\;\cos 2 \chi t
 +\frac{2}{3}~{\cal{A}}\;\chi\;\dot{\theta}_{res}\;\erbold^{\,2}~P_2\left(\cos\varphi\right)~\cos \chi t\right]\\
 \nonumber\\
 &\,&\qquad+\;\rho\,\nabla\;\left[~\frac{1}{6}\;\chi^2\,{\cal{A}}^2\,\erbold^{\,2}\;\cos 2 \chi t
 \;+\;\frac{2}{3}~{\cal{A}}\;\chi\;\dot{\theta}_{res}\,\erbold^{\,2}~\cos \chi t\,\right]~~.\;\;\;
  \ea
 For brevity, we shall write this as
 \ba
 -\;\rho~\omegabold\times\left(\omegabold\times\rbold\right)^{(oscill)}=\,-\,\rho\,\nabla W_2^{(cent)}
 \,-\,\rho\,\nabla W_{\textstyle{_{\rm{rad}}}}\;~~,\;\;\;
 \ea
 where the centrifugal input into the alternating part of the degree-2 perturbation is
 \ba
 W_2^{(cent)}=\;\frac{1}{6}\;\chi^2\,{\cal{A}}^2\,\erbold^{\,2}~P_2\left(\cos\varphi\right)\;\cos 2 \chi t
 \;+\;\frac{2}{3}~{\cal{A}}\;\chi\;\dot{\theta}_{res}\;\erbold^{\,2}~P_2\left(\cos\varphi\right)~\cos \chi t
 \,\;,
 \label{}
 \ea
 while the purely radial perturbation is
 \ba
 W_{\textstyle{_{\rm{rad}}}}=\;-\;\frac{1}{6}\;\chi^2\,{\cal{A}}^2\,\erbold^{\,2}\;\cos 2 \chi t
 \;-\;\frac{2}{3}~{\cal{A}}\;\chi\;\dot{\theta}_{res}\,\erbold^{\,2}~\cos \chi t
 \,\;.
 \label{}
 \ea

 \section{Expressions for the radial displacement and velocity}

 \subsection{Displacement, velocity, elastic moduli\label{AppendixRad}}

 The solution of a system comprising the radial equation (\ref{vt}) and the continuity equation $\;\partial \rho(\rbold,\,t)/\partial t\,+\,\nabla\cdot\left[\rho(\rbold)\,\dot{\ubold}_{\textstyle{_{\rm{rad}}}}\right]\,=\,0\;$, where $\,\rho(\rbold)\,$ is now the $\,${\it{mean}}$\,$ density, is presented in Matsuyama \& Bills (2010, Appendix A3).$\,$\footnote{~A similar expression from Yoder (1982, eqn A9) contains a misprint, as can be easily understood by checking the dimensions. Presenting this result, Yoder refers to Love (1944, p. 255). In reality, Love never wrote down that expression, but only provided general guidance on how to derive it.} \,For the radial deformation, they obtained:
 \ba
 u_{\textstyle{_{\rm{rad}}}}\,=\;\frac{\rho\,R^{\,3}\,\omegabold^2}{15\,\left(K\,+\,4\,\mu/3\right)}\;\left[\,\left(\frac{5}{3}\,+\,\frac{8\,\mu}{9\,K}\right)\,\frac{r}{R}\;-\;\frac{r^3}{R^{\,3}}\,\right]\,\;,
 \label{urad}
 \ea
 the corresponding radial velocity being
 \ba
 \stackrel{\bf\centerdot}{u}_{\textstyle{_{\rm{rad}}}}
 \;=\;\frac{2\,\rho\,R^{\,3}\,\omegabold\,\dotomegabold}{15\,\left(K\,+\,4\,\mu/3\right)}\;\left[\,\left(\frac{5}{3}\,+\,\frac{8\,\mu}{9\,K}\right)\,\frac{r}{R}\;-\;\frac{r^3}{R^{\,3}}\,\right]\,\;.
 \label{}
 \ea
 Consider a frequency $\chi\,$ of the longitudinal libration, see equation (\ref{zulu}). Then, from the expression (\ref{shaka}) we see that the $\,\chi$-harmonic of libration produces, in the radial deformation velocity, both the frequency $\,\chi\,$ and a double thereof:
 \ba
 \nonumber
 \stackrel{\bf\centerdot}{u}_{\textstyle{_{\rm{rad}}}}&=&^{(1)}\stackrel{\bf\centerdot}{u}_{\textstyle{_{\rm{rad}}}}\;+\;^{({2})}\stackrel{\bf\centerdot}{u}_{\textstyle{_{\rm{rad}}}}\\
 \label{}\\
 \nonumber
  &=& -\;\chi^2\,{\cal{A}}\,\stackrel{\,\bf\centerdot}{\theta}_{\rm{res}}\;\frac{\,2\,\rho\,R^{\,3}\,\sin \chi t\,}{15\,\left(K\,+\,4\,\mu/3\right)}\;\left[\,\left(\frac{5}{3}\,+\,\frac{8\,\mu}{9\,K}\right)\,\frac{r}{R}\;-\;\frac{r^3}{R^{\,3}}\,\right]\\
 \nonumber\\
 \nonumber
  &\,& -\;\chi^2\,{\cal{A}}\;\frac{\chi\,{\cal{A}}}{2}\;\frac{\,2\,\rho\,R^{\,3}\,\sin 2\chi t\,}{15\,\left(K\,+\,4\,\mu/3\right)}\;\left[\,\left(\frac{5}{3}\,+\,\frac{8\,\mu}{9\,K}\right)\,\frac{r}{R}\;-\;\frac{r^3}{R^{\,3}}\,\right]\,\;.
 \ea
 %  while the radial force, according to equation (\ref{radrad}), reads as
 %  \ba
 %  \Fbold_{\textstyle{_{\rm{rad}}}}\,=\;-\;\nabla W_{\textstyle{_{\rm{rad}}}}\;=\;\frac{{\cal{A}}^2\,\rho\;\erbold}{3}\;\cos 2 \chi t\,\;.
 %  \label{}
 %  \ea
 In complex notation, the harmonic (\ref{zulu}) of the longitudinal forced libration is given by
 \ba
 \gamma\;=\;{\cal{A}}\;e^{i \chi t}\;\;,\qquad\omega\;\equiv\;\stackrel{\bf\centerdot}{\theta\,}\;=\;\stackrel{\bf\centerdot}{\theta\,}_{res}\,+\;i\;\chi\;{\cal{A}}\;e^{i\chi t}\;\;,\qquad
 \stackrel{\bf\centerdot}{\omega\,}\;=\;-\;\chi^2\;{\cal{A}}\;e^{i\chi t}\,\;,
 \label{over}
 \ea
 whence the complex velocity is
 \ba
 \nonumber
 \stackrel{\bf\centerdot}{{u}}_{\textstyle{_{\rm{rad}}}}&=&
  i\;e^{i\chi t}\;\chi^2\;{\cal{A}}\;\stackrel{\,\bf\centerdot}{\theta}_{res}\;\frac{\,2\,\rho\,R^{\,3}\,}{15\;\bar{K}(\chi)\;}\;
 \frac
 {\,\left(\frac{\textstyle 5}{\textstyle 3}\,+\,\frac{\textstyle 8\,\bar{\mu}(\chi)}{\textstyle 9\,\stackrel{\_}{K}(\chi)}\right)\,\frac{\textstyle r}{\textstyle R}\;-\;\frac{\textstyle r^3}{\textstyle R^{\,3}}\,}
 {\,1\,+\,\frac{\textstyle 4}{\textstyle 3}\;\frac{\textstyle\bar{\mu}(\chi)}{\textstyle\stackrel{\_}{K}(\chi)} }\\
 \label{velocity}\\
 \nonumber
 &+&  i\;e^{2i\chi t}\;\chi^2\;{\cal{A}}\;\frac{\chi\,{\cal{A}}}{2}\,\frac{\,2\,\rho\,R^{\,3}\,}{15\;\bar{K}(2\chi)\;}\;
 \frac
 {\,\left(\frac{\textstyle 5}{\textstyle 3}\,+\,\frac{\textstyle 8\,\bar{\mu}(2\chi)}{\textstyle 9\,\stackrel{\_}{K}(2\chi)}\right)\,\frac{\textstyle r}{\textstyle R}\;-\;\frac{\textstyle r^3}{\textstyle R^{\,3}}\,}
 {\,1\,+\,\frac{\textstyle 4}{\textstyle 3}\;\frac{\textstyle\bar{\mu}(2\chi)}{\textstyle\stackrel{\_}{K}(2\chi)} }
 \,\;,
 \ea
 %  \ba
 %  \bar{F}_{\textstyle{_{\rm{rad}}}}(2\chi)\,=\;\frac{|\,\bar{\cal{A}}(\chi)\,|^2\,\rho\;r}{3}\;e^{2 i \chi t}\,\;,
 %  \label{fcomp}
 %  \ea
 $\bar{K}(\chi)\,$ and $\,\bar{\mu}(\chi)\,$ being the Fourier components of the operators of bulk and shear moduli.  For a Maxwell body, they read as
 \ba
 \bar{\mu}(\chi)\,=\,{\mu}\;\frac{\inc\,\chi\,\eta/\mu}{1\,+\,\inc\chi\eta/\mu}\,\quad,\qquad
 \bar{K}(\chi)\,=\,K\;\frac{\inc\,\chi\,\zeta/K}{1\,+\,\inc\chi\zeta/K}\,\quad,
 \label{LL42}
 \ea
 where $\,\mu=\mu(0)\,$ and $\,K=K(0)\,$ are the unrelaxed elasticity moduli, while $\,\eta\,$ and $\,\zeta\,$ are the shear and bulk viscosities.

 %  For small-magnitude libration,
 %  \ba
 %  \stackrel{\,\bf\centerdot}{\theta}_{res}\;\gg\;\frac{\chi\,{\cal{A}}(\chi)}{2}\,\;.
 %  \label{}
 %  \ea
 %  Therefore, in this case we can neglect the second harmonic:
 %  \ba
 %  \stackrel{\bf\centerdot}{{u}}_{\textstyle{_{\rm{rad}}}}\;\approx\;
 %  i\;e^{i\chi t}\;\chi^2\;\bar{\cal{A}}(\chi)\;\stackrel{\,\bf\centerdot}{\theta}_{res}\;\frac{\,2\,\rho\,R^{\,3}\,}{15\;\bar{K}(\chi)\;}\;
 %  \frac{\,\left(\frac{\textstyle 5}{\textstyle 3}\,+\,\frac{\textstyle 8\,\bar{\mu}(\chi)}{\textstyle 9\,\stackrel{\_}{K}(\chi)}\right)\,
 %  \frac{\textstyle r}{\textstyle R}\;-\;\frac{\textstyle r^3}{\textstyle R^{\,3}}\,}
 %  {\,1\,+\,\frac{\textstyle 4}{\textstyle 3}\;\frac{\textstyle\bar{\mu}(\chi)}{\textstyle\stackrel{\_}{K}(\chi)} }
 %  \label{velocity}
 %  \ea

 The expression for the velocity can be simplified to
 \ba
 \nonumber
 \stackrel{\bf\centerdot}{{u}}_{\textstyle{_{\rm{rad}}}}&=&
 i\;e^{i\chi t}\;\chi^2\;{\cal{A}}\;\stackrel{\,\bf\centerdot}{\theta}_{res}\;\frac{\,2\,\rho\,R^{\,3}\,}{15\;\bar{K}(\chi)}\;
 \left[\,\frac{5}{3}\;\frac{r}{R}\;-\;\frac{r^3}{R^{\,3}}\,\right]\;+\;O\left(\;{\,\bar{\mu}(\chi)}/{\bar{K}(\chi)\,}\;\right)\\
 \label{velocity1}\\
  &+&i\;e^{2i\chi t}\;\chi^2\;{\cal{A}}\; \frac{\chi\;{\cal{A}}}{2} \;\frac{\,2\,\rho\,R^{\,3}\,}{15\;\bar{K}(2\chi)}\;
 \left[\,\frac{5}{3}\;\frac{r}{R}\;-\;\frac{r^3}{R^{\,3}}\,\right]\;+\;O\left(\;{\,\bar{\mu}(2\chi)}/{\bar{K}(2\chi)\,}\;\right)\,\;,
 \nonumber
 \ea
 provided the Fourier components of the operator moduli obey
 \ba
 \left|\frac{\,\bar{\mu}(\chi)}{\bar{K}(\chi)\,}\right|\;\ll\;1\qquad\mbox{and}\qquad\left|\frac{\,\bar{\mu}(2\chi)}{\bar{K}(2\chi)\,}\right|\;\ll\;1\;\,.
 \label{}
 \ea
 For a Maxwell body, this condition becomes (see equation \ref{LL42}):
 \ba
 \left|\frac{\,\bar{\mu}(\chi)\,}{\,\bar{K}(\chi)\,}\right|\;=\;\frac{\,\eta\,}{\,\zeta\,}\;\,\sqrt{\frac{1\,+\,(\chi\zeta/K)^2}{1\;+\;(\chi\eta/\mu)^2}\,}\;\,\ll\,1\,\;,
 \label{ineq}
 \ea
 and a similar inequality for $\,2\chi\,$. Here $\mu\,$ and $\,K\,$ are the unrelaxed elasticity moduli, and $\,\eta\,$ and $\,\zeta\,$ the shear and bulk viscosities.

  When the frequency $\,\chi\,$ is higher than both $\,\mu/\eta\,$ and $\,K/\zeta\,$ (which are the inverse Maxwell times), the requirement (\ref{ineq}) reduces to
  \ba
  \frac{\mu}{K}\,\ll\,1\,\;,
  \label{}
  \ea
  a condition that is barely met by cracked silicate rocks, and is not met by sandstones.

  When $\,\chi\,$ is lower than both the thresholds $\,\mu/\eta\,$ and $\,K/\zeta\,$, the inequality (\ref{ineq}) becomes
  \ba
  \frac{\eta}{\zeta}\,\ll\,1\,\;.
  \label{}
  \ea
  Fulfilment of this condition should not be taken for granted. It depends on the porosity and the fraction of partial melt.$\,$\footnote{~While for most viscoelastic materials the bulk viscosity $\,\zeta\,$ is hundreds to thousands of times larger than its shear counterpart $\,\eta\,$, this is not always so for porous media close to the melting point, because the presence of partial melt changes the situation radically. As melt fraction increases from zero to the critical melt fraction value, rapid decreases occur in both the shear and bulk viscosities. While the shear viscosity of silicates falls by a factor of about five, the bulk viscosity of silicates can fall by a factor from hundreds to thousands, and can become only slightly larger than the shear viscosity when the melt fraction becomes appreciable, see Figure 9 in Takei and Holtzman (2009).

  For ices, the situation similar. As explained by Kalousová et al. (2014, eqn 7 and the paragraph thereafter)

  While for solid ice the bulk viscosity $\,\zeta\,\gg\,\eta\,$, for porous ice the bulk viscosity can be approximated as $\,\zeta\,\approx\,\eta/\phi\,$ where $\,\phi\,$ is the ice porosity. This for the porosity of about 10\%, the bulk viscosity is only some ten times higher than the shear viscosity.
  \label{Kalous}}

 All in all, the legitimacy of the approximation (\ref{velocity1}) depends on the libration frequency (as compared to the inverse Maxwell times) and on the hypothesised porosity of the body. With this caveat in mind, we shall accept this approximation, to simplify our algebra in the subsequent calculation of the power.

  \subsection{How many terms to keep in series (\ref{power_radial})?\label{check}}

 Consider forced libration in longitude about a spin-orbit resonance parameterised by a rational $\,z$. When (a) the obliquity is low and (b) only quadrupole terms matter, only the resonances with integer and semi-integer values of  $\,z\,$ are important, see Section \ref{definition}. Then the libration,
 \begin{equation}
 \gamma(t)\;=\;\sum_{j=1}^{\infty} {\cal{A}}_j\;\sin(\chi_j\,t)\,\;,
 \label{eq1}
 \end{equation}
 has its Fourier magnitudes expressed as (e.g., Frouard \& Efroimsky 2017\,a)
 \ba
  {\cal{A}}_j\;=\;\omega_0^2\;\frac{G_{20(j+2z-2)}(e) - G_{20(-j+2z-2)}(e)}{2\;\omega_0^2\;
 G_{20(2z-2)}(e)\,-\,\chi_j^2 }
  \;\approx\;-\;\omega_0^2\;\frac{G_{20(j+2z-2)}(e) - G_{20(-j+2z-2)}(e)}{\chi_j^2 }\,\;,\quad
 \label{}
 \ea
 with the eccentricity polynomials linked to the Hansen coefficients as $\,G_{lpq}(e)=X_{l-2p+q}^{\,-(l+1),~l-2p}(e)\,_{\textstyle{_{\textstyle{.}}}}\,$ The auxiliary quantity $\,\omega_0^2\,$ is given by equation (\ref{omega0}).

 For a Maxwell body, the $\,j$-th term of series (\ref{power_radial}) is proportional to the factor
 \ba
 \Xi_j\,=\;\chi_j^5\;{\cal{A}}_j^4\;|\,\bar{J}(\chi_j)\,|\;\sin\delta(\chi_j)\;=\;\frac{\omega_0^4}{\eta}\;\frac{\omega_0^4}{\chi_j^4}
 \;\left[\,G_{20(j+2z-2)}(e) - G_{20(-j+2z-2)}(e)\,\right]^4
 \,\;.
 \label{}
 \ea
 This relation gives us a clue to understanding of how many terms should be kept in the series expansion (\ref{power_radial}) for the power.
 Owing to the rule $\,G_{lpq}(e)\,\propto\,e^{|q|}\,$, we usually \footnote{~A rare exception shows itself in the case of a 3:2 resonance ($\,z=3/2\,$). From equation (\ref{3:2}) we have: $\,\Xi_{1}=\,O(e^0)\,$, $\,\Xi_{2}=\,O(e^4)\,$, $\,\Xi_{3}=\,O(e^{\bf{16}})\,$, $\,\Xi_{4}=\,O(e^{12})\,$. To understand the reason for this irregularity, recall that the general rule $\,G_{lpq}(e)\,\propto\,e^{|q|}\,$ has an occasional exception: $\,G_{20,-2}(e)\,=\,0\,$. So the leading term (of the order $e^8\,$) in the expression for $\,\Xi_{3}\,$ vanishes, the next nonzero term being of the order of $\,e^{16}\,$.} expect that
 \ba
 \frac{\,\Xi_{j+1}}{\Xi_j\,}\;\propto\;e^4\;\left(\frac{\chi_j}{\chi_{j+1}}\right)^4\,\;,
 \label{caveat}
 \ea
 so expansion (\ref{power_radial}) is a power series in $\,e\,$. Unfortunately, the numerical coefficients entering the expressions for $\,G_{lpq}(e)\,$ grow rapidly with the increase of $\,q\,$, wherefore all series involving these functions are notorious for their slow convergence for $\,e\,>\,0.5\,$. For lower eccentricities, though, we may approximate series (\ref{power_radial}) with its first term.

 If libration is forced then, by equation (\ref{follows}), its frequencies are $\,\chi_j\,=\,j\,n\,$.
  $\,$So relation (\ref{caveat}) becomes simply $\;\,
 \frac{\textstyle \,\Xi_{j+1}}{\textstyle \Xi_j\,}\;\propto\;e^4\;\left(\frac{\textstyle j}{\textstyle j+1}\right)^4\,\;$. For not very high eccentricities, series (\ref{power_radial}) is approximated with its first term equal to
 \ba
 \mbox{Forced libration, ~~$\;e\,\lesssim\,0.5\;\,:$}\qquad \langle\,P\,\rangle_{\textstyle{_{\rm{rad}}}}\;\approx
 % \;\frac{8\,\pi}{945\;\zeta}\;R^{\,7}\;\rho^2\;n^4\;{\cal{A}}(n)^4\;\approx\;
 \;\frac{R^{\,7}\;\rho^2}{75\;\zeta}\;n^4\;{\cal{A}}_1^4
 \,\;.
 \label{}
 \ea

 \section{Expressions for the toroidal force, displacement and strain\label{tor}}

 \subsection{Preparatory work. Vector formulae\label{tor1}}

 To begin, express $\,\dotomegabold\,$ and $\,\rbold\,$ through the unit vectors $\,\hat{\bf{e}}_{\smalldotomegabold}\,$ and $\,\hat{\bf{e}}_{\smallerbold}\;$:
  \ba
  \dotomegabold\,=\,\hat{\bf{e}}_{\smalldotomegabold}\,|\,\dotomegabold\,|\qquad\mbox{and}\qquad\rbold\,=\,\hat{\bf{e}}_{\smallerbold}\,r\,\;.
  \label{}
  \ea
 We also shall need the obvious vector formula for $\,\nabla \,\equiv\,\nabla_{\smallerbold}\;$:
 \bs
 \ba
 -\,\hat{\bf{e}}_{\smalldotomegabold}\times\rbold\,=\,\rbold\times\nabla(\hat{\bf{e}}_{\smalldotomegabold}\cdot\rbold)\,\;.
 \label{}
 \ea
 Owing to $\;\rbold\times\,$, the absolute value $\,r\,=\,|\rbold|\,$ can be taken outside the action of the gradient:~\footnote{~The following relation is valid for an arbitrary scalar function $\,f\,=\,f(\rbold)\,$:
 $$\,\nabla(f\,r)\,=\,r\,\nabla f\,+\,f\,\hat{\bf{e}}_{\smallerbold}\,\;.$$
 Applying of $\;\rbold\times\,$ to both sides results in:
 $$\,\rbold\times\nabla(f\,r)\,=\,r\,\rbold\times\nabla f\,\;.$$
 In our case, $\,f(\rbold)\,=\,\hat{\bf{e}}_{\smalldotomegabold}\cdot\,\hat{\bf{e}}_{\smallerbold}\,$.}
 \ba
 -\,\hat{\bf{e}}_{\smalldotomegabold}\times\rbold\,=\,r\,\rbold\times\nabla(\hat{\bf{e}}_{\smalldotomegabold}\cdot\hat{\bf{e}}_{\smallerbold})\,\;,
 \label{}
 \ea
 or, equivalently:
 \ba
 -\,\hat{\bf{e}}_{\smalldotomegabold}\times\hat{\bf{e}}_{\smallerbold}\,=\,\rbold\times\nabla(\hat{\bf{e}}_{\smalldotomegabold}\cdot\hat{\bf{e}}_{\smallerbold})\,\;.
 \label{formula}
 \ea
 \es
 As the absolute value $\,|\,\dotomegabold\,|\,$ is $\,\rbold$-independent, multiplying of the above with $\,|\,\dotomegabold\,|\,$ yields:
 \ba
 -\,\dotomegabold\times\rbold\,=\,\rbold\times\nabla(\dotomegabold\cdot\rbold)\;=\;
 r\;|\,\dotomegabold\,|\;\rbold\times\nabla(\hat{\bf{e}}_{\smalldotomegabold}\cdot\hat{\bf{e}}_{\smallerbold})\,\;.
 \label{fo}
 \ea
 \label{}

 \subsection{The toroidal force\label{tor2}}

 With aid of formula (\ref{fo}), the toroidal force
 \bs
 \ba
 \Fbold_{\textstyle{_{\rm{tor}}}}\,=\;-\,\rho\,\dotomegabold\times\rbold\;=\;-\;\rho\;r\;|\,\dotomegabold\,|\;\hat{\bf{e}}_{\smalldotomegabold}\times\hat{\bf{e}}_{\smallerbold}
 \label{force}
 \ea
 can be expressed through the dot product of the unit vectors:
  \ba
  \Fbold_{\textstyle{_{\rm{tor}}}}\,=\;\rho\;\rbold\times\nabla(\dotomegabold\cdot\rbold)\;=\;\rho\;|\,\dotomegabold\,|\;r\;\rbold\times\nabla
 (
  \hat{\bf{e}}_{\smalldotomegabold}
  \cdot\,
  \hat{\bf{e}}_{\smallerbold}\;
  )\;\,.
 \label{}
 \ea
 This can also be written as
  \ba
  \Fbold_{\textstyle{_{\rm{tor}}}}
  \,=\;
   \rho\;|\,\dotomegabold\,|\;r\;\rbold\times\nabla
 P_1(\hat{\bf{e}}_{\smalldotomegabold}\cdot\,\hat{\bf{e}}_{\smallerbold})
  \;\,.
 \label{}
 \ea
 Let the spin rate $\,\omegabold\,$ of the body be aimed along its maximal-inertia axis $\,z\,$. Under libration in longitude, the spin acceleration $\,\dotomegabold\,$ is then pointing in the same direction as $\,\omegabold\,$. Denoting with $\,\varphi\,$ the colatitude of the point $\,\rbold\,$, we arrive at
  \ba
  \Fbold_{\textstyle{_{\rm{tor}}}}\,=\;\rho\;|\,\dotomegabold\,|\;r\;\rbold\times \nabla P_1(\cos\varphi)\,\;.
  \label{forc}
  \ea
 \es

 \subsection{Toroidal deformation\label{tor3}}

 Although in our technical calculations we shall employ the initial expression (\ref{force}), it was worthwhile writing the force in the form of (\ref{forc}), in order to emphasise that it is a toroidal perturbation of degree 1. So we expect the deformation, too, to be a degree-1 function:
 \ba
 \ubold_{\textstyle{_{\rm{tor}}}}\;=\;D(r,\,t)\;
 \rbold
 %  \hat{\bf{e}}_{\smallerbold}
 \times\nabla P_1(\cos\varphi)\,\;.
 \label{}
 \ea
 With aid of the vector formula (\ref{formula}), we rewrite this as
 \ba
 \ubold_{\textstyle{_{\rm{tor}}}}\;=\;-\;D(r,\,t)\;
 \hat{\bf{e}}_{\smalldotomegabold}
 \times\hat{\bf{e}}_{\smallerbold}\,\;,
 \label{}
 \ea
 where $\,\hat{\bf{e}}_{\smalldotomegabold}\,$ and $\,\hat{\bf{e}}_{\smallerbold}\;$ are the unit vectors pointing along $\,\dotomegabold\,$ and $\,\rbold\,$, correspondingly.

 \subsection{The force and displacement expressed through the unit vectors\label{tor4}}

 When the angular acceleration vector $\,\dotomegabold\,$ is aimed along the polar axis $\,z\,$ of the body ($\,\hat{\bf{e}}_{\smalldotomegabold} = \hat{\bf{e}}_{z}\,$), the product $\,\hat{\bf{e}}_{\smalldotomegabold}\times\hat{\bf{e}}_{\smallerbold}\,$ points in the direction of longitude decrease: $\,\hat{\bf{e}}_{\smalldotomegabold}
 \times\hat{\bf{e}}_{\smallerbold}\,=\,-\,\hat{\bf{e}}_{\smalllambdabold}\,\sin\varphi\,$, $\,$whence
 \ba
 \Fbold_{\textstyle{_{\rm{tor}}}}\,=\;\rho\;|\,\dotomegabold\,|\;r\;\sin\varphi\;\,
 \hat{\bf{e}}_{\smalllambdabold}\,\;,
 \label{}
 \ea
 and
 \ba
 \ubold_{\textstyle{_{\rm{tor}}}}\;=\;D(r,\,t)\;\sin\varphi\;\,
 \hat{\bf{e}}_{\smalllambdabold}\,\;,
 \label{displacement}
 \ea
 $\varphi\,$ being the colatitude of the point $\,\rbold\,$.

 \subsection{The toroidal strain}

 In the spherical coordinates, the strain reads as (Sokolnikoff 1946, eqn 48.17):
 \ba
 \nonumber
 \varepsilon_{rr}\,=\,u_{r\,,\,r}~~~,\qquad
 \varepsilon_{\varphi\varphi}\,=\,\frac{u_{\varphi\,,\,\varphi}\,+\,u_{r}}{r}~~,\qquad
 \varepsilon_{\lambda\lambda}\,=\,\frac{u_{\lambda\,,\,\lambda}\,+\,u_{r}\,\sin\varphi\,+\,u_{\varphi}\,\cos\varphi}{r\;\sin\varphi}~~,\qquad ~\\
 \nonumber\\
 \label{strain}
 \varepsilon_{r\varphi}\,=\,\frac{1}{2}\,\left(\frac{u_{r\,,\,\varphi}}{r}\,+\,u_{\varphi\,,\,r}\,-\,\frac{u_\varphi}{r}\right)~~~,\qquad\qquad\qquad\qquad\qquad\qquad\qquad\qquad\qquad\qquad\qquad  ~\\
 \nonumber\\
 \nonumber
 \varepsilon_{\varphi\lambda}\,=\,
 \frac{1}{2\,r}\,\left(\frac{u_{\varphi\,,\,\lambda}}{\sin\varphi}\,+\,u_{\lambda\,,\,\varphi}\,-\,u_{\lambda}\,\cot\varphi\right)~~,\qquad
 \varepsilon_{\lambda r}\,=\,
 \frac{1}{2}\,\left(\frac{u_{r\,,\,\lambda}}{r\;\sin\varphi}\,+\,u_{\lambda\,,\,r}\,-\,\frac{u_{\lambda}}{r}\right)~~.~\,\quad
 \ea
 where we employed the self-evident notation $\,u_{r\,,\,r}\,\equiv\,{\partial u_r}/{\partial r}\,$ and alike.
 Also mind that these formulae are written via the polar angle $\,\varphi=\,\pi/2-\,\phi\,$, not the latitude $\,\phi\,$.

 The divergence of the strain has the form (Sokolnikoff 1946, eqn 48.18):
 \bs
 \ba
 \er\,\cdot\,(\nabla\varepsilon)\,=\,\varepsilon_{rr\,,\,r}\,+\,\frac{\varepsilon_{r \varphi\,,\,\varphi}}{r}
 \,+\,\frac{\varepsilon_{r \lambda\,,\,\lambda}}{r\,\sin\varphi}\,+\,
 \frac{2\,\varepsilon_{rr}\,-\,\varepsilon_{\varphi\varphi}\,-\,\varepsilon_{\lambda\lambda}\,+\,\varepsilon_{r \varphi}\,\cot\varphi}{r}\,\;,
 \label{}
 \ea
 \ba
 \evarphi\,\cdot\,(\nabla\varepsilon)\,=\,\varepsilon_{\varphi r\,,\,r}\,+\,\frac{\varepsilon_{\varphi \varphi\,,\,\varphi}}{r}
 \,+\,\frac{\varepsilon_{\varphi\lambda\,,\,\lambda}}{r\,\sin\varphi}\,+\,
 \frac{3\,\varepsilon_{\varphi r}\,+\,(\varepsilon_{\varphi\varphi}\,-\,\varepsilon_{\lambda\lambda})\,\cot\varphi}{r}\;\;,
 \label{}
 \ea
 \ba
 \elambda\,\cdot\,(\nabla\varepsilon)\,=\,\varepsilon_{\lambda r\,,\,r}\,+\,\frac{\varepsilon_{\lambda \varphi\,,\,\varphi}}{r}
 \,+\,\frac{\varepsilon_{\lambda\lambda,\,\lambda}}{r\,\sin\varphi}
 \,+\;\frac{3\,\varepsilon_{r\lambda}\,+\,2\,\varepsilon_{\varphi\lambda}\,\cot\varphi}{r}\;\;.
 \label{}
 \ea
 \label{div}
 \es
 Insertion of expression (\ref{displacement}) into formulae (\ref{strain}) yields:
 \ba
 % \varepsilon_{\varphi \lambda}\,=\;\frac{D(r,\,t)}{2\,r}\;\left(0\;+\;\cos\varphi\;-\;\sin\varphi\;\cot\varphi\right)\;=\;0 ~\\
 {\varepsilon}_{{{\lambda r}}}\,=\;\frac{\sin\varphi}{2}\;\left(\frac{\partial D}{\partial r}\;-\;\frac{D}{r}\right)\;=\;\frac{\sin\varphi}{2}\;r\;\frac{\partial }{\partial r}\left(\frac{D}{r}\right)\;\,,
 \label{}
 \ea
 the other components being zero. Plugging of this expression into equations (\ref{div}) entails:
  \bs
 \ba
 \er\,\cdot\,(\nabla\varepsilon_{\,tor}\,)\,=\,\evarphi\,\cdot\,(\nabla\varepsilon_{\,tor}\,)\,=\,0\;\;,
 \label{}
 \ea
 \ba
 \nonumber
 \elambda\,\cdot\,(\nabla\varepsilon_{\,tor}\,)&=&\frac{\sin\varphi}{2}\;\left\{
 \frac{\partial }{\partial r}\;\left[
 r\;\frac{\partial }{\partial r}\left(\frac{D}{r}\right)
 \right]
 \;+\;
 3\;\frac{\partial }{\partial r}\left(\frac{D}{r}\right)
 \right\}~\\
 \nonumber\\
 &=&
 \frac{\sin\varphi}{2}\;\frac{1}{r^3}
 \frac{\partial }{\partial r}\;\left[
 r^4\;\frac{\partial }{\partial r}\left(\frac{D}{r}\right)
 \right]
 \;\;,
 \label{}
 \ea
 \label{divv}
 \es
 where we reinstalled the subscript $\,$`$\,${\it{tor}}$\,$'$\,$, to emphasise that our formulae pertain to the toroidal part of the strain.

 \subsection{How many terms to keep in series (\ref{comment})?\label{bezobrazie}}

 Carrying out analysis similar to that in Section \ref{check}, we find that for a Maxwell body the $\,j$-th term in the power (\ref{comment}) is proportional to the factor
 \ba
 \Xi_j\,=\;\chi_j^5\;{\cal{A}}(\chi_j)^2\;|\,\bar{J}(\chi_j)\,|\;\sin\delta(\chi_j)\;=\;\frac{\omega_0^4}{\eta}
 \;\left[\,G_{20(j+2z-2)}(e) - G_{20(-j+2z-2)}(e)\,\right]^2
 \,\;.
 \label{tobechecked}
 \ea
 Hence we expect that, usually,$\,$\footnote{~See, however, the caveat after equation (\ref{caveat}).}
 \ba
 \frac{\,\Xi_{j+1}}{\Xi_j\,}\;\propto\;e^2\,\;.
 \label{}
 \ea
 Using the same line of reasoning as in Section \ref{check}, we see that for not very high eccentricities series (\ref{comment}) can be approximated by its first term.

 In the special case of forced libration, the frequencies are given by $\,\chi_j\,=\,j\,n\,$, so the first term becomes:
 \ba
 \mbox{Forced libration, ~~$\;e\,<\,0.5\;\,:$}\qquad  \langle\,P\,\rangle_{\textstyle{_{\rm{tor}}}}\;\approx
 % \;\frac{4\,\pi}{105\,\eta}\;R^{\,7}\;\rho^2\;\sum_{j=1}^{\infty} \chi_j^4\;{\cal{A}}(\chi_j)^2\;\approx\;
 \;\frac{R^{\,7}\;\rho^2}{8\,\eta}\;n^4\;{\cal{A}}(n)^2
 \,\;.
 \label{}
 \ea

 \section{Tidal reaction of a near-spherical Maxwell body
 % Derivation of the relation $\;\chi\,{{k}_2(\chi)\,\sin\epsilon_2(\chi)}\,\approx\,G\,R^{\,2}\,\rho^2/{\eta}\;$\\ valid for a near-spherical Maxwell body
 \label{chi}}

\begin{figure}[htbp]
 \vspace{2.5mm}
 \centering
 \includegraphics[angle=0,width=0.68\textwidth]{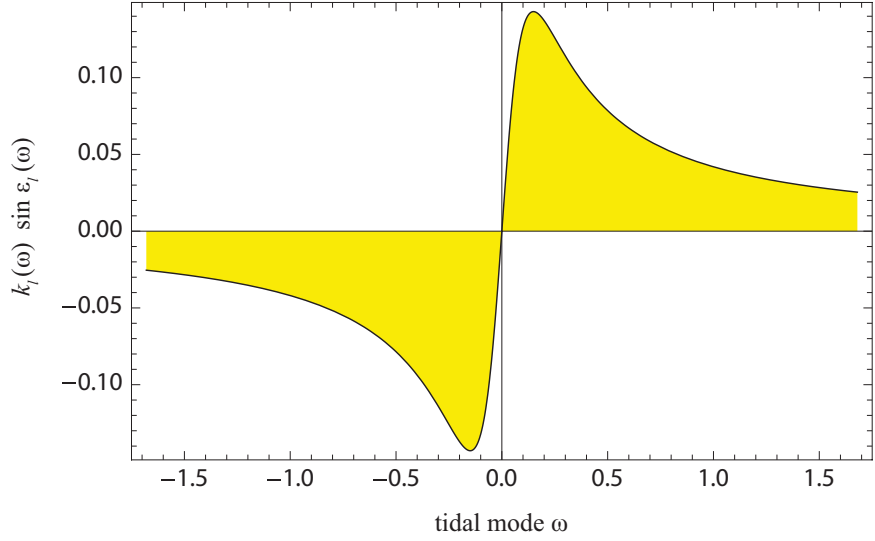}
 \caption{\small{.~A typical shape of the quality function $\,{\textstyle k_{l}}/{\textstyle Q_l}\,=\,k_l(\omega)\,\sin\epsilon_l(\omega)\,$. (From Noyelles et al. 2014.) ~Here $\,k_l\,$ and $\,\epsilon_l\,$ are the degree-$l\,$ Love number and phase lag. The quadrupole approximation corresponds to $\,l=2\,$. The letter $\,\omega\,$ is a shortened notation for the tidal Fourier mode (which is $\,\omega_{\textstyle{_{lmpq}}}\,$, under no libration, or $\,\beta_{lmpqs}\,$ when libration is present.$\,$)
 The physical forcing frequency $\,\chi\,$ is the absolute value of the tidal Fourier mode: $\,\chi\,=\,|\,\omega\,|\,$. ~~
 \label{Fig1}}}
 \end{figure}

  For a near-spherical body $\,${\underline{\it{of \,an \,arbitrary \,linear \,rheology\,}}}, $\,$the frequency-dependence of $\,k_l/Q_l\,=\,{k}_l\,\sin\epsilon_l\,$ has the form of a sharp kink, with the maxima located at an extremely low frequency, see Figure \ref{Fig1}. In that figure, the letter $\,\omega\,$ is a shortened notation for the tidal Fourier mode (which is $\,\omega_{\textstyle{_{lmpq}}}\,$, under no libration; or is $\,\beta_{lmpqs}\,$ when libration is present). The physical forcing frequency $\,\chi\,$ is the absolute value of the tidal Fourier mode: $\,\chi\,=\,|\,\omega\,|\,$.
 Between the peaks, the dependence is near-linear and goes through zero continuously. (This is indeed as it should be, lest we get a singularity in the tidal torque on crossing a spin-orbit resonance). At not so low frequencies, the shape of the function, as well as the magnitude and location of the peaks, depends on the rheological model.

 For a Maxwell body, the extrema of $\,k_2/Q_2\,=\,k_2(\omega)\,\sin\epsilon_2(\omega)\,$ are located at
 \ba
 \omega_{peak}\,=\;\pm\;\frac{\textstyle{8\,\pi\,G\,\rho^2\,R^2}}{\textstyle{57\;\eta}}\quad,
 \label{peak}
 \ea
 where $\,G\,$ is the Newton gravitational constant, while  $\eta\,$ and $\,\rho\,$ denote the shear viscosity and mean density, correspondingly,~---~see Efroimsky (2015, eqn 68).

 The peaks' amplitude is virtually insensitive to the viscosity $\,\eta\,$, while the spread between the extrema does depend on $\,\eta\,$. According to equation (\ref{peak}), for a higher viscosity the peaks reside closer to the spin-orbit resonance. Accordingly, as the viscosity assumes lower values, the peak frequency spreads out.$\,$\footnote{~In theory, the peak can eventually reach and bypass the orbital frequency $\,n\,$ as the viscosity assumes lower values. In realistic situations, however, this would require extraordinarily low viscosities.}

 Outside of the inter-peak interval, the function $\,k_l(\omega)\,\sin\epsilon_l(\omega)\,$ behaves as the inverse $\,Q\,$ in seismology. Specifically, for a  Maxwell material, $\,k_l\,\sin\epsilon_l\,$ scales as the inverse frequency:
 \ba
 \frac{{k}_2(\chi)\;\sin\epsilon_2(\chi)}{\,\frac{\textstyle 4\pi}{\textstyle 19}\;G\;R^{\,2}\;\rho^2}\;\approx\;|\,\bar{J}(\chi)\,|\,\sin\delta(\chi)\;=\;\frac{1}{\eta\,\chi}\,\;,
 \label{88888}
 \ea
 where we switched from the Fourier tidal mode $\,\omega\,$ to the physical forcing frequency $\,\chi\,=\,|\,\omega\,|\,$. In this expression,
 $|\,\bar{J}(\chi)\,|\,\sin\delta(\chi)\,=\,\frac{\textstyle 1}{\textstyle \chi\,\eta}\,$ is the negative imaginary part of the Maxwell complex compliance $\,\bar{J}\,$ at the frequency $\,\chi\,$, see equations (31) and (67) in Efroimsky (2015).

 Owing to equation (\ref{88888}), the product of $\,k_l(\chi)\,\sin\epsilon_l(\chi)\,$ by the frequency $\,\chi\,$ is about constant (provided we stay away from the narrow inter-peak interval of extremely low frequencies):
 \ba
 \chi\;\frac{{k}_2(\chi)\;\sin\epsilon_2(\chi)}{G\;R^{\,2}\;\rho^2}\;\approx\;\frac{4\pi}{19}\;\frac{1}{\eta}\;\approx\;\frac{2}{3}\;\frac{1}{\eta}\,\;.
 \label{me}
 \ea
 Staying away from the inter-peak interval of extremely low frequencies means, according to equation (\ref{peak}), that the viscosity obeys
 \ba
 \eta\;>\;\frac{\textstyle{8\,\pi\,G\,\rho^2\,R^2}}{\textstyle{57\;\chi}}\quad,
 \label{}
 \ea
 where $\,\chi\,$ is the forcing frequency.

 In the special case when the role of the forcing frequency $\,\chi\,$ is played by the main-mode frequency $\,n\,$ of forced libration, the above two formulae become, correspondingly:
 \ba
 n\;\frac{{k}_2(n)\;\sin\epsilon_2(n)}{G\;R^{\,2}\;\rho^2}\;\approx\;\frac{4\pi}{19}\;\frac{1}{\eta}\;\approx\;\frac{2}{3}\;\frac{1}{\eta}\,\;,
 \label{manner}
 \ea
 and
  \ba
 \eta\;>\;\frac{\textstyle{8\,\pi\,G\,\rho^2\,R^2}}{\textstyle{57\;n}}\quad.
 \label{}
 \ea
 To get a touch of the numbers, consider Enceladus. With $\;\rho=1.61\times 10^3$~kg/m$^3$, $\;R=2.52\times 10^5$ m, $\;n=5.31\times 10^{-5}$ rad/s, and $\;G=6.67\times 10^{-11}$ m$^3$ kg$^{-1}$ s$^{-2}$, \,the above inequality yields:
 \ba
 \eta > 10^{11}\;\mbox{Pa~s}\,\;,
 \label{gaft}
 \ea
 a condition satisfied very safely by realistic ices. (Recall that the viscosity of ice near melting point is about $\,10^{14}$ Pa s.)

 \section{Power exerted by the gravitational tides\\ in the 1:1 spin-orbit resonance\label{powergrav}}

 Here and everywhere, we say \,{\it{gravitational tides}}, to distinguish them from the tides generated by the alternating component of the quadrupole part of the centrifugal force.
  In this section, we write down a truncated version of the expression (\ref{long}) for the power, in the 1:1 spin-orbit state. This means, that when the tidal modes $\,\beta_{lmpqs}\,$ emerging in that expression are calculated through formula (\ref{betaapprox}), the value of $\,z=1\,$ must be used there. We shall keep only the quadrupole ($\,l=2\,$) terms containing the zeroth, first or second powers of $\,\sin i\,$, $\,e\,$ or $\,{\cal{A}}_1\,$. In the $\,z=1\,$ spin state, the magnitude $\,{\cal{A}}_1\,$ of the main harmonic of the forced libration in longitude is rendered by equation (\ref{10}).

 \subsection{The main input, one unrelated to libration or obliquity}

 Apply the following simplifying assumptions:
 \begin{itemize}
 \item[1.~] the perturber's orbit is coplanar to the equator of the perturbed body: $\,i=0\,$;
 \item[2.~] the terms of power 4 and higher in the eccentricity $\,e\,$ are omitted;
 \item[3.~] only the quadrupole ($\,l=2\,$) contributions are kept;~\footnote{~While in most settings the $\,l=2\,$ contributions are sufficient, exceptions are known. For example, the orbital dynamics of Phobos is influenced by the $\,l=3\,$ and, possibly, $\,l=4\,$ terms (Bills et al. 2005). For close binary asteroids, terms of even higher degrees may be relevant (Taylor and Margot 2010).}
 \item[4.~] the periapse rate can be neglected as compared to the mean motion;
 \item[5.~] the tidally perturbed body is synchronised with the perturber (so $\,z\,=\,1\,$)\,;
 \item[6.~] there is no libration: $\,{\cal{A}}_1=0\,$.
 \end{itemize}
 The assumptions [1 - 3] require us to keep in expression (\ref{long}) only the terms with~\footnote{~The semidiurnal term with $\, (lmpqs)\,=\,(22000)\,$ does not enter this group, because it contains a multiplier $\,\beta_{lmpqs}\,=\,\beta_{22000}\,$ vanishing in the 1:1 resonance. (This is simply the Kaula tidal mode $\,\omega_{lmpq}\,=\,\omega_{2200}\,$.)\label{for}}
 \ba
 (lmpq^{\,\prime}qs)\;=\;(201,-1,-1,0)\;,\;\,(201110)\;,\;\,(220,-1,-1,0)\;,\;\,(220110)\,\;,
 \label{terms}
 \ea
 where we set $\,q^{\,\prime}=q-s\,$ because $\,J_0(0)=1\,$.
 With also the assumptions [4 - 5] taken into consideration, we see from formula (\ref{betaapprox}) that for all the four terms (\ref{terms}) the physical forcing frequency $\,\chi_{lmpq0}\,\equiv\,|\beta_{lmpq0}|\,$ takes the value of $\,n\,$. So, for synchronous spin, the factors $\,k_2\,\sin\epsilon_2\,$ assume the same value $\,k_2(n)\,\sin\epsilon_2(n)\,$ in  these four terms. Together, these terms yield the commonly used main input that bears no dependence upon libration (Peale \& Cassen 1978, Makarov \& Efroimsky 2014)\,:
 \ba
 ^{(1:1)}\langle P\rangle_{\rm{tide}}^{\,\rm(main)}\,=\;\frac{21}{2}\;\frac{G~{M^{\,*}}^{\,{2}}~R^{\,5}}{a^6}\,n\,e^2\;k_2(n)\;\sin\epsilon_2(n)~~.
 \label{198}
 \label{the}
 \ea

 \subsection{The contribution from forced libration\label{contrlibr}}

 For $\,{\cal{A}}_1\neq 0\,$, the terms with $\,lmpqs\,=\,(2200s)\,$, $\,s\neq 0\,$, $\,$come into play in expression (\ref{long}). Using the asymptotic approximations for the Bessel functions~\footnote{~For small-amplitude libration ($\,m\,{\cal A}\,\ll\,1\,$), the asymptotic expression has the form of
  \ba
 J_s(m\,{\cal{A}})\,\approx\;
 \left\{
 \begin{array}{ll}
 \frac{\textstyle 1}{\textstyle s!}\;\left(\frac{\textstyle m\,{\cal{A}}}{\textstyle 2}\right)^{\rm s}\quad            & \mbox{for~integer}~s\,>\,0\\
 ~\\
 1\;-\;\left(\frac{\textstyle m\,{\cal{A}}}{\textstyle 2}\right)^2             & \mbox{for~integer}~s\,=\,0\\
 ~\\
 \frac{\textstyle (-1)^{\rm s}}{\textstyle (-s)!}\;\left(\frac{\textstyle m\,{\cal{A}}}{\textstyle 2}\right)^{\rm -s}\quad & \mbox{for~integer}~s\,<\,0\;\;\;.
 \end{array}
 \right.
 \nonumber
 \ea
 \label{asymptotic}
 }
  and keeping in mind that $\,G_{lpq}(e)\,\propto\,e^{|q|}\,$, it is possible to show that the leading additional terms caused by libration have a form similar to (\ref{198}), with $\,e{\cal{A}}_1\,$ or $\,{\cal{A}}_1^2\,$ standing in the place of $\,e^2\,$. This way, the forced-libration-caused contribution $\,^{(1:1)}\langle P\rangle_{\rm{tide}}^{(forced)}\,$
  contains two leading-order terms. One is
  \ba
  \nonumber
  ^{(1:1)}\langle P\rangle_{\rm{tide}}^{(lmpq^{\,\prime}qs)=(220,-1,01)}
  +\,^{(1:1)}\langle P\rangle_{\rm{tide}}^{(lmpq^{\,\prime}qs)=(22010,-1)}
  \,+\, \qquad\qquad\qquad\qquad\qquad\qquad\quad\,\quad
  \qquad\;\;\\
  \label{pervyj}\\
  \nonumber
  ^{(1:1)}\langle P\rangle_{\rm{tide}}^{(lmpq^{\,\prime}qs)=(2200,-1,0)}
  +\,^{(1:1)}\langle P\rangle_{\rm{tide}}^{(lmpq^{\,\prime}qs)=(220010)}=\,-\,6\,\frac{\textstyle G\,M^{*\,2}\,R^{\,5}}{\textstyle a^6}\;n\;e\;{\cal{A}}_1\; k_2(n)\;\sin\epsilon_2(n)\;,
  \nonumber
  \ea
  while the other is
  \ba
  ^{(1:1)}\langle P\rangle_{\rm{tide}}^{(lmpq^{\,\prime}qs)=(220001)}
  +\,^{(1:1)}\langle P\rangle_{\rm{tide}}^{(lmpq^{\,\prime}qs)=(22000,-1)}=\,\frac{3}{2}\,\frac{\textstyle G\,M^{*\,2}\,R^{\,5}}{\textstyle a^6}\;n\;{\cal{A}}^2_1\; k_2(n)\;\sin\epsilon_2(n)\,\;.\qquad
  \label{vtoroj}
  \ea

  \subsection{The contribution due to the equatorial obliquity to orbit  }

 In expression (\ref{long}), we now single out the terms containing the inclination functions with subscripts $\,(lmp)\,=\,(201)\,,\,(220)\,,\,(210)\,,\,(211)\;$ only:
 \ba
 \nonumber
 F_{201}(\inc)\,=\,-\,\frac{1}{2}\,+\,\frac{3}{4}\,\sin^2i\,+\,O(\inc^4)~~,~~~F_{220}(\inc)\,=\,3\,-\,\frac{3}{2}\,\sin^2i\,+\,O(\inc^4)~~\\
 \nonumber\\
 F_{210}(\inc)\,=\,\frac{3}{2}\;\sin\inc\,+\,O(\inc^2)\;\;,~~~F_{211}(\inc)\,=\,-\,\frac{3}{2}\;\sin\inc\,+\,O(\inc^2)~~,~~~
 \nonumber
 \ea
 because all the other $\,F_{2mp}(\inc)\,$ are of the order of $\,O(\inc^2)\,$ or higher. Of the zero order in $\,e\,$ are the terms with
 $\,(lpq)\,=\,(200)\,,\,(220)\,,\,(210)\;$:
 \ba
 G_{200}(e)\;=\;1\;+\;O(e^2)\,\;,~~~G_{220}(e)\;=\;1\;+\;O(e^2)\,\;,~~~G_{210}(e)\;=\;1\;+\;O(e^2)\,\;.~~~
 \nonumber
 \ea
 From these formulae for $\,F_{lmp}\,$ and $\,G_{lpq}\,$, we see that of interest in the expansion (\ref{long}) are the terms with $\,(lmpq^{\,\prime}qs)$=$(210000)$, $(211000)$, $(220000)$. For the latter term, however, the frequency $\,\beta_{lmpqs}=\beta_{22000}=\omega_{2200}\,$ vanishes, so we end up with the former two terms only:
 \ba
 \nonumber
 ^{(1:1)}\langle P\rangle_{\rm{tide}}^{\rm(obliquity)}&=&^{(1:1)}\langle P\rangle_{\rm{tide}}^{(lmpq^{\,\prime}qs)=(210000)}
  +\,^{(1:1)}\langle P\rangle_{\rm{tide}}^{(lmpq^{\,\prime}qs)=(211000)}\\
  \label{incl}\\
  &=&\frac{3}{2}\,\frac{\textstyle G\,M^{*\,2}\,R^{\,5}}{\textstyle a^6}\;n\;\sin^2 i\; k_2(n)\;\sin\epsilon_2(n)\,\;.\qquad
 \nonumber
 \ea

 \subsection{The total power exerted by the gravitational tides\\ in the synchronous spin state}

 Combining the main input (\ref{the}) with the additions (\ref{pervyj}), (\ref{vtoroj}), and (\ref{incl}), we write:
  \ba
  \nonumber
  ^{(1:1)}\langle P\rangle_{\rm{tide}}&=&^{(1:1)}\langle P\rangle_{\rm{tide}}^{\rm(main)}\,+\;^{(1:1)}\langle P\rangle_{\rm{tide}}^{\rm(forced)}\,+\;^{(1:1)}\langle P\rangle_{\rm{tide}}^{\rm(obliquity)}\\
  \nonumber\\
  &=&\frac{G~{M^{\,*}}^{\,{2}}~R^{\,5}}{a^6}\,n\,k_2(n)\;\sin\epsilon_2(n)\;
  \left[\,\frac{21}{2}\;e^2\,-\;6\;{\cal{A}}_1\;{e}\;+\;\frac{3}{2}\;{\cal{A}}^2_1\;+\;\frac{3}{2}\;\sin^2 i\,\right]
  \nonumber\\
  \label{spinstate}\\
  &=&\frac{21}{2}\;\frac{G~{M^{\,*}}^{\,{2}}~R^{\,5}}{a^6}\,n\,e^2\,k_2(n)\;\sin\epsilon_2(n)\;
  \left[\,1\,-\;\frac{4}{7}\;\frac{{\cal{A}}_1}{e}\;+\;\frac{1}{7}\;\frac{{\cal{A}}^2_1}{e^2}\;+\;\frac{4}{7}\;\frac{\sin^2 i}{e^2}\,\right]\,\;.\;\qquad\;
  \nonumber
  \ea

 \section{Power exerted by the gravitational tides\\ in the 3:2 spin-orbit resonance\label{gravpower}\label{ken}}

 Our goal here is to write down a truncated version of the expression (\ref{long}) for the power. The tidal modes $\,\beta_{lmpqs}\,$ emerging in that expression are given by formula (\ref{betaapprox}), with $\,z=3/2\,$ inserted therein. In our approximation, we are interested in the quadrupole ($\,l=2\,$) terms containing the zeroth, first or second powers of $\,\sin i\,$, $\,e\,$, $\,{\cal{A}}_1\,$.
 Everywhere in this section, by $\,{\cal{A}}_1\,$ we imply the magnitude of the principal harmonic of forced libration,
  $\,^{(3:2)}{\cal{A}}_1\,$, which is given by equation (\ref{11}).

 \subsection{The main input and the obliquity-caused input}

 We start out with the inputs not related to libration. This means that in the expression (\ref{long}) we take into consideration only the terms with $\,s=0\,$ and $\,q^{\,\prime}=q\,$. These terms are:
 \ba
 \nonumber
 & &\sum_{q\,q^{\,\prime}\,s}G_{lpq}(e)\,G_{lpq^{\,\prime}}\,J_{q^{\,\prime}-q+s}(m{\cal{A}}_1)\,J_{s}(m{\cal{A}}_1)\,\beta_{lmpqs}\,k_l(\beta_{lmpqs})\,\sin\epsilon_l(\beta_{lmpqs})\\
 \nonumber
 &=&\sum_q G_{lpq}^2(e)\,J^2_0(0)\,\beta_{lmpqs}\,k_l(\beta_{lmpqs})\,\sin\epsilon_l(\beta_{lmpqs})\\
 &=&\sum_q G_{lpq}^2(e)\,\beta_{lmpqs}\,k_l(\beta_{lmpqs})\,\sin\epsilon_l(\beta_{lmpqs})\,\;.
 \label{}
 \ea

 \subsubsection{Terms with $\,m=0\,$}

 We need the summands that are at most quadratic in the sine of the obliquity. Hence, out of the terms with $\,m=0\,$ in the expression (\ref{long}), we shall keep only those with $\,\{lmp\}=\{201\}\,$. This is justified the following expressions for the inclination functions $\,F_{lmp}(i)\;$:
 \ba
 \nonumber
 F_{200}(i)&=&-\;\frac{3}{8}\;\sin i\quad\Longrightarrow\quad F^2_{200}(i)\,=\;O(i^4)\quad,\\
 F_{201}(i)&=&-\;\frac{1}{2}\;+\;\frac{3}{4}\;\sin^2 i\quad\Longrightarrow\quad F^2_{201}(i)\,=\;\frac{1}{4}\;-\;\frac{3}{4}\;\sin^2i\;+\;O(i^4)\quad,
 \label{}\\
 F_{202}(i)&=&-\;\frac{3}{8}\;\sin i\quad\Longrightarrow\quad F^2_{202}(i)\,=\;O(i^4)\,\;.
 \nonumber
 \ea
 Of the terms with $\,\{lmpqs\}\,=\,\{201q0\}\,$, we keep only those with $\,q\,=\,-1,\,0,\,1\,$. The squares of the appropriate eccentricity functions $\,G_{lpq}(e)\,$ contain terms of the zeroth, first, and second order in $\,e\,$:
 \ba
 \nonumber
 G_{21,-1}(e)&=&\frac{3}{2}\;e\;+\;O(e^3)\quad\Longrightarrow\quad G^2_{21,-1}(e)\,=\;\frac{9}{4}\;e^2\;+\;O(e^4)\quad,\\
 G_{210}(e)&=&(1\,-\,e^2)^{-3/2}\quad\Longrightarrow\quad G^2_{210}(e)\,=\;1\;+\;3\;e^2\;+\;O(e^4)\quad,
 \label{needed}\\
 G_{211}(e)&=&\frac{3}{2}\;e\;+\;O(e^3)\quad\Longrightarrow\quad G^2_{211}(e)\,=\;\frac{9}{4}\;e^2\;+\;O(e^4)\,\;,
 \nonumber
 \ea
 while the squares of the functions with $\,|q|\geq 2\,$ are of the order of $\,e^4\,$ or higher.

 The numerical factor emerging in the expression (\ref{long}) is
 \ba
 \frac{(l-m)!}{(l+m)!}\;\left(2\,-\,\delta_{0m}\right)\;=\;1\,\;.
 \label{}
 \ea
 Inserting $\,z=3/2\,$, $\,m=0\,$ and $\,p=1\,$ in the expression (\ref{betaapprox}), we obtain the tidal modes:
 \ba
 \beta_{lmpqs}\,=\;\beta_{201q0}\,=\;q\,n\,\;.
 \label{}
 \ea
 Therefrom we see that the term with $\,q=0\,$ can be thrown out, because $\,\beta_{20100}\,$ vanishes. (Recall that in the expression (\ref{long}) an $\,lmpqs\,$ term is proportional to $\,\beta_{lmpqs}\,$.) Hence the only $\,m=0\,$ terms that we keep are those with $\,\{lmpqs\}\,=\,\{201,-1,0\}\,$ and $\,\{20110\}\,$. Together, they make the following input into the sum (\ref{long}):
 \ba
 \nonumber
 \frac{G\,{M^{\,*}}^{\,2}}{a}\,\left(\frac{R}{a}\right)^5
 F^2_{201}(i)\,\left[\,G^2_{21,-1}(e)\,\beta_{201,-1,0}\,k_2(\beta_{201,-1,0})\,\sin\epsilon_2(\beta_{201,-1,0}) \right.\qquad\qquad\qquad\qquad\\
 \nonumber
 \left.  +\;G^2_{211}(e)\,\beta_{20110}\,k_2(\beta_{20110})\,\sin\epsilon_2(\beta_{20110})
 \,\right]\qquad\qquad\qquad\qquad\qquad\,\quad\\
  \label{m=0}\\
 =\;\frac{G\,{M^{\,*}}^{\,2}}{a}\,\left(\frac{R}{a}\right)^5\;\frac{9}{8}\;e^2\,\left(1\,-\,3\,\sin^2 i\right)\;\beta\,k_2(\beta)\,\sin\epsilon_2(\beta)\bigg|_{\beta=n}\,\;,   \qquad\;\qquad\;\qquad
 \nonumber
 \ea
 where we took into account that the product $\;\beta\,k_2(\beta)\,\sin\epsilon_2(\beta)\;$ in an even function of $\,\beta\,$, so the $\,q=-1\,$ and $q=1$ terms are equal here.

 \subsubsection{Terms with $\,m=1\,$ and $\,p=0\,$\label{sect}}

 From the expressions for the $\,F_{lmp}\,$ functions,
 \ba
 \nonumber
 F_{210}(i)&=&\frac{3}{2}\;\sin i\;\left(1+\cos i\right) \quad\Longrightarrow\quad F^2_{210}(i)\,=\;\frac{9}{4}\;\sin^2i\;+\;O(i^3)\quad,\\
 F_{211}(i)&=&-\;\frac{3}{2}\;\sin i\;\cos i\quad\Longrightarrow\quad F^2_{211}(i)\,=\;\frac{9}{4}\;\sin^2i\;+\;O(i^3)\quad,
 \label{guba}\\
 F_{212}(i)&=&-\;\frac{3}{4}\;\sin i\;\left(1-\cos i\right)\quad\Longrightarrow\quad F^2_{212}(i)\,=\;O(i^6)\,\;,
 \nonumber
 \ea
 we see that only the terms with $\,p=0\,$ and $\,p=1\,$ matter. On both these occasions, we need only the terms with $\,q\,=\,-1,\,0,\,1\,$,
 because the squares of the functions with $\,|q|\geq 2\,$ are of the order of $\, e^4\,$ or higher.

 In the case of $\,\{m,p\}\,=\,\{1,0\}\,$, the squares of the eccentricity functions $\,G_{lpq}(e)\,$ are:
 \ba
 \nonumber
 G_{20,-1}(e)&=&-\;\frac{1}{2}\;e\;+\;O(e^3)\quad\Longrightarrow\quad G^2_{20,-1}(e)\,=\;\frac{1}{4}\;e^2\;+\;O(e^4)\quad,\\
 G_{200}(e)&=&1\;-\;\frac{5}{2}\;e^2\;+\;O(e^4)\quad\Longrightarrow\quad G^2_{200}(e)\,=\;1\;-\;5\;e^2\;+\;O(e^4)\quad,
 \label{gir}\\
 G_{201}(e)&=&\frac{7}{2}\;e\;+\;O(e^3)\quad\Longrightarrow\quad G^2_{201}(e)\,=\;\frac{49}{4}\;e^2\;+\;O(e^4)\,\;,
 \nonumber
 \ea
 while the numerical factor in the expression (\ref{long}) is
 \ba
 \frac{(l-m)!}{(l+m)!}\;\left(2\,-\,\delta_{0m}\right)\;=\;\frac{1}{3}\,\;.
 \label{numa}
 \ea
 Inserting $\,z=3/2\,$, $\,m=1\,$ and $\,p=0\,$ in the expression (\ref{betaapprox}), we get the tidal modes:
 \ba
 \beta_{lmpqs}\,=\;\beta_{210q0}\,=\;\left(\,\frac{1}{2}\,+\,q\right)\,n\,\;.
 \label{}
 \ea
 Hence the terms with $\,\{lmpqs\}\,=\,\{210,-1,0\}\,$ $\,\{21000\}\,$, and $\,\{21010\}\,$ together give us:
 \ba
 \nonumber
 \frac{G\,{M^{\,*}}^{\,2}}{a}\,\left(\frac{R}{a}\right)^5
 \,\frac{1}{3}\;F^2_{210}(i)\,
 \left[\,
 G^2_{20,-1}(e)\,\beta_{210,-1,0}\,k_2(\beta_{210,-1,0})\,\sin\epsilon_2(\beta_{210,-1,0})\;+ \right. \qquad\qquad\\
 \nonumber
 G^2_{200}(e)\,\beta_{21000}\,k_2(\beta_{21000})\,\sin\epsilon_2(\beta_{21000})\;+
 %  \qquad\qquad  ~\\  \nonumber
 \left.  G^2_{201}(e)\,\beta_{21010}\,k_2(\beta_{21010})\,\sin\epsilon_2(\beta_{21010})
 \,\right]\;=
 %  \qquad\,\quad
 ~\\
 \nonumber\\
 \nonumber
 \frac{G\,{M^{\,*}}^{\,2}}{a}\,\left(\frac{R}{a}\right)^5\;\frac{3}{4}\;\sin^2 i\,\left[\,\frac{1}{4}\,e^2\;
  \beta\,k_2(\beta)\,\sin\epsilon_2(\beta)\bigg|_{\beta=-n/2}\right.\qquad\qquad\qquad\qquad\qquad\\
  \nonumber\\
 \left.
 \,+\;\left(1\,-\,5\,e^2\right)\;\beta\,k_2(\beta)\,\sin\epsilon_2(\beta)\bigg|_{\beta=n/2}
 \,+\;
  \frac{49}{4}\,e^2\;
  \beta\,k_2(\beta)\,\sin\epsilon_2(\beta)\bigg|_{\beta=3n/2}
 \;\right]\,\;,
  \label{Reco}
 \ea
 Recalling that the product $\,\beta\,k_2(\beta)\,\sin\epsilon_2(\beta)\,$ is an even function of $\,\beta\,$, we rewrite the above as
 \ba
 \nonumber
  \frac{G\,{M^{\,*}}^{\,2}}{a}\,\left(\frac{R}{a}\right)^5\;\frac{3}{4}\;\sin^2 i\;
  \beta\,k_2(\beta)\,\sin\epsilon_2(\beta)\bigg|_{\beta=n/2}\;+\qquad\qquad\qquad\qquad\qquad\qquad\qquad\qquad\qquad\;\qquad\\
   \label{m=1,p=0}\\
 \frac{G\,{M^{\,*}}^{\,2}}{a}\,\left(\frac{R}{a}\right)^5\;\frac{3}{4}\;e^2\;\sin^2 i\,\left[\,
 -\;\frac{19}{4}\;\beta\,k_2(\beta)\,\sin\epsilon_2(\beta)\bigg|_{\beta=n/2}
 \,+\;
  \frac{49}{4}\;
  \beta\,k_2(\beta)\,\sin\epsilon_2(\beta)\bigg|_{\beta=3n/2}
 \;\right]\,\;.\qquad
 \nonumber
 \ea

 \subsubsection{Terms with $\,m=1\,$ and $\,p=1\,$\label{nect}}

 As we saw in the preceding subsection, for $\,m=1\,$ only the terms with $\,p=0\,$ and $\,p=1\,$ matter. On both occasions, only the terms with $\,q\,=\,-1,\,0,\,1\,$ are
 important, because the squares of the eccentricity functions with $\,|q|\geq 2\,$ are of the order of $\, e^4\,$ or higher. The case of $\,p=0\,$ was addressed in that subsection. Now, we consider the situation with $\,p=1\,$. The needed eccentricity functions are given by the expression (\ref{needed}) above, while the numerical factor is given by (\ref{numa}).  Inserting $\,z=3/2\,$, $\,m=1\,$ and $\,p=1\,$ in the expression (\ref{betaapprox}), we write down the tidal modes:
 \ba
 \beta_{lmpqs}\,=\;\beta_{211q0}\,=\;\left(\,-\,\frac{1}{2}\,+\,q\right)\,n\,\;.
 \label{}
 \ea

 Then the terms with $\,\{lmpqs\}\,=\,\{211,-1,0\}\,$ $\,\{21100\}\,$, and $\,\{21110\}\,$ make, together, the following contribution in the sum (\ref{long}):
 \ba
 \nonumber
 \frac{G\,{M^{\,*}}^{\,2}}{a}\,\left(\frac{R}{a}\right)^5
 \,\frac{1}{3}\;F^2_{211}(i)\,
 \left[\,
 G^2_{21,-1}(e)\,\beta_{211,-1,0}\,k_2(\beta_{211,-1,0})\,\sin\epsilon_2(\beta_{211,-1,0})\;+ \right. \qquad\qquad\\
 \nonumber
 G^2_{210}(e)\,\beta_{21100}\,k_2(\beta_{21100})\,\sin\epsilon_2(\beta_{21100})\;+
 %  \qquad\qquad  ~\\  \nonumber
 \left.  G^2_{211}(e)\,\beta_{21110}\,k_2(\beta_{21110})\,\sin\epsilon_2(\beta_{21110})
 \,\right]\;=
 %  \qquad\,\quad
 ~\\
 \nonumber\\
 \nonumber
 \frac{G\,{M^{\,*}}^{\,2}}{a}\,\left(\frac{R}{a}\right)^5\;\frac{3}{4}\;\sin^2 i\,\left[\,\frac{9}{4}\,e^2\;
  \beta\,k_2(\beta)\,\sin\epsilon_2(\beta)\bigg|_{\beta=-5n/2}\right.\qquad\qquad\qquad\qquad\qquad\\
  \nonumber\\
 \left.
 \,+\;\left(1\,+\,3\,e^2\right)\;\beta\,k_2(\beta)\,\sin\epsilon_2(\beta)\bigg|_{\beta=-3n/2}
 \,+\;
  \frac{9}{4}\,e^2\;
  \beta\,k_2(\beta)\,\sin\epsilon_2(\beta)\bigg|_{\beta=-n/2}
 \;\right]
 \nonumber
 \ea
 \ba
 % \nonumber
 =\;\frac{G\,{M^{\,*}}^{\,2}}{a}\,\left(\frac{R}{a}\right)^5\;\frac{3}{4}\;\sin^2 i\,\;
 \beta\,k_2(\beta)\,\sin\epsilon_2(\beta)\bigg|_{\beta=-3n/2}\;+\;
  \frac{G\,{M^{\,*}}^{\,2}}{a}\,\left(\frac{R}{a}\right)^5\;\frac{3}{4}\;e^2\;\sin^2 i\,\;\;\qquad\;\qquad\;
    \label{m=1,p=1}
    \label{Neco}
 \ea
 \ba
 \nonumber
 \left[\,\frac{9}{4}\;\beta\,k_2(\beta)\,\sin\epsilon_2(\beta)\bigg|_{\beta=-5n/2}
 % \right. \qquad\qquad\\ \nonumber\\  && \left.
 \;+\;3\,\beta\,k_2(\beta)\,\sin\epsilon_2(\beta)\bigg|_{\beta=-3n/2}
 +\;\frac{9}{4}\;\beta\,k_2(\beta)\,\sin\epsilon_2(\beta)\bigg|_{\beta=n/2}
 \right]\,\;.\;
 \nonumber
 \ea
 % where we took into account that the product $\,\beta\,k_2(\beta)\,\sin\epsilon_2(\beta)\,$ is an even function of $\,\beta\,$.

 \subsubsection{Terms with $\,m=2\,$\label{be}}

 Writing down the relevant $\,F_{lmp}\,$ functions,
 \ba
 \nonumber
 F_{220}(i)&=&\frac{3}{4}\;\left(1+\cos i\right)^2 \quad\Longrightarrow\quad F^2_{220}(i)\,=\;9\;-\;9\;\sin^2i\;+\;O(i^4)\quad,\\
 F_{221}(i)&=&-\;\frac{3}{2}\;\sin^2 i \quad\Longrightarrow\quad F^2_{221}(i)\,=\;O(i^4)\quad,
 \label{1198}\\
 F_{222}(i)&=&-\;\frac{3}{4}\;\left(1-\cos i\right)^2\quad\Longrightarrow\quad F^2_{222}(i)\,=\;O(i^8)\,\;,
 \nonumber
 \ea
 we observe that only the terms with $\,p=0\,$ are important. As before, of those we need only the ones with $\,q\,=\,-1,\,0,\,1\,$,
 because the squares of the functions with $\,|q|\geq 2\,$ are of the order of $\, e^4\,$ or higher. The corresponding eccentricity functions are given by the expression (\ref{gir}), while the numerical factor is equal to
 \ba
 \frac{(l-m)!}{(l+m)!}\;\left(2\,-\,\delta_{0m}\right)\;=\;\frac{1}{12}\,\;.
 \label{numma}
 \ea
 Inserting $\,z=3/2\,$, $\,m=2\,$ and $\,p=0\,$ in the expression (\ref{betaapprox}), we get the tidal modes:
 \ba
 \beta_{lmpqs}\,=\;\beta_{220q0}\,=\;\left(\,-\,1\,+\,q\right)\,n\,\;,
 \label{}
 \ea
 which become zero for $\,q=1\,$, so the appropriate term contributes nothing in the sum (\ref{long}). The terms with $\,\{lmpqs\}\,=\,\{220,-1,0\}\,$ and $\,\{22000\}\,$ make the following contribution:
  \ba
 \nonumber
 \frac{G\,{M^{\,*}}^{\,2}}{a}\,\left(\frac{R}{a}\right)^5
 \,\frac{1}{12}\;F^2_{220}(i)\,
 \left[\,
 G^2_{20,-1}(e)\,\beta_{220,-1,0}\,k_2(\beta_{220,-1,0})\,\sin\epsilon_2(\beta_{220,-1,0}) \right. \qquad\qquad\\
 \nonumber
 \left. +\;G^2_{200}(e)\,\beta_{22000}\,k_2(\beta_{22000})\,\sin\epsilon_2(\beta_{22000})\,\right]\;=  \quad\quad\qquad\,\quad
 ~\\
 \label{begin}\\
 \nonumber
 \frac{G{M^{*}}^{\,2}}{a}\left(\frac{R}{a}\right)^5\,\frac{3}{4}\,\cos^2i\,\left[\frac{1}{4}\,e^2\;
  \beta\,k_2(\beta)\,\sin\epsilon_2(\beta)\bigg|_{\beta=-2n}
 %  \right.\qquad\qquad\qquad\qquad\qquad\\ \nonumber\\ \left.
 +\,\left(1-5e^2\right)\;\beta\,k_2(\beta)\,\sin\epsilon_2(\beta)\bigg|_{\beta=-n}
 \;\right]\,\;.\qquad
  \nonumber
 \ea
 Separating the powers of $\,e\,$ and $\,\sin i\,$, we write this as
 \ba
 \nonumber
 &\,&\frac{G{M^{*}}^{\,2}}{a}\left(\frac{R}{a}\right)^5\,\frac{3}{4}\;\beta\,k_2(\beta)\,\sin\epsilon_2(\beta)\bigg|_{\beta=-n}
 -\;\frac{G{M^{*}}^{\,2}}{a}\left(\frac{R}{a}\right)^5\,\frac{3}{4}\;\sin^2i\;\beta\,k_2(\beta)\,\sin\epsilon_2(\beta)\bigg|_{\beta=-n}\\
 \nonumber\\
 \nonumber\\
 %  \nonumber
 &+&\frac{G{M^{*}}^{\,2}}{a}\left(\frac{R}{a}\right)^5\,\frac{3}{4}\;e^2\;\left[\,\frac{1}{4}\;\beta\,k_2(\beta)\,\sin\epsilon_2(\beta)\bigg|_{\beta=-2n}
 -\;5\;\beta\,k_2(\beta)\,\sin\epsilon_2(\beta)\bigg|_{\beta=-n}
 \,\right]
 \label{m=2}\\
 \nonumber\\
 \nonumber\\
 \nonumber
 &+&\frac{G{M^{*}}^{\,2}}{a}\left(\frac{R}{a}\right)^5\,\frac{3}{4}\;e^2\;\sin^2 i\;\left[\;-\;\frac{1}{4}\;\beta\,k_2(\beta)\,\sin\epsilon_2(\beta)\bigg|_{\beta=-2n}
 +\;5\;\beta\,k_2(\beta)\,\sin\epsilon_2(\beta)\bigg|_{\beta=-n}
 \,\right]\,\;.
 \ea

 \subsubsection{The total power exerted by the main and obliquity-caused terms}

 Summing up the expressions (\ref{m=0}), (\ref{m=1,p=0}), (\ref{m=1,p=1}), and (\ref{m=2}), we arrive at
 \ba
 \nonumber
 && ^{(3:2)}P_{\rm{tide}}^{\rm(main)}\,+\,^{(3:2)}P_{\rm{tide}}^{\rm({obliquity})}\,= \\
 \label{}\\
 \nonumber
 &&\frac{G{M^{*}}^{\,2}}{a}\left(\frac{R}{a}\right)^5\,\frac{3}{4}\;\beta\,k_2(\beta)\,\sin\epsilon_2(\beta)\bigg|_{\beta=n}+\\
 \nonumber\\
 \nonumber
  &&\frac{G{M^{*}}^{\,2}}{a}\left(\frac{R}{a}\right)^5\,\frac{3}{4}\;e^2\;\left[\;-\;\frac{7}{2}\;\beta\,k_2(\beta)\,\sin\epsilon_2(\beta)\bigg|_{\beta=n}
 +\;\frac{1}{4}\;\beta\,k_2(\beta)\,\sin\epsilon_2(\beta)\bigg|_{\beta=2n}
  \,\right]+\\
    \nonumber\\
 \nonumber
 &&\frac{G{M^{*}}^{\,2}}{a}\left(\frac{R}{a}\right)^5\,\frac{3}{4}\;\sin^2i\;\left[\,\beta\,k_2(\beta)\,\sin\epsilon_2(\beta)\bigg|_{\beta=n/2}
 -\;\beta\,k_2(\beta)\,\sin\epsilon_2(\beta)\bigg|_{\beta=n}
 +\;\beta\,k_2(\beta)\,\sin\epsilon_2(\beta)\bigg|_{\beta=3n/2}
 \,\right]+\\
  \nonumber\\
 \nonumber
 &&\frac{G{M^{*}}^{\,2}}{a}\left(\frac{R}{a}\right)^5\,\frac{3}{4}\;e^2\;\sin^2 i\;\left[\;-\;\frac{5}{2}\;\beta\,k_2(\beta)\,\sin\epsilon_2(\beta)\bigg|_{\beta=n/2}
 +\;\frac{1}{2}\;\beta\,k_2(\beta)\,\sin\epsilon_2(\beta)\bigg|_{\beta=n}
 \right.\\
 \nonumber\\
 % \nonumber
 &&\left.
 +\;\frac{61}{4}\;\beta\,k_2(\beta)\,\sin\epsilon_2(\beta)\bigg|_{\beta=3n/2}-\;\frac{1}{4}\;\beta\,k_2(\beta)\,\sin\epsilon_2(\beta)\bigg|_{\beta=2n}
 +\;\frac{9}{4}\;\beta\,k_2(\beta)\,\sin\epsilon_2(\beta)\bigg|_{\beta=5n/2}
  \,\right]
  \label{tot}
 \ea
 Writing this expression, we were keeping in mind that the product $\,\beta\,k_2(\beta)\,\sin\epsilon_2(\beta)\,$ is an even function of $\,\beta\,$. So in the subscripts we freely substituted $\;\beta=-n\;$ with $\;\beta=n\;$, etc.

 In the special case of the Maxwell model, and under the additional assumption that all the involved tidal modes are located to the right of the peak shown in Figure \ref{Fig1},
 the product $\;\beta\,k_2(\beta)\,\sin\epsilon_2(\beta)\;$ is frequency-independent and is equal to $\;2GR^{\,2}\rho^2/3\eta\;$, $\,$see equation (\ref{me}) in Appendix \ref{chi}. In this case, the above expression becomes
 \bs
 \ba
 \nonumber
 && ^{(3:2)}P_{\rm{tide}}^{\rm(main)}\,+\,^{(3:2)}P_{\rm{tide}}^{\rm({obliquity})}\,=\\
 \label{baron}\\
 \nonumber
 && \;\; \frac{G{M^{*}}^{\,2}}{a}\left(\frac{R}{a}\right)^5\,\frac{3}{4}\,\left[\,\left(1\;-\;\frac{13}{4}\;e^2\right)\;+\;\left(1\;+\;
 \frac{61}{4}\;e^2\right)\;\sin^2 i
  \,\right]\;\beta\,k_2(\beta)\,\sin\epsilon_2(\beta)\,\;,\qquad\;\qquad
  \ea
 which agrees with the equation (16) from Makarov \& Efroimsky (2014).

 The further employment of the relation (\ref{manner}) leads us to
  \ba
 ^{(3:2)}P_{\rm{tide}}^{\rm(main)}\,+\,^{(3:2)}P_{\rm{tide}}^{\rm({obliquity})}\,= \;\frac{1}{2\eta}\;R^{\,7}\,\rho^2\,n^4\,\left[\,\left(1\;-\;\frac{13}{4}\;e^2\right)\;+\;\left(1\;+\;
 \frac{61}{4}\;e^2\right)\;\sin^2 i
  \,\right]\;\;,\qquad
  \label{graf}
 \ea
 \es
 where we approximated the osculating mean motion $\,\sqrt{G(M^*+M)/a^3\,}\approx\sqrt{GM^*/a^3\,}\,$ with the anomalistic mean motion $\,n\equiv\,\stackrel{\bf{\centerdot}}{\cal{M}\,}$.

 \subsection{The forced-libration-caused input}

 Since in the expression for the power (\ref{long}) the libration magnitude $\,{\cal{A}}\,$ stands multiplied with $\,m\,$, the contribution from the terms with $\,m=0\,$ is obviously zero. Thence, as we saw in Sections \ref{sect} - \ref{be}, of importance will be the terms with $\,\left\{lmp\right\}$ = $\left\{210\right\}$, $\left\{211\right\}$, $\left\{220\right\}$.

 \subsubsection{The contribution from the terms with $\,\left\{lmp\right\}$ = $\left\{210\right\}$}

 The expressions for $\,F^2_{210}(i)\,$ and the numerical factor are given by the formulae (\ref{guba}) and (\ref{numa}), correspondingly.

 \begin{itemize}
 \item[] {{Terms with $\,q\;=\;-\;1\;$}} are given in Table \ref{210-1}.
 %  \vspace{2mm} ~\\
 The corresponding modes are
 \ba
 \beta_{lmpqs}\,=\;\beta_{210,-1,s}\,=\;\left(\,\frac{1}{2}\,+\,q\,-\,s\right)\,n\,=\,\left(\,-\,\frac{1}{2}\,-\,s\right)\,n\,\;.
 \label{}
 \ea
 \end{itemize}
 \begin{table}[htbp]
\begin{center}
\begin{tabular}{llccccc}
\hline
\hline
             &  & $\;q^{\,\prime}\;=\;-\;1\;$ & $\;q^{\,\prime}\;=\;0\;$                                             & $\;q^{\,\prime}\;=\;1\;$                   & $\;q^{\,\prime}\;=\;2\;$                   &\\
$~\;s\;$    & $\beta_{210,-1,s}$ &               &                                          &                                                     &                           &\\
\hline
\hline
$-2~\quad$  & $\;\;\;\,\;\frac{\textstyle 3}{\textstyle 2}\,n$ & $O(e^2{\cal{A}}_1^4)$ & $O(e{\cal{A}}_1^3)$ & $-\;\frac{\textstyle 3}{\textstyle 128}\,e^2\,{\cal{A}}_1^2\;\sin^2 i$&$\,O(e^3{\cal{A}}_1^3)\,$ &\\
\hline
$-1~\quad$  & $\;\;\,\;\;\frac{\textstyle 1}{\textstyle 2}\,n$ & $\;\,\frac{\textstyle 3}{\textstyle 64}\,e^2\,{\cal{A}}_1^2\;\sin^2 i$ & $\;\;\frac{\textstyle 3}{\textstyle 16}\,e\,{\cal{A}}_1\;\sin^2 i$ & $\;\;\;\frac{\textstyle 3}{\textstyle 64}\,e^2\,{\cal{A}}_1^2\;\sin^2 i$ &$\,O(e^3{\cal{A}}_1^3)\,$ &\\
\hline
$~\;0\quad$ & $\;-\;\frac{\textstyle 1}{\textstyle 2}\,n$ & $-\,\frac{\textstyle 3}{\textstyle 32}\,e^2\,{\cal{A}}_1^2\;\sin^2 i\;\,$  &  $\;-\,\frac{\textstyle 3}{\textstyle 16}\,e\,{\cal{A}}_1\;\sin^2 i\;$ &  $-\;\frac{\textstyle 3}{\textstyle 128}\,e^2\,{\cal{A}}_1^2\;\sin^2 i$                         &$\,O(e^3{\cal{A}}_1^3)\,$ &\\
\hline
$~\;1\quad$ & $\;-\;\frac{\textstyle 3}{\textstyle 2}\,n$ & $\;\,\frac{\textstyle 3}{\textstyle 64}\,e^2\,{\cal{A}}_1^2\;\sin^2 i$ &         $O(e{\cal{A}}_1^3)$                  &         $O(e^2{\cal{A}}_1^4)$                  &$\,O(e^3{\cal{A}}_1^5)\,$ &\\
\hline
$~\;2\quad$ & $\;-\;\frac{\textstyle 5}{\textstyle 2}\,n$ & $O(e^2{\cal{A}}_1^4)$                      &         $O(e{\cal{A}}_1^5)$                  &           $O(e^2{\cal{A}}_1^6)$                & $\,O(e^3{\cal{A}}_1^7)\,$ &\\
\hline
\hline
\end{tabular}
\end{center}
\caption{. Terms with $\,\left\{lmpq\right\}$ = $\left\{210,-1\right\}$. Only inputs linear or quadratic in $\,{\cal{A}}_1\,$ are kept.
The entry with $\,s=0\,$ and $\,q^{\,\prime}\,=\,-\,1\,$ should contain also an $\,{\cal{A}}_1$-independent term $\;\frac{\textstyle 3}{\textstyle 16}\,e^2\,\sin^2i\;$, but we have dropped this term here, because it has already been taken care of in Section \ref{sect}, see the first term in equation (\ref{Reco}).}
\label{210-1}
\end{table}

 \pagebreak

 \begin{itemize}
 \item[] {{Terms with $\,q\;=\;0\;$}} are given in Table \ref{2100}.
  % \vspace{2mm} ~\\
 The corresponding modes are
 \ba
 \beta_{lmpqs}\,=\;\beta_{2100s}\,=\;\left(\,\frac{1}{2}\,+\,q\,-\,s\right)\,n\,=\,\left(\,\frac{1}{2}\,-\,s\right)\,n\,\;.
 \label{}
 \ea
  \end{itemize}
  \begin{table}[htbp]
\begin{center}
\begin{tabular}{llccccc}
\hline
\hline
             &  & $\;q^{\,\prime}\;=\;-\;1\;$ & $\;q^{\,\prime}\;=\;0\;$                                             & $\;q^{\,\prime}\;=\;1\;$                   & $\;q^{\,\prime}\;=\;2\;$                   &\\
$~\;s\;$    & $\beta_{2100s}$ &                                                         &    &                                                 &                           &\\
\hline
\hline
$-2~\quad$  & $\;\;-\;\frac{\textstyle 3}{\textstyle 2}\,n$ & $O(e{\cal{A}}_1^5)$ & $O({\cal{A}}_1^4)$ & $O(e{\cal{A}}_1^3)$ & $\;\;\;\frac{\textstyle 51}{\textstyle 64}e^2{\cal{A}}_1^2\sin^2i$ &\\
\hline
$-1~\quad$  & $\;\;-\;\frac{\textstyle 1}{\textstyle 2}\,n$ & $O(e{\cal{A}}_1^3)$  & $\;\;\;\frac{\textstyle 3}{\textstyle 16}\,(1-5e^2)\,{\cal{A}}_1^2\;\sin^2 i$ & $ \;-\;\frac{\textstyle 21}{\textstyle 16}\,e\,{\cal{A}}_1\,\sin^2i$ & $-\,\frac{\textstyle 51}{\textstyle 32}\;e^2{\cal{A}}_1^2\sin^2i$ &\\
\hline
$~\;0\quad$ & $\;\;\;\,\;\frac{\textstyle 1}{\textstyle 2}\,n$ & $\,\;\;\;\frac{\textstyle 3}{\textstyle 16}\,e\,{\cal{A}}_1\;\sin^2 i$  &  $\;-\;\frac{\textstyle 3}{\textstyle 8}\,(1-5e^2)\,{\cal{A}}_1^2\;\sin^2 i\;$ &  $ \;\;\;\;\frac{\textstyle 21}{\textstyle 16}\,e\,{\cal{A}}_1\,\sin^2i$                         & $\;\;\;\frac{\textstyle 51}{\textstyle 64}e^2{\cal{A}}_1^2\sin^2i$ &\\
\hline
$~\;1\quad$ & $\;-\;\frac{\textstyle 1}{\textstyle 2}\,n$ & $-\,\frac{\textstyle 3}{\textstyle 16}\,e\,{\cal{A}}_1\;\sin^2 i$ &         $\;\;\;\frac{\textstyle 3}{\textstyle 16}\,(1-5e^2)\,{\cal{A}}_1^2\;\sin^2 i$                  &            $O({\cal{A}}_1^3)$               & $O(e^2{\cal{A}}_1^4)$ &\\
\hline
$~\;2\quad$ & $\;-\;\frac{\textstyle 3}{\textstyle 2}\,n$ & $O(e{\cal{A}}_1^3)$                      &         $O({\cal{A}}_1^4)$                  &           $O(e{\cal{A}}_1^5)$                & $O(e^2{\cal{A}}_1^6)$ &\\
\hline
\hline
\end{tabular}
\end{center}
\caption{. Terms with $\,\left\{lmpq\right\}$ = $\left\{2100\right\}$. Only inputs linear or quadratic in $\,{\cal{A}}_1\,$ are kept. The entry with $\,s=0\,$ and $\,q^{\,\prime}\,=\,0\,$ should contain also an $\,{\cal{A}}_1$-independent term $\;\frac{\textstyle 3}{\textstyle 4}\,(1-5e^2)\,\sin^2 i\;$, but we have dropped this term here, because it has already been taken care of in Section \ref{sect}, see the second term in equation (\ref{Reco}).}
\label{2100}
\end{table}

  \begin{itemize}
 \item[] {{Terms with $\,q\;=\;1\;$}} are given in Table \ref{2101}. The corresponding modes are
 \ba
 \beta_{lmpqs}\,=\;\beta_{2101s}\,=\;\left(\,\frac{1}{2}\,+\,q\,-\,s\right)\,n\,=\,\left(\,\frac{3}{2}\,-\,s\right)\,n\,\;.
 \label{}
 \ea
  \end{itemize}
   \begin{table}[htbp]
\begin{center}
\begin{tabular}{llccccc}
\hline
\hline
             &  & $\;q^{\,\prime}\;=\;-\;1\;$ & $\;q^{\,\prime}\;=\;0\;$                                             & $\;q^{\,\prime}\;=\;1\;$                   &$\;q^{\,\prime}\;=\;2\;$                   &\\
$~\;s\;$    & $\beta_{2101s}$ &         &                                                &                                                     &                           &\\
\hline
\hline
$-2~\quad$  & $\;\;\;\;\,\frac{\textstyle 7}{\textstyle 2}\,n$ & $O(e^2{\cal{A}}_1^6)$ & $O(e{\cal{A}}_1^5)$ & $O(e^2{\cal{A}}_1^4)$&   $O(e^3{\cal{A}}_1^3)$&\\
\hline
$-1~\quad$  & $\;\;\,\;\frac{\textstyle 5}{\textstyle 2}\,n$ & $O(e^2{\cal{A}}_1^4)$ & $O(e{\cal{A}}_1^3)$ & $\;\;\;\frac{\textstyle 147}{\textstyle 64}\,e^2\,{\cal{A}}_1^2\,\sin^2i$ &$O(e^3{\cal{A}}_1)$&\\
\hline
$~\;0\quad$ & $\;\;\,\;\frac{\textstyle 3}{\textstyle 2}\,n$ & $-\,\frac{\textstyle 21}{\textstyle 128}\,e^2\,{\cal{A}}_1^2\;\sin^2 i\;\,$  &  $\;-\,\frac{\textstyle 21}{\textstyle 16}\,e\,{\cal{A}}_1\;\sin^2 i\;\;$ &  $-\;\frac{\textstyle 147}{\textstyle 32}\,e^2\,{\cal{A}}_1^2\,\sin^2i$                         &$O(e^3{\cal{A}}_1)$&\\
\hline
$~\;1\quad$ & $\;\;\,\;\frac{\textstyle 1}{\textstyle 2}\,n$ & $\;\,\frac{\textstyle 21}{\textstyle 64}\,e^2\,{\cal{A}}_1^2\;\sin^2 i$ &         $\;\;\,\frac{\textstyle 21}{\textstyle 16}\,e\,{\cal{A}}_1\;\sin^2 i\;$                  &         $\;\;\;\frac{\textstyle 147}{\textstyle 64}\,e^2\,{\cal{A}}_1^2\,\sin^2i$                  &$O(e^3{\cal{A}}_1^3)$&\\
\hline
$~\;2\quad$ & $\;-\;\frac{\textstyle 1}{\textstyle 2}\,n$ & $-\,\frac{\textstyle 21}{\textstyle 128}\,e^2\,{\cal{A}}_1^2\;\sin^2 i\;\,$                      &         $O(e{\cal{A}}_1^3)$                  &           $O(e^2{\cal{A}}_1^4)$                &$O(e^3{\cal{A}}_1^5)$&\\
\hline
\hline
\end{tabular}
\end{center}
\caption{. Terms with $\,\left\{lmpq\right\}$ = $\left\{2101\right\}$. Only inputs linear or quadratic in $\,{\cal{A}}_1\,$ are kept.
 The entry with $\,s=0\,$ and $\,q^{\,\prime}\,=\,1\,$ should contain also an $\,{\cal{A}}_1$-independent term $\;\frac{\textstyle 147}{\textstyle 16}\,e^2\,\sin^2i\;$, but we have dropped this term here, because it has already been taken care of in Section \ref{sect}, see the third term in equation (\ref{Reco}).}
\label{2101}
\end{table}

 \subsubsection{The contribution from the terms with $\,\left\{lmp\right\}$ = $\left\{211\right\}$}

 The expressions for $\,F^2_{211}(i)\,$ and the numerical factor are given by the formulae (\ref{guba}) and (\ref{numa}), correspondingly.

 \begin{itemize}
 \item[] {{Terms with $\,q\;=\;-\;1\;$}} are given in Table \ref{211-1}.
 % \vspace{2mm} ~\\
 For these terms, the modes are
 \ba
 \beta_{lmpqs}\,=\;\beta_{211,-1,s}\,=\;\left(\,-\,\frac{3}{2}\,+\,q\,-\,s\right)\,n\,=\,\left(\,-\;\frac{5}{2}\,-\,s\right)\,n\,\;.
 \label{}
 \ea
 \end{itemize}
   \begin{table}[htbp]
\begin{center}
\begin{tabular}{llccccc}
\hline
\hline
             &  & $\;q^{\,\prime}\;=\;-\;1\;$ & $\;q^{\,\prime}\;=\;0\;$                                             & $\;q^{\,\prime}\;=\;1\;$                   &$\;q^{\,\prime}\;=\;2\;$                   &\\
$~\;s\;$    & $\beta_{211,-1,s}$ &   &                                                      &                                                     &                           &\\
\hline
\hline
$-2~\quad$  & $\;-\;\frac{\textstyle 1}{\textstyle 2}\,n$ & $O(e^2{\cal{A}}_1^4)$ & $O(e{\cal{A}}_1^3)$ & $\;\;\;\frac{\textstyle 27}{\textstyle 128}\,e^2\,{\cal{A}}_1^2\,\sin^2i$&   $O(e^3{\cal{A}}_1^3)$&\\
\hline
$-1~\quad$  & $\;-\;\frac{\textstyle 3}{\textstyle 2}\,n$ & $\;\,\frac{\textstyle 27}{\textstyle 64}\,e^2\,{\cal{A}}_1^2\;\sin^2 i$  & $-\;\frac{\textstyle 27}{\textstyle 16}\,e\,{\cal{A}}_1\;\sin^2 i\;$ & $-\;\frac{\textstyle 27}{\textstyle 64}\,e^2\,{\cal{A}}_1^2\,\sin^2i\;$ &$O(e^3{\cal{A}}_1^3)$&\\
\hline
$~\;0\quad$ & $\;-\;\frac{\textstyle 5}{\textstyle 2}\,n$ & $-\,\frac{\textstyle 27}{\textstyle 32}\,e^2\,{\cal{A}}_1^2\;\sin^2 i\;\,$  &  $\;\;\,\frac{\textstyle 27}{\textstyle 16}\,e\,{\cal{A}}_1\;\sin^2 i$ &  $\;\;\;\frac{\textstyle 27}{\textstyle 128}\,e^2\,{\cal{A}}_1^2\,\sin^2i$                         &$O(e^3{\cal{A}}_1^3)$&\\
\hline
$~\;1\quad$ & $\;-\;\frac{\textstyle 7}{\textstyle 2}\,n$ & $\;\,\frac{\textstyle 27}{\textstyle 64}\,e^2\,{\cal{A}}_1^2\;\sin^2 i$ &         $O(e{\cal{A}}_1^3)$                 &         $O(e^2{\cal{A}}_1^4)$                  &$O(e^3{\cal{A}}_1^5)$&\\
\hline
$~\;2\quad$ & $\;-\;\frac{\textstyle 9}{\textstyle 2}\,n$ & $O(e^2{\cal{A}}_1^4)$ &         $O(e{\cal{A}}_1^5)$                  &           $O(e^2{\cal{A}}_1^6)$                &$O(e^3{\cal{A}}_1^7)$&\\
\hline
\hline
\end{tabular}
\end{center}
\caption{. Terms with $\,\left\{lmpq\right\}$ = $\left\{211,-1\right\}$. Only inputs linear or quadratic in $\,{\cal{A}}_1\,$  are kept.
The entry with $\,s=0\,$ and $\,q^{\,\prime}\,=\,-\,1\,$ should contain also an $\,{\cal{A}}_1$-independent term $\;\frac{\textstyle 27}{\textstyle 16}\,e^2\,\sin^2i\;$, but we have dropped this term here, because it has already been taken care of in Section \ref{nect}, see the first term in equation (\ref{Neco}).
}
\label{211-1}
\end{table}

\pagebreak

 \begin{itemize}
 \item[] {{Terms with $\,q\;=\;0\;$}} are given in Table \ref{2110}.
 % \vspace{2mm} ~\\
 For these terms, the modes are
 \ba
 \beta_{lmpqs}\,=\;\beta_{2110s}\,=\;\left(\,-\,\frac{3}{2}\,+\,q\,-\,s\right)\,n\,=\,\left(\,-\;\frac{3}{2}\,-\,s\right)\,n\,\;.
 \label{}
 \ea
  \end{itemize}
   \begin{table}[htbp]
\begin{center}
\begin{tabular}{llccccc}
\hline
\hline
             &  & $\;q^{\,\prime}\;=\;-\;1\;$ & $\;q^{\,\prime}\;=\;0\;$                                             & $\;q^{\,\prime}\;=\;1\;$                   & $\;q^{\,\prime}\;=\;2\;$                   &\\
$~\;s\;$    & $\beta_{2110s}$ &                                                         &                         &                            &                           &\\
\hline
\hline
$-2~\quad$  & $\;\;\;\;\;\frac{\textstyle 1}{\textstyle 2}\,n$ & $O(e{\cal{A}}_1^5)$ & $O({\cal{A}}_1^4)$ & $O(e{\cal{A}}_1^3)$&$\;\;\;\frac{\textstyle 27}{\textstyle 128}e^2{\cal{A}}_1^2\,\sin^2i$                &\\
\hline
$-1~\quad$  & $\;-\;\frac{\textstyle 1}{\textstyle 2}\,n$ & $O(e{\cal{A}}_1^3)$  & $\;\;\frac{\textstyle 3}{\textstyle 16}\,(1+3e^2)\,{\cal{A}}_1^2\;\sin^2 i\;$ & $-\;\frac{\textstyle 9}{\textstyle 16}\,e\,{\cal{A}}_1\,\sin^2i\;$ &$-\;\frac{\textstyle 27}{\textstyle 64}e^2{\cal{A}}_1^2\,\sin^2i$                &\\
\hline
$~\;0\quad$ & $\;-\;\frac{\textstyle 3}{\textstyle 2}\,n$ & $-\,\frac{\textstyle 9}{\textstyle 16}\,e\,{\cal{A}}_1\;\sin^2 i\;\,$  &  $-\;\frac{\textstyle 3}{\textstyle 8}\,(1\,+\,3e^2)\,{\cal{A}}_1^2\;\sin^2 i\;\;$ &  $\;\;\;\frac{\textstyle 9}{\textstyle 16}\,e\,{\cal{A}}_1\,\sin^2i$                         &$\;\;\;\frac{\textstyle 27}{\textstyle 128}e^2{\cal{A}}_1^2\,\sin^2i$                &\\
\hline
$~\;1\quad$ & $\;-\;\frac{\textstyle 5}{\textstyle 2}\,n$ & $\;\,\frac{\textstyle 9}{\textstyle 16}\,e\,{\cal{A}}_1\;\sin^2 i$ &         $\;\;\frac{\textstyle 3}{\textstyle 16}\,(1+3e^2)\,{\cal{A}}_1^2\;\sin^2 i\;$                 &         $O(e{\cal{A}}_1^3)$                  &$O(e^2{\cal{A}}_1^4)$                &\\
\hline
$~\;2\quad$ & $\;-\;\frac{\textstyle 7}{\textstyle 2}\,n$ & $O(e{\cal{A}}_1^3)$ &         $O({\cal{A}}_1^4)$                  &           $O(e{\cal{A}}_1^5)$                &$O(e^2{\cal{A}}_1^6)$                &\\
\hline
\hline
\end{tabular}
\end{center}
\caption{. Terms with $\,\left\{lmpq\right\}$ = $\left\{2110\right\}$. Only inputs linear or quadratic in $\,{\cal{A}}_1\,$ are kept.  The entry with $\,s=0\,$ and $\,q^{\,\prime}\,=\,0\,$ should contain also an $\,{\cal{A}}_1$-independent term $\;\frac{\textstyle 3}{\textstyle 4}\,(1+3e^2)\,\sin^2i\;$, but we have dropped this term here, because it has already been taken care of in Section \ref{nect}, see the second term in equation (\ref{Neco}).}
\label{2110}
\end{table}

\pagebreak

  \begin{itemize}
 \item[] {{Terms with $\,q\;=\;1\;$}} are given in Table \ref{2111}.
 % \vspace{2mm} ~\\
 For these terms, the modes are
 \ba
 \beta_{lmpqs}\,=\;\beta_{2111s}\,=\;\left(\,-\,\frac{3}{2}\,+\,q\,-\,s\right)\,n\,=\,\left(\,-\;\frac{1}{2}\,-\,s\right)\,n\,\;.
 \label{}
 \ea
  \end{itemize}
   \begin{table}[htbp]
\begin{center}
\begin{tabular}{llccccc}
\hline
\hline
             &  & $\;q^{\,\prime}\;=\;-\;1\;$ & $\;q^{\,\prime}\;=\;0\;$                                             & $\;q^{\,\prime}\;=\;1\;$                   &$\;q^{\,\prime}\;=\;2\;$                   &\\
$~\;s\;$    & $\beta_{2111s}$ &     &                                                    &                                                     &                           &\\
\hline
\hline
$-2~\quad$  & $\;\;\;\;\;\frac{\textstyle 3}{\textstyle 2}\,n$ & $O(e^2{\cal{A}}_1^6)$ & $O(e{\cal{A}}_1^5)$ & $O(e^2{\cal{A}}_1^4)$&  $O(e^3{\cal{A}}_1^3)$&\\
\hline
$-1~\quad$  & $\;\;\;\;\;\frac{\textstyle 1}{\textstyle 2}\,n$ & $O(e^2{\cal{A}}_1^4)$  & $O(e{\cal{A}}_1^3)$ & $\;\;\;\frac{\textstyle 27}{\textstyle 64}\,e^2\,{\cal{A}}_1^2\,\sin^2i$ &$O(e^3{\cal{A}}_1)$&\\
\hline
$~\;0\quad$ & $\;-\;\frac{\textstyle 1}{\textstyle 2}\,n$ & $\;\,\frac{\textstyle 27}{\textstyle 128}\,e^2\,{\cal{A}}_1^2\;\sin^2 i$  &  $-\;\frac{\textstyle 9}{\textstyle 16}\,e\,{\cal{A}}_1\;\sin^2 i\;\;$ &  $-\;\frac{\textstyle 27}{\textstyle 32}\,e^2\,{\cal{A}}_1^2\,\sin^2i\;$                         &$O(e^3{\cal{A}}_1)$&\\
\hline
$~\;1\quad$ & $\;-\;\frac{\textstyle 3}{\textstyle 2}\,n$ & $-\;\frac{\textstyle 27}{\textstyle 64}\;e^2\;{\cal{A}}_1^2\;\sin^2 i\;\,$ &         $\;\;\frac{\textstyle 9}{\textstyle 16}\,e\,{\cal{A}}_1\;\sin^2 i\;$                 &         $\;\;\;\frac{\textstyle 27}{\textstyle 64}\,e^2\,{\cal{A}}_1^2\,\sin^2i$                  &$O(e^3{\cal{A}}_1^3)$&\\
\hline
$~\;2\quad$ & $\;-\;\frac{\textstyle 5}{\textstyle 2}\,n$ & $\;\,\frac{\textstyle 27}{\textstyle 128}\,e^2\,{\cal{A}}_1^2\;\sin^2 i$ &         $O(e{\cal{A}}_1^3)$                  &           $O(e^2{\cal{A}}_1^4)$                &$O(e^3{\cal{A}}_1^5)$&\\
\hline
\hline
\end{tabular}
\end{center}
\caption{. Terms with $\,\left\{lmpq\right\}$ = $\left\{2111\right\}$. Only inputs linear or quadratic in $\,{\cal{A}}_1\,$ are kept.
 The entry with $\,s=0\,$ and $\,q^{\,\prime}\,=\,1\,$ should contain also an $\,{\cal{A}}_1$-independent term $\;\frac{\textstyle 27}{\textstyle 16}\,e^2\,\sin^2i\;$, but we have dropped this term here, because it has already been taken care of in Section \ref{nect}, see the third term in equation (\ref{Neco}).}
\label{2111}
\end{table}

 \subsubsection{The contribution from the terms with $\,\left\{lmp\right\}$ = $\left\{220\right\}$}

 The expressions for $\,F^2_{220}(i)\,$ and the numerical factor are given by the formulae (\ref{1198}) and (\ref{numma}), correspondingly.

 \begin{itemize}
 \item[] {{Terms with $\,q\;=\;-\;1\;$}} are given in Table \ref{220-1}.
 % \vspace{2mm} ~\\
 The corresponding modes are
  \ba
 \beta_{lmpqs}\,=\;\beta_{220qs}\,=\;(\,-\,1\,+\,q\,-\,s)\,n\,=\,(\,-\,2\,-\,s)\,n\,\;,
 \label{}
 \ea
 so the terms with $\,s=-2\,$ can be omitted.
 \end{itemize}
\begin{table}[htbp]
\begin{center}
\begin{tabular}{llccccc}
\hline
\hline
             &  & $\;q^{\,\prime}\;=\;-\;1\;$ & $\;q^{\,\prime}\;=\;0\;$                                             & $\;q^{\,\prime}\;=\;1\;$                   &$\;q^{\,\prime}\;=\;2\;$                   &\\
$~\;s\;$    & $\beta_{220,-1,s}$ &                                                         &                                                     &          &                 &\\
\hline
\hline
$-1~\quad$  & $\;-\;n$ & $\;\,\frac{\textstyle 3}{\textstyle 16}\cos^2i\,e^2 {\cal{A}}_1^2\;$ & $\;\frac{\textstyle 3}{\textstyle 8}\cos^2i\,e {\cal{A}}_1\;$ & \textcolor{red}{$\frac{\textstyle21}{\textstyle16}\cos^2i\,e^2 {\cal{A}}_1^2$} & $O(e^3{\cal{A}}_1^3)$                &\\
\hline
$~\;0\quad$ & $-\,2n$ & $-\,\frac{\textstyle 3}{\textstyle 8}\cos^2i\,e^2 {\cal{A}}_1^2\;\;\;$  & $-\,\frac{\textstyle 3}{\textstyle 8}\cos^2i\,e {\cal{A}}_1\;$  &  \textcolor{red}{$-\,\frac{\textstyle 21}{\textstyle 32}\cos^2i\,e^2 {\cal{A}}_1^2\;\;\;$}                         & $O(e^3{\cal{A}}_1^3)$                &\\
\hline
$~\;1\quad$ & $-\,3n$ & $\;\,\frac{\textstyle 3}{\textstyle 16}\cos^2i\,e^2 {\cal{A}}_1^2\;$ &         $O(e{\cal{A}}_1^3)$                  &         $O(e^2{\cal{A}}_1^4)$                  & $O(e^3{\cal{A}}_1^5)$                &\\
\hline
$~\;2\quad$ & $-\,4n$ & $O(e^2{\cal{A}}_1^4)$                      &         $O(e{\cal{A}}_1^5)$                  &           $O(e^2{\cal{A}}_1^6)$                & $O(e^3{\cal{A}}_1^7)$                &\\
\hline
\hline
\end{tabular}
\end{center}
\caption{. Terms with $\,\left\{lmpq\right\}$ = $\left\{220,-1\right\}$. Only inputs  linear or quadratic in $\,{\cal{A}}_1\,$ are kept. The terms which do not mutually cancel are shown in red. The entry with $\,s=0\,$ and $\,q^{\,\prime}\,=\,-\,1\,$ should contain also an $\,{\cal{A}}_1$-independent term $\;\frac{\textstyle 3}{\textstyle 16}\,e^2\,(1-\sin^2i)\;$, but we have dropped this term here, because it has already been taken care of in Section \ref{be}, see the first term in equation (\ref{begin}).}
 \label{220-1}
\end{table}

 \pagebreak

 \begin{itemize}
 \item[] {{Terms with $\,q\;=\;0\;$}} are given in Table \ref{2200}.
 % \vspace{2mm} ~\\
 For these terms, the modes are
  \ba
 \beta_{lmpqs}\,=\;\beta_{2200s}\,=\;(\,-\,1\,+\,q\,-\,s)\,n\,=\,(\,-\,1\,-\,s)\,n\,\;,
 \label{}
 \ea
 so the terms with $\,s=-1\,$ can be omitted.
 \end{itemize}
\begin{table}[htbp]
\begin{center}
\begin{tabular}{llccccc}
\hline
\hline
            &  & $\;q^{\,\prime}\;=\;-\;1\;$ & $\;q^{\,\prime}\;=\;0\;$ & $\;q^{\,\prime}\;=\;1\;$  & $\;q^{\,\prime}\;=\;2\;$  &\\
$~\;s\;$    & $\beta_{2200s}\;\;$  &                                                     &                                                     &                &           &\\
\hline
\hline
$-2~\quad$ & $\;\;n$ & $O(e^3{\cal{A}}_1^5)$ & $O(e^0{\cal{A}}_1^4)$ & $O(e{\cal{A}}_1^3)$ &  \textcolor{red}{$\frac{\textstyle 51}{\textstyle 16}\cos^2i\,e^2 {\cal{A}}_1^2\;$}  &\\
\hline
$~\;0\quad$ & $-\,n$ & $\;\,\;\frac{\textstyle 3}{\textstyle 8}\cos^2i\,e {\cal{A}}_1\;$  & \textcolor{red}{$-\,\frac{\textstyle 3}{\textstyle 2}\cos^2i\, (1-5e^2) {\cal{A}}_1^2\;$}  &
   \textcolor{red}{$\frac{\textstyle 21}{\textstyle 8}\cos^2i\,e {\cal{A}}_1\;$}                         & \textcolor{red}{$\frac{\textstyle 51}{\textstyle 16}\cos^2i\,e^2 {\cal{A}}_1^2\;$}   &\\
\hline
$~\;1\quad$ & $-\,2n$ & $-\,\frac{\textstyle 3}{\textstyle 8}\cos^2i\,e {\cal{A}}_1\;$ &   \textcolor{red}{$\frac{\textstyle 3}{\textstyle 4}\cos^2i\, (1-5e^2) {\cal{A}}_1^2$}                &         $O(e{\cal{A}}_1^3)$                  & $O(e^2{\cal{A}}_1^4)$  &\\
\hline
$~\;2\quad$ & $-\,3n$ &     $O(e{\cal{A}}_1^3)$                      &         $O(e^0{\cal{A}}_1^4)$                  &           $O(e{\cal{A}}_1^5)$                & $O(e^2{\cal{A}}_1^6)$  &\\
\hline
\hline
\end{tabular}
\end{center}
\caption{. Terms with $\,\left\{lmpq\right\}$ = $\left\{2200\right\}$. Only inputs  linear or quadratic in $\,{\cal{A}}_1\,$  are kept. The terms which do not mutually cancel are shown in red. The entry with $\,s=0\,$ and $\,q^{\,\prime}\,=\,0\,$ should contain also an $\,{\cal{A}}_1$-independent term $\;\frac{\textstyle 3}{\textstyle 4}\,(1-5e^2)\,(1-\sin^2i)\;$, but we have dropped this term here, because it has already been taken care of in Section \ref{be}, see the second term in equation (\ref{begin}). }
 \label{2200}
\end{table}

 \begin{itemize}
 \item[] {{Terms with $\,q\;=\;1\;$}} are given in Table \ref{2201}.
 % \vspace{2mm} ~\\
 For these terms, the modes are
 \ba
 \beta_{lmpqs}\,=\;\beta_{2201s}\,=\;(\,-\,1\,+\,q\,-\,s)\,n\,=\,-\,s\,n\,\;,
 \label{}
 \ea
 so the terms with $\,s=0\,$ can be omitted.
 \end{itemize}
\begin{table}[htbp]
\begin{center}
\begin{tabular}{llccccc}
\hline
\hline
  & & $\;q^{\,\prime}\;=\;-\;1\;$ & $\;q^{\,\prime}\;=\;0\;$ & $\;q^{\,\prime}\;=\;1\;$  & $\;q^{\,\prime}\;=\;2\;$ &\\
$~\;s\;$  &  $\beta_{2201s}\;\;$ &                                                         &                                                     &       &                    &\\
\hline
$-2~\quad$ & $\;\,2n$ & $O(e^2{\cal{A}}_1^6)$ & $O(e{\cal{A}}_1^5)$ & $O(e^2{\cal{A}}_1^4)$ & $O(e^3{\cal{A}}_1^3)$ &\\
\hline
$-1\quad$ & $\;\;\;n$ & $O(e^2{\cal{A}}_1^4)$  & $O(e{\cal{A}}_1^3)$  &  \textcolor{red}{$\;\frac{\textstyle 147}{\textstyle 16}\cos^2i\,e^2 {\cal{A}}_1^2\;$} & $O(e^3{\cal{A}}_1)$                         &\\
\hline
$~\;1\quad$ & $-\,n$ & \textcolor{red}{$\frac{\textstyle 21}{\textstyle 16}\cos^2i\,e^2 {\cal{A}}_1^2$} & ~\textcolor{red}{$\;\frac{\textstyle 21}{\textstyle 8}\cos^2i\,e {\cal{A}}_1\,$} & \textcolor{red}{$\;\frac{\textstyle 147}{\textstyle 16}\cos^2i\,e^2 {\cal{A}}_1^2\;$} & $O(e^3{\cal{A}}_1^3)$ &\\
\hline
$~\;2\quad$ & $-\,2n$ &   \textcolor{red}{$-\,\frac{\textstyle 21}{\textstyle 32}\cos^2i\,e^2 {\cal{A}}_1^2\;\;$}                        &         $O(e{\cal{A}}_1^3)$                  &           $O(e^2{\cal{A}}_1^4)$                & $O(e^3{\cal{A}}_1^5)$ &\\
\hline
\hline
\end{tabular}
\end{center}
\caption{. Terms with $\,\left\{lmpq\right\}$ = $\left\{2201\right\}$. Only inputs linear or quadratic in $\,{\cal{A}}_1\,$  are kept. The terms which do not mutually cancel are shown in red.}
 \label{2201}
\end{table}

 \begin{itemize}
 \item[] {{Terms with $\,q\;=\;2\;$}} are given in Table \ref{2202}.
 % \vspace{2mm} ~\\
 For these terms, the modes are
 \ba
 \beta_{lmpqs}\,=\;\beta_{2202s}\,=\;(\,-\,1\,+\,q\,-\,s)\,n\,=\,(1\,-\,s)\,n\,\;,
 \label{}
 \ea
 so the terms with $\,s=1\,$ can be omitted.
 \end{itemize}
\begin{table}[htbp]
\begin{center}
\begin{tabular}{llccccc}
\hline
\hline
  & & $\;q^{\,\prime}\;=\;-\;1\;$ & $\;q^{\,\prime}\;=\;0\;$ & $\;q^{\,\prime}\;=\;1\;$  & $\;q^{\,\prime}\;=\;2\;$  &\\
$~\;s\;$  &  $\beta_{2202s}\;\;$ &                                                         &                                                     &      &                     &\\
\hline
$-2~\quad$ & $\;\,3n$ & $O(e^3{\cal{A}}_1^7)$ & $O(e^2{\cal{A}}_1^6)$ & $O(e^3{\cal{A}}_1^5)$ & $O(e^4{\cal{A}}_1^4)$ &\\
\hline
$-1\quad$ & $\;\;2n$ & $O(e^3{\cal{A}}_1^5)$  & $O(e^2{\cal{A}}_1^4)$  &  $O(e^3{\cal{A}}_1^3)$                          &$O(e^4{\cal{A}}_1^2)$ &\\
\hline
$~\;0\quad$ & $\;\;\;n$ & $O(e^3{\cal{A}}_1^3)$ & ~\textcolor{red}{$\;\frac{\textstyle 51}{\textstyle 16}\cos^2i\,e^2 {\cal{A}}_1^2\,$} & $O(e^3{\cal{A}}_1)$ &$O(e^4{\cal{A}}_1^0)$ &\\
\hline
$~\;2\quad$ & $-\,n\;\;$ &   $O(e^3{\cal{A}}_1^3)$                      &         ~\textcolor{red}{$\;\frac{\textstyle 51}{\textstyle 16}\cos^2i\,e^2 {\cal{A}}_1^2\,$}                  &           $O(e^3{\cal{A}}_1^3)$                &$O(e^4{\cal{A}}_1^4)$ &\\
\hline
\hline
\end{tabular}
\end{center}
\caption{. Terms with $\,\left\{lmpq\right\}$ = $\left\{2202\right\}$. Only inputs linear or quadratic in $\,{\cal{A}}_1\,$  are kept. The terms which do not mutually cancel are shown in red.}
 \label{2202}
\end{table}

   %  \textcolor{red}{$\;\frac{\textstyle 147}{\textstyle 16}\cos^2i\,e^2 {\cal{A}}_1^2\;$}

 \subsection{The total power exerted by the gravitational tides,\\ under forced libration in the 3:2 spin state}

  When the body is Maxwell and the frequencies are not too low, the product $\,\beta_{lmpqs}\,k_l(\beta_{lmpqs})\,\sin\epsilon_l(\beta_{lmpqs})\,$ is constant (i.e., independent of $\,\beta_{lmpqs}\,$)~---~see equation (\ref{me}) in Appendix \ref{chi}. Moreover, for $\,z=3/2\,$ and $\,m=1\,$, none of the frequencies $\,\beta_{21pqs}\,$ vanish. So all products
  $\,\beta_{l1pqs}\,k_l(\beta_{l1pqs})\,\sin\epsilon_l(\beta_{l1pqs})\,$ assume the same value. In this situation, in the subsum of the expression (\ref{long}), which consists of the terms with $\,m=1\,$ only, we can separate out the sum $\;\sum_{s=-\infty}^{\infty} J_{q^{\,\prime}-q+s}(m{\cal{A}}_1)\,J_s(m{\cal{A}}_1)\;$ as a factor, and can use the equality
  \footnote{~To derive this equality, set $\,u=v\,$ in Neumann's Addition Theorem (Abramowitz \& Stegun 1972, Eqn 9.1.75~---~and also mind an important comment after that equation).}
  \ba
  \sum_{s=-\infty}^{\infty}J_{n+s}(m{\cal{A}}_1)\,J_{s}(m{\cal{A}}_1)\,=\,\delta_{n0}\quad\mbox{where}\quad n\,=\,q^{\,\prime}-q\,\;.
  \label{equality}
  \ea
  Then the said subsum of the terms with $\,m=1\,$ becomes simply
  \ba
  \nonumber
  \frac{G M^{*\,2}}{3\;a}\bigg( \frac{R}{a} \bigg)^{5}
     \sum_{l,p\,q,\,q^{\,\prime}}
  %  \sum_{l=2}^{\infty}  \sum_{q^{\prime}=-\infty}^{\infty}  \sum_{p=0}^l \sum_{q=-\infty}^{\infty}
  F_{l1p}^{\,2} (i)  G_{lpq^{\,\prime}}(e) G_{lpq}(e)
 %  ~\\ ~\\ \,\;\;
     \left[\sum_{s=-\infty}^{\infty} J_{(q^{\prime}-q+s)} ({\cal{A}}_1) J_s ({\cal{A}}_1)\right]
     \beta_{21pqs} \,k_l(\beta_{21pqs}) \sin\epsilon_l(\beta_{21pqs})
 %    \beta\,k_l(\beta)\,\sin\epsilon_l(\beta)
  \nonumber\\
  \nonumber\\
  =\;\frac{G M^{*\,2}}{3\;a}\,\bigg( \frac{R}{a} \bigg)^{5}  \sum_{l=2}^{\infty} \sum_{p=0}^l \sum_{q=-\infty}^{\infty} F_{l1p}^{\,2} (i)  \, G^2_{lpq}(e)\,\beta\,k_l(\beta)\,\sin\epsilon_l(\beta)
     \;\,.\qquad\qquad\qquad\qquad
\label{subsum}
\ea
  where the products $\,\beta\,k_l(\beta)\,\sin\epsilon_l(\beta)\,\equiv\,\beta_{l1pqs}\,k_l(\beta_{l1pqs})\,\sin\epsilon_l(\beta_{l1pqs})\,$ assume the same value for all terms. The important thing here is that, owing to the equality (\ref{equality}), the subsum (\ref{subsum}) bears no dependence on the libration magnitude $\,{\cal{A}}_1\,$. This is an exact result which is valid in any order of $\,e\,$ or $\,\sin i\,$.

  Thus we see that, under libration about the $\,z=3/2\,$ spin-orbit resonance at not too low a frequency, terms with $\,m=1\,$ make no input into the libration-generated part of the tidal power. This agrees well with our Tables \ref{210-1} - \ref{2111}: as can be seen from these tables, libration-caused terms with $\,m=1\,$ mutually cancel. We would emphasise again that this cancelation takes place owing to that fact that the product $\,\beta\,k_l(\beta)\,\sin\epsilon_l(\beta)\,$ takes the same value for all these terms. The $\,m=1\,$ terms would not mutually cancel for a different tidal model. Nor would they cancel for $\,m=2\,$, because for some of those terms the tidal modes $\,\beta\,$ vanish. This can be seen from Tables \ref{220-1} - \ref{2201}.

 Table \ref{220-1} contributes the term:
 \ba
 \frac{G\,{M^{\,\*}}^2}{a}\left(\frac{R}{a}\right)^5\,\frac{\textstyle 21}{\textstyle 32}\,\cos^2i\,e^2 {\cal{A}}_1^2\,\beta\,k_l(\beta)\,\sin\epsilon_l(\beta)
 % \,\approx\,\frac{G\,{M^{\,\*}}^2}{a}\left(\frac{R}{a}\right)^5\;\frac{\textstyle 21}{\textstyle 32}e^2 {\cal{A}}_1^2\,\beta\,k_l(\beta)\,\sin\epsilon_l(\beta)
 \label{}
 \ea

 Table \ref{2200} contributes the terms:
 \ba
 \nonumber
 && \frac{G\,{M^{\,\*}}^2}{a}\left(\frac{R}{a}\right)^5\,\frac{\textstyle 21}{\textstyle 8}\,\cos^2i\;e {\cal{A}}_1\,\beta\,k_l(\beta)\,\sin\epsilon_l(\beta)
 ~\\
  &+& \frac{G\,{M^{\,\*}}^2}{a}\left(\frac{R}{a}\right)^5\,\frac{\textstyle 51}{\textstyle 8}\,\cos^2i\;e^2 {\cal{A}}_1^2\,\beta\,k_l(\beta)\,\sin\epsilon_l(\beta)\label{}\\
 \nonumber
 &-& \frac{G\,{M^{\,\*}}^2}{a}\left(\frac{R}{a}\right)^5\,\frac{\textstyle 3}{\textstyle 4}\,\cos^2i\;(1-5e^2) {\cal{A}}_1^2\,\beta\,k_l(\beta)\,\sin\epsilon_l(\beta)
 \ea

 Table \ref{2201} contributes the terms:
 \ba
 \nonumber
 && \frac{G\,{M^{\,\*}}^2}{a}\left(\frac{R}{a}\right)^5\,\frac{\textstyle 21}{\textstyle 8}\,\cos^2 i\;e {\cal{A}}_1\,\beta\,k_l(\beta)\,\sin\epsilon_l(\beta)
 ~\\
 \label{}\\
 \nonumber
 &+&\frac{G\,{M^{\,\*}}^2}{a}\left(\frac{R}{a}\right)^5\left(\frac{\textstyle 21}{\textstyle 32}\,+\,\frac{147}{8} \right)\cos^2i\;e^2 {\cal{A}}_1^2\,\beta\,k_l(\beta)\,\sin\epsilon_l(\beta)
 \ea

  Table \ref{2202} contributes the term:
 \ba
 \frac{G\,{M^{\,\*}}^2}{a}\left(\frac{R}{a}\right)^5\,\frac{\textstyle 51}{\textstyle 8}\,\cos^2i\,e^2 {\cal{A}}_1^2\,\beta\,k_l(\beta)\,\sin\epsilon_l(\beta)
 % \,\approx\,\frac{G\,{M^{\,\*}}^2}{a}\left(\frac{R}{a}\right)^5\;\frac{\textstyle 21}{\textstyle 32}e^2 {\cal{A}}_1^2\,\beta\,k_l(\beta)\,\sin\epsilon_l(\beta)
 \,\;,
 \label{}
 \ea
 all these expressions being valid in the said approximation that the body is Maxwell and the frequencies are not too low.

 Then, within the said tidal model, the combined input into the power, generated by forced libration, is:
 \ba
 ^{(3:2)}P_{\rm{tide}}^{\rm(forced)}&=& \frac{G\,{M^{\,\*}}^2}{a}\left(\frac{R}{a}\right)^5\;\frac{3}{4}\,\cos^2i\,\left[\,7\,e\,{\cal{A}}_1\,-\,\left(1\,-\,\frac{\textstyle 159}{\textstyle 4}e^2\right) {\cal{A}}_1^2 \right]\,\beta\,k_l(\beta)\,\sin\epsilon_l(\beta)\,\;.
 \label{}
 \ea

 Summing this with the expression (\ref{baron}), we write down the total tidal power generated in the 3:2 spin-orbit resonance:
 \ba
 \nonumber
  && ^{(3:2)}P_{\rm{tide}}^{\rm(main)}\,+\,^{(3:2)}P_{\rm{tide}}^{\rm({obliquity})}\,+\,^{(3:2)}P_{\rm{tide}}^{\rm(forced)}\,=\\
 \nonumber\\
  && \;\; \qquad\qquad
  \frac{G{M^{*}}^{\,2}}{a}\left(\frac{R}{a}\right)^5\,\frac{3}{4}\,\left[\,\left(1\;-\;\frac{13}{4}\;e^2\right)\;+\;\left(1\;+\;
 \frac{61}{4}\;e^2\right)\;\sin^2 i
  % \,\right]\;\beta\,k_2(\beta)\,\sin\epsilon_2(\beta)
 \right. ~\\ \nonumber\\
 && \left.
 \qquad  \qquad\qquad
 \;+\;
  % \frac{G\,{M^{\,\*}}^2}{a}\left(\frac{R}{a}\right)^5\;\frac{3}{4}\left[\,
  7\,e\,{\cal{A}}_1\,\cos^2i\,-\,\left(1\,-\,\frac{\textstyle 193}{\textstyle 4}e^2\right) {\cal{A}}_1^2\,\cos^2i
   \right]\,\beta\,k_l(\beta)\,\sin\epsilon_l(\beta)
  \,\;,\qquad\;\qquad
  \nonumber
  \ea
 or, with the expression (\ref{manner}) substituted:
  \ba
 \nonumber
  && ^{(3:2)}P_{\rm{tide}}^{\rm(main)}\,+\,^{(3:2)}P_{\rm{tide}}^{\rm({obliquity})}\,+\,^{(3:2)}P_{\rm{tide}}^{\rm(forced)}\,=\\
 \nonumber\\
   &&
  \qquad\qquad
  \frac{1}{2\eta}\;R^7\,\rho^2\,n^4\;\left[\,\left(1\;-\;\frac{13}{4}\;e^2\right)\;+\;\left(1\;+\;
 \frac{61}{4}\;e^2\right)\;\sin^2 i
  % \,\right]\;\beta\,k_2(\beta)\,\sin\epsilon_2(\beta)
 \right.  \label{altogether}\\ \nonumber\\
 && \left.
 \qquad\qquad\qquad
 \;+\;
  % \frac{G\,{M^{\,\*}}^2}{a}\left(\frac{R}{a}\right)^5\;\frac{3}{4}\left[\,
  7\,e\,{\cal{A}}_1\,\cos^2i\,-\,\left(1\,-\,\frac{\textstyle 193}{\textstyle 4}e^2\right) {\cal{A}}_1^2\,\cos^2i
   \right]\,\beta\,k_l(\beta)\,\sin\epsilon_l(\beta)
  \,\;,\qquad\;\qquad
  \nonumber
  \ea
 where the osculating mean motion $\,\sqrt{G(M^*+M)/a^3\,}\approx\sqrt{GM^*/a^3\,}\,$ was approximated with its anomalistic counterpart
 $\,n\equiv\,\stackrel{\bf{\centerdot}}{\cal{M}\,}$.

 Also recall that everywhere in this section we implied $\,{\cal{A}}_1\,=\;^{(3:2)}{\cal{A}}_1\,$.

  \subsection{Addition owing to free libration\label{pizda}}

 Here $\,{\cal{A}}\,$ will stand for the magnitude of free libration with the frequency $\,\chi\,$.

 The calculation of power is analogous to that carried out above for forced libration, except that now we borrow from Tables \ref{210-1} -- \ref{2200}$\,$ only the terms with $\,q^{\,\prime} = q\,$, see Frouard \& Efroimsky (2017\,a). Summing up these terms, we obtain:
 \ba
 \nonumber
 ^{(3:2)}P_{\rm{tide}}^{\rm(free)}\,=\;\frac{G{M^{*}}^{\,2}}{a}\left(\frac{R}{a}\right)^5\,\frac{3}{4}\,\cos^2 i\;\left(1+\frac{39}{2}
 e^2\right)\,{\cal{A}}^2\,\beta\,k_l(\beta)\,\sin\epsilon_l(\beta)\,\;.\qquad\;
  \label{}
  \ea
 With the expression (\ref{manner}) substituted, this becomes:
 \ba
 ^{(3:2)}P_{\rm{tide}}^{\rm(free)}\,=\;
 \frac{1}{2\eta}\;R^7\,\rho^2\,n^4\,
 \left(1\,+\,\frac{39}{2}
 \,e^2\right)\,{\cal{A}}^2\,\cos^2i\;\;.\qquad
 \label{zaebalo}
 \ea

 \end{document}